\documentclass[11pt,epsf]{article}
\usepackage{hyperref}
\usepackage{graphics}
\usepackage{amssymb}
\usepackage{epsf}
\usepackage{rotating}
\usepackage{pstricks}
\usepackage{subfigure}

\def\l{\lambda}

\def\g{\gamma}

\def\ds#1{#1\kern-1ex\hbox{/}}
\def\dsh{h\kern-1.2ex /}

\newcommand{\bea}{\begin{eqnarray}}
\newcommand{\eea}{\end{eqnarray}}

\def\beq{\begin{equation}}
\def\eeq{\end{equation}}

\def\beqn{\begin{eqnarray}}
\def\eeqn{\end{eqnarray}}
\def\ba{\begin{eqnarray}}
\def\ea{\end{eqnarray}}

\newcommand{\beqa}{\begin{eqnarray}}
\newcommand{\eeqa}{\end{eqnarray}}

\begin{document}
\begin{center}
\vspace{1.cm}

{\bf \large
A Light Supersymmetric Axion in an Anomalous Abelian Extension of the Standard Model \\}

\vspace{1.5cm}
{  Claudio Corian\`{o}, Marco Guzzi, Antonio Mariano and Simone Morelli }

\vspace{1cm}

{\it Dipartimento di Fisica, Universit\`{a} del Salento \\
and  INFN Sezione di Lecce,  Via Arnesano 73100 Lecce, Italy\footnote{claudio.coriano@le.infn.it, marco.guzzi@le.infn.it, antonio.mariano@le.infn.it,
 simone.morelli@le.infn.it}
}\\
\vspace{.5cm}
%{\it $^b$ Department of Physics and Institute of Plasma Physics \\
%University of Crete, 71003 Heraklion, Greece}\\
\vspace{.5cm}

\begin{abstract}
We present a supersymmetric extension of the Standard Model (USSM-A) 
with an anomalous $U(1)$ and St\"uckelberg axions for anomaly cancellation, 
generalizing similar non-supersymmetric constructions.
The model, built by a bottom-up approach, is expected to capture the low-energy 
supersymmetric description of axionic symmetries in theories with gauged anomalous 
abelian interactions, previously explored in the non-supersymmetric case for scenarios 
with intersecting branes. The choice of a USSM-like superpotential, with one extra singlet 
superfield and an extra abelian symmetry, allows a physical axion-like particle in the spectrum. 
We describe some general features of this construction and in particular the modification of 
the dark-matter sector which involves both the axion and several neutralinos with an axino component. 
The axion is expected to be very light in the absence of phases in the superpotential but could 
acquire a mass which can also be in the few GeV range or larger. In particular, the gauging of 
the anomalous symmetry allows independent mass/coupling interaction to the gauge fields 
of this particle, a feature which is absent in traditional (invisible) axion models. 
We comment on the general implications
of our study for the signature of moduli from string theory due to 
the presence of these anomalous symmetries.
\end{abstract}
\end{center}
\newpage
%\tableofcontents

\section{Introduction}
Extensions of the Standard Model (SM)  describing axion-like particles 
- and with supersymmetry as a basic low energy scenario - are an 
interesting area of investigation which has the potentiality to provide an answer to a series of unsolved theoretical issues. Among them are those concerning the possible presence of anomalous extra neutral gauge interactions at current and future colliders in some special channels, especially in the search for an anomalous 
extra $Z^\prime$. This investigation could also clarify 
the role of weakly coupled pseudoscalars in the early universe.
For this reason several studies addressing the experimental detection of pseudoscalars  at future experiments \cite{Ahlers:2007qf,Ahlers:2007rd,DeAngelis:2007yu,DeAngelis:2008sk,DeAngelis:2007dy,Berezhiani:2000gh}, has received an impressive impulse in the recent literature.

One of the distinctive features of these extensions is the 
presence of extra abelian interactions which are anomalous.
We just recall that anomalous $U(1)$'s are quite common in several string 
constructions and that the mechanism of anomaly cancellation, if realized 
at low energy by a Wess-Zumino counterterm (WZ), may cause the presence of 
a physical axion in the spectrum. This result points directly towards the 
possibility of having a new dark matter candidate (see also \cite{Khlopov}), which is certainly one
 of the most appealing features of this class of theories \cite{Coriano:2005js}.

One of the first successful
realization of the non-supersymmetric version of these models comes from special 
vacua of string/brane theory (orientifold vacua), in the form of stacks 
of intersecting branes, which induce a gauge structure given by the product 
of $U(N)\sim SU(N)\times U(1)$ factors, where $N$ is the number of branes of 
each stack (see \cite{Kiritsis:2003mc} for an overview). Among the $U(1)$ factors, one of them is identified with the 
SM hypercharge ($U(1)_Y$), while the remaining ones are anomalous and involve 
St\"uckelberg axions for anomaly cancellation. 
 In effective string models the abelian structure is in general characterized by the presence of several $U(1)$ factors, described in the hypercharge basis by direct products of the form 
$G_1\equiv U(1)_Y\times U(1)_1\times ... \times U(1)_p$, with an anomaly-free hypercharge generator and $p$ 
anomalous $U(1)$'s which are accompanied by axions $b_i$, with $i=1,2,...p$. The anomalous $U(1)$'s in this construction are in a broken phase, called the "St\"uckelberg phase". 
In particular, after electroweak 
symmetry breaking (EWSB), one of the axions becomes physical \cite{Coriano:2005js} and is characterized by independent mass/coupling relations, where the coupling appears in an ordinary $b F\tilde{F}$ interaction with the gauge fields, providing a generalization of the Peccei-Quinn (PQ) axion. One shortcoming of this description, at this time, is the absence of a supersymmetric extension of it with the appearance of a physical 
axion. The generalization to the supersymmetric case of these theories is 
interesting on several grounds. For instance, it allows to study an 
entire new class of extensions of the MSSM in the presence of a gauging 
of the axionic symmetries \cite{Anastasopoulos:2008jt}  and, at the same time, represents an intermediate step toward the unification with gravity of the same models, within 
certain formulations of supergravity \cite{DeRydt:2007vg,Zagermann:2008gy}. The formulation of \cite{Anastasopoulos:2008jt}, which is specific for a MSSM superpotential parallels a previous general study of the same topic contained in 
\cite{DeRydt:2007vg}. 

Therefore, these types of 
constructions provide a consistent framework for the study of the effects of moduli of 
string/brane theory within scenarios with large extra dimensions or via supergravity, together with their low energy implications in cosmology and in collider physics \cite{Armillis:2008vp}. 
Recently, an extension of the MSSM containing 
an anomalous $U(1)$, made massive by a St\"uckelberg supermultiplet \cite{Kors:2004ri} has been 
introduced in \cite{Anastasopoulos:2008jt}. This has been based on the superpotential of the MSSM with an extra abelian symmetry. One of the features 
of this construction is the absence of a Higgs-axion mixing, since the bosonic 
component of the St\"uckelberg multiplet remains an ordinary goldstone mode. 
Therefore, the final theory is characterized by a physical axino but not by a physical axion. 
The objective of our analysis is to show that a similar construction 
can be performed in more general ways, thereby generating a model 
with a physical axion-like particle. This provides a complete supersymmetric generalization of the (gauged) PQ axion.  We will work out the requirements that are needed in order to make this extension possible, detailing 
some of the arguments that have been presented in short form in \cite{Coriano:2008aw} and 
analyzing the main features of the effective action of such a theory, that we call the USSM-A due to the anomalous 
$U(1)$ (A) and to the specific choice of the USSM superpotential.

Our work is organized as follows. We briefly describe the class of models that we are going to investigate, outlining their basic structure, together with their supersymmetric generalizations.
Along the way, we will underline the differences between our construction and the previous construction of  \cite{Anastasopoulos:2008jt}. We show how a physical axion is bound to appear in the spectrum and describe all the sectors of this theory. We derive the corresponding generalized Ward identities and characterize the Chern-Simons interactions of this class of models bringing up one typical example of application. We study the neutralino sector of the model and present a brief numerical analysis of its spectrum.  Most of our attention in this work focuses on the basic characterization of this model, stressing on the mechanism that allows a physical axion in the spectrum. We conclude with some comments on possible extensions of this analysis to more general potentials characterized by moduli in different scenarios derived from string theory.

\section{Supersymmetric Extensions of the Standard Model and extra $U(1)$'s} 

Abelian (anomaly-free) supersymmetric extensions of the SM have been discussed
in several previous works \cite{Kim:1983dt,Jain:1994hd,Nir:1995bu,Suematsu:1994qm,Cvetic:1997ky,Cvetic:1995rj}. 
In \cite{Cvetic:1997ky} the authors explore an extension
of the Minimal Supersymmetric SM (MSSM) with an extra SM singlet chiral superfield
$\hat{S}$, with chiral charges chosen so to allow trilinear couplings of $\hat{S}$
to the two Higgs doublets $\hat{H}_{1},\hat{H}_{2}$ in the superpotential.
The $\mu$ term, in this case, is generated by the vev of the scalar component of
$\hat{S}$, precisely by the $\hat{S}\hat{H}_1\cdot \hat{H}_2$ interaction.
The structure of this model, usually called USSM, shares some similarity with the nearly-Minimal
Supersymmetric SM (nMSSM) \cite{Balazs:2007pf} and the next-to-Minimal Supersymmetric SM (NMSSM) \cite{Ellwanger:2000fh}. In all of these three models the extra scalar $\hat{S}$ is introduced for the same purpose but in the nMSSM and NMSSM this field is a singlet under the complete gauge group (which is the same as the SM) while in the USSM the field is charged under the extra $U(1)$. We recall that the nMSSM and the NMSSM differ at the level of the superpotential 
in the structure of the pure $\hat{S}$ contribution, which is either linear (nMSSM) or cubic (NMSSM). 
 
In the approach of \cite{Cvetic:1997ky} this appears to be a necessary requirement
since a scalar superfield, singlet under the complete 
$SU(3)\times SU(2) \times U(1)_Y\times U(1)_{B}$ gauge group, 
while solving the $\mu$ problem,  however, does not allow a consistent 
pattern of EWSB, leaving the extra 
$Z^\prime$  of the neutral sector massless. This construction 
is realized with an anomaly-free chiral spectrum.

\subsection{MSSM and USSM with an anomalous $U(1)'$}

In \cite{Anastasopoulos:2008jt} the authors investigate a supersymmetric 
extension of the SM with an extra $U(1)$, based on the superpotential of the 
MSSM. They make an important step forward in the analysis of this class of 
theories, using a bottom-up approach, that is by  1) fixing the effective 
action of the anomalous abelian symmetry using the St\"uckelberg supermultiplet 
to give mass to the anomalous gauge boson and 2 ) using Wess-Zumino counterterms 
to balance the mixed and cubic $U(1)_{B}$ anomalies of the theory. A third element 
of the construction is the possible presence of Chern-Simons interactions
\cite{Coriano:2005js} which find their way to low energy from string theory
\cite{Anastasopoulos:2006cz}, and which amount to a re-distribution of the
anomaly starting from a symmetric distribution on each
leg of the anomaly vertex. This re-distribution is allowed whenever the
symmetry of the vertex does not allow to uniquely define the breaking of
the Ward identities separately on each of its legs.  The meaning of this
freedom, from the point of view of the effective field theory,
is that each model allows a set of
additional (defining) Ward identities for the distribution of the anomaly
which are a specific feature of anomalous models
in which the trilinear gauge interactions are not identically zero (in the
massless fermion phase, the chiral phase).

In the first supersymmetric version of these models \cite{Anastasopoulos:2008jt},
the ordinary MSSM Lagrangean is naturally extended by the 
St\"uckelberg multiplet which provides a kinetic term
for the same multiplet while rendering the extra $Z'$ massive. 
The defining phase of the model is, therefore, the St\"uckelberg phase. 
In this construction the bosonic partner of the axino, which is the 
fermionic component of the multiplet, remains a goldstone mode after 
EWSB and is therefore unphysical.

\subsection{Inducing Higgs-axion mixing}

At the origin of the physical axion is the mechanism of Higgs-axion mixing. For this to take place one 
needs a Higgs sector which is charged under the anomalous $U(1)_{B}$ so that the mass of the anomalous gauge boson comes from a combination of the Higgs and St\"uckelberg mechanisms. In the case of the 
MSSM this mixing does not occur even if the two Higgses are charged under the anomalous U(1). The presence of a $\mu$ term in the superpotential forces the two charges of the two Higgs doublets to take opposite values, thereby guaranteeing also the cancellation of the extra anomalies due to the circulating higgsinos, but is not enough to give mass to the anomalous gauge boson. In other words, in the absence of a St\"uckelberg multiplet the mass matrix of the gauge boson has still an additional null eigenvalue. The true mechanism of mass generation of the anomalous $Z^{\prime}$, therefore, is just the St\"uckelberg, which in this situation is a goldstone mode. In fact, one reobtains a massive Yang-Mills theory just by going to the unitary gauge and eliminating the axion.

\section{The structure of the model}

A simple way out in order to have Higgs-axion mixing and a light axion 
in the physical spectrum consists to
use a modified superpotential as in \cite{Cvetic:1997ky}, but now with an 
anomalous gauge structure, and to combine it with the Lagrangean of the 
St\"uckelberg supermultiplet. In other words, we move from the superpotential 
of MSSM-type to the one typical of the USSM, introducing an extra scalar 
superfield $\hat{S}$ which is non-singlet under an extra $U(1)_{B}$, 
maintaining the anomalous structure induced by the extra neutral current. 
This specific assumption allows to 
remove the second massless eigenvalue in the mass matrix of the gauge bosons 
and allows to induce Higgs-axion mixing once that the St\"uckelberg mechanism 
is invoked to contribute to the mass of the extra $Z^{\prime}$. The conditions 
that need to be verified in order to have a physical axion in the spectrum are 
obtained from an analysis of the CP-odd sector of the theory and involve both 
the potential and the derivative couplings (mixings) of the 
massive gauge bosons with their goldstones
$(Z_I \partial G_{Z_I})$ extracted from the broken phase. In general, 
the presence of extra singlet superfields in the superpotential 
allows such a mixing and we will illustrate this requirement 
in one of the sections below. The analysis that we will present in the next 
sections has the goal to clarify this point, starting from the MSSM case, where 
none of the CP-odd states acquires an axion-like coupling.

These new features do not affect the chargino sector with respect to the MSSM.

\section{The superpotential}

The construction of models characterized by a physical 
axion in their spectrum requires an appropriate superpotential. 
In order to obtain this, we consider 
the introduction of an extra SM singlet $\hat{S}$.
For this reason, the superpotential of the model investigated is given by
\begin{eqnarray}
{\cal W}&=&\lambda\hat{S}\hat{H}_{1}\cdot\hat{H}_{2}+y_{e}\hat{H}_{1}
\cdot\hat{L}\hat{R}+y_{d}\hat{H}_{1}\cdot\hat{Q}\hat{D}_{R}
+y_{u}\hat{H}_{2}\cdot\hat{Q}\hat{U}_{R},
\end{eqnarray}
which coincides with the model of \cite{Cvetic:1997ky}, called the USSM. 
We refer to Table \ref{tab01} for a list of the charge 
assignment of the chiral superfields of our model; the scalar superfields
corresponding to $SU(3)$, $SU(2)$, $U(1)_{Y}$ and $U(1)_{B}$ are, respectively, 
$\hat{G}^{a}(x,\theta,\bar{\theta})$ (with a=1,2\dots,8),
$\hat{W}^{i}(x,\theta,\bar{\theta})$ (with i=1,2,3),$\hat{Y}(x,\theta,\bar{\theta})$
and $\hat{B}(x,\theta,\bar{\theta})$ and they fall in the usual adjoint representations 
of the gauge group factors.
 
We have denoted the charges by $Q_{f,X}$, where $X$ denotes the hypercharge (Y), 
the charged $W^{\pm}$ bosons (W), the non abelian gluons ($G$) and the anomalous 
gauge boson ($B$). At the same time we denote with $B_X$ the charges of the $X$ 
superfield respect to the anomalous $U(1)$.
Unlike the NMSSM and the nMSSM, ${\cal W}$ does not contain
linear and cubic terms in $\hat{S}$ in order to
preserve the gauge invariance in the presence of a non vanishing 
$B_S$ charge. This requirement is strictly necessary if the 
extra scalar $\hat{S}$ is only a SM singlet. 
Gauge invariance gives the conditions 
\begin{table}
\begin{center}
\begin{tabular}{|l||c|c|c|c|}
\hline
Superfields &SU(3)& SU(2)& $U(1)_{Y}$ & $U(1)_{B}$\\
\hline
$\hat{\bf b}(x,\theta,\bar{\theta})$& {\bf 1} & {\bf 1} & 0 & $ -- $\\
$\hat{S}(x,\theta,\bar{\theta})$& {\bf 1} & {\bf 1} & 0 & $B_{S}$\\
$\hat{L}(x,\theta,\bar{\theta})$& {\bf 1} & {\bf 2} & -1/2 & $B_{L}$\\
$\hat{R}(x,\theta,\bar{\theta})$& {\bf 1} & {\bf 1} & 1 & $B_{R}$\\
$\hat{Q}(x,\theta,\bar{\theta})$& {\bf 3} & {\bf 2} & 1/6 & $B_{Q}$\\
$\hat{U}_{R}(x,\theta,\bar{\theta})$& $\bar{{\bf 3}}$ & {\bf 1} & -2/3 & $B_{U_R}$\\
$\hat{D}_{R}(x,\theta,\bar{\theta})$& $\bar{{\bf 3}}$ & {\bf 1} & +1/3 & $B_{D_R}$\\
$\hat{H}_{1}(x,\theta,\bar{\theta})$& {\bf 1} & {\bf 2} & -1/2 & $B_{H_{1}}$\\
$\hat{H}_{2}(x,\theta,\bar{\theta})$& {\bf 1} & {\bf 2} & 1/2 & $B_{H_{2}}$\\
\hline
\end{tabular}
\end{center}
\caption{Charge assignment of the model; boldface numbers indicate the dimensions of the corresponding representations.}
\label{tab01}
\end{table}
\begin{eqnarray}
B_{H_{1}}+B_{H_{2}}+B_{S}&=&0\nonumber\\
B_{H_{1}}+B_{L}+B_{R}&=&0\nonumber\\
B_{H_{1}}+B_{Q}+B_{D_R}&=&0\nonumber\\
B_{H_{2}}+B_{Q}+B_{U_R}&=&0,
\label{condcariche}
\end{eqnarray}
which will be used below. It is not hard to show that the possibility of declaring $\hat{S}$ to be a singlet under the entire gauge group ($B_S=0$) $SU(3)\times SU(2)\times G_1$ leaves an extra gauge boson massless beside the photon,  after EWSB and as such it is not acceptable.

\subsection{Anomaly cancellation: defining the model}
We start by identifying the anomalous contributions of the 
model, whose gauge structure is of the form $SU(3)\times SU(2) \times U(1)_Y\times U(1)_B$.

The anomalous trilinear gauge interactions are all 
the ones involving the extra anomalous $U(1)_{B}$, namely
$\left\{U(1)_{B},U(1)_{B},U(1)_{B}\right\}$, $\left\{U(1)_{B},U(1)_{Y},U(1)_{Y}\right\}$,
$\left\{U(1)_{B},U(1)_{B},U(1)_{Y}\right\}$, $\left\{U(1)_{B},SU(2),SU(2)\right\}$,
$\left\{U(1)_{B},SU(3),SU(3)\right\}$.
In terms of the charges we can write each sector as follows
\ba
&&{\mathcal A}_{BBB} = \sum_f Q_{f,B}^3
\nonumber\\
&&{\mathcal A}_{BYY} = \sum_f Q_{f,B}\,Q_{f,Y}^2
\nonumber\\
&&{\mathcal A}_{BBY} = \sum_f Q_{f,B}^2\,Q_{f,Y}
\nonumber\\
&&{\mathcal A}_{BWW} = \sum_f Q_{f,B}\textrm{Tr}\left[\tau^{i} \tau^{j}\right]
\nonumber\\
&&{\mathcal A}_{BGG} = \sum_f Q_{f,B}\textrm{Tr}\left[T^{a} T^{b}\right],
\ea 
where $T^a$ are the generators of $SU(3)$ and $\tau^i$ the Pauli matrices. Compared to the analysis of \cite{Anastasopoulos:2008jt}, here we have anomalous trilinear interactions 
also in the sector involving the $SU(3)$ mixed anomaly due to the non vanishing charge $B_S$. 
Using the constraints coming from the Yukawa couplings and the conditions of gauge invariance, 
the expressions of the anomalies take the form
\ba
{\cal A}_{BBB}&=&3 (6 B_Q^3 + 3 B_{U_R}^3 + 3 B_{D_R}^3) + 
(6 B_L^3 + 3 B_{R}^3) + (2 B_{H_1}^3 + 2 B_{H_2}^3 + B_S^3)
\nonumber\\
&&=-3B_{H_1}^3 - 3(3 B_L + 18 B_Q - 7 B_S) B_{H_1}^2 - 3 
(3 B_L^2 +(18 B_Q - 7 B_S) B_S ) B_{H_1} 
\nonumber\\
&&\hspace{0.5cm}
+ 3 B_L^3 + B_S (27 B_Q^2 - 27 B_S B_Q + 8 B_S^2) 
\nonumber\\
{\cal A}_{BYY}&=& 
3 (6 B_Q Y_Q^2 + 3 B_{U_R} Y_{U_R}^2 + 3 B_{D_R} Y_{D_R}^2) + 
(6 B_L Y_L^2 + 3 B_{R} Y_{R}^2)
\nonumber\\ 
&&+ (2 B_{H_1} + 2 B_{H_2})Y_{H_1}^2
\nonumber\\ 
&&=\frac{1}{2}(-3 B_L - 9 B_Q + 7 B_S)
\nonumber\\ 
{\cal A}_{BBY}&=&
3 (6 B_Q^2 Y_Q + 3 B_{U_R}^2 Y_{U_R} + 3 B_{D_R}^2 Y_{D_R}) + 
(6 B_L^2 Y_L + 3 B_{R}^2 Y_{R})
\nonumber\\ 
&&+ (2 B_{H_1}^2 - 2 B_{H_2}^2)Y_{H_1}
\nonumber\\ 
&&=2 B_{H_1} (3 B_L + 9 B_Q - 5 B_S) + (12 B_Q - 5 B_S) B_S
\nonumber\\ 
{\cal A}_{BWW}&=&\frac{1}{2}
(18 B_Q + 6 B_L + 2 B_{H_1} + 2 B_{H_2})
=3 B_L + 9 B_Q - B_S
\nonumber\\
{\cal A}_{BGG}&=&
\frac{1}{2}(6 B_Q + 3 B_{U_R} + 3 B_{D_R})
=\frac{3}{2} B_S,
\label{anomalies1}
\ea
where $Y_Q, Y_L$ are the hypercharges of the left-handed doublets of the quarks and leptons respectively,
while $Y_{U_R}, Y_{D_R}, Y_R$ are the hypercharges of the $\hat{U}_R,\hat{D}_R,\hat{R}$ superfields
which correspond to the hypercharges of the right-handed quarks and leptons, with the opposite sign.

In the absence of a specific charge assignment coming from 
a string (or other) construction, these equations can be interpreted as defining 
conditions of a specific model. The role of string theory or of any other construction 
is to fix the charges, but for the rest the basic structure 
remains determined by the approach outlined below, and as such is truly general. 

\section{The St\"uckelberg multiplet}
In supersymmetric models the cancellation of the anomaly using the Wess-Zumino (WZ) counterterm can be obtained by the introduction of a St\"uckelberg supermultiplet, associated with the extra $U(1)$. The multiplet contributes to the supersymmetric version of the St\"uckelberg mechanism \cite{Kors:2004ri} and in the WZ interaction that describes the coupling of the supermultiplet to the gauge supermultiplet. We recall that in anomaly-free theories the St\"uckelberg mechanism has the feature of contributing to the mass of the anomalous gauge boson, eventually also in combination with the Higgs sector \cite{Feldman:2006wb,Feldman:2007wj}. 
This construction holds both in the non-supersymmetric and in the supersymmetric case.

Obviously, the presence of a mixing between the Higgs and St\"uckelberg components 
in the potential of more generic models 
in an anomaly-free theory, produces a new CP-odd component in the scalar sector, 
but deprived of axion-like couplings. 
On the contrary, these couplings appear in the case in 
which the two mechanisms (the Higgs and the St\"uckelberg) 
involve an anomalous $U(1)$, due to the presence of Wess-Zumino 
terms, for specific superpotentials. These interactions are 
induced in the effective action by the mechanism of anomaly cancellation.

The Lagrangean describing the St\"uckelberg supermultiplet is given by \cite{Kors:2004ri}
\begin{eqnarray}
{\cal L}_{st}=\int d^{4}\theta\left[2 M_{st}\hat{B}+\hat{{\bf b}}+\hat{{\bf b}}^{\dagger} \right]^{2}
\end{eqnarray}
where $\hat{B}$ is the abelian scalar superfield associated to the extra $U(1)_{B}$, 
$\hat{{\bf b}}$ is a left-chiral superfield and $M_{st}$ is the St\"uckelberg mass.

The former Lagrangean is invariant under the following gauge transformations
\begin{eqnarray}
\hat{B}&\rightarrow&\hat{B'}+i\left(\hat{\Lambda}-\hat{\Lambda}^{\dagger} \right)\nonumber\\
\hat{{\bf b}}&\rightarrow&\hat{{\bf b}}'-i 2 M_{st}\hat{\Lambda}
\label{tdigaugegen}
\end{eqnarray}
where $\hat{\Lambda}$ is a generic left-chiral superfield.
Introducing the component fields expansion we obtain
\begin{eqnarray}
\hat{B}&=&-\theta\sigma^{\mu}\bar{\theta}B_{\mu}
+i \theta\theta\bar{\theta}\bar{\lambda}_{B}
-i\bar{\theta}\bar{\theta}\theta\lambda_{B}
+\frac{1}{2}\theta\theta\bar{\theta}\bar{\theta}D_{B}\\
\hat{{\bf b}}&=& b + i\sqrt{2}\theta\psi_{{\bf b}}
-i\theta\sigma^{\mu}\bar{\theta}\partial_{\mu} b
+\frac{\sqrt{2}}{2}\theta\theta\bar{\theta}\bar{\sigma}^{\mu}\partial_{\mu}\psi_{{\bf b}}
-\frac{1}{4}\theta\theta\bar{\theta}\bar{\theta}\Box b+\theta\theta F_{{\bf b}},
\end{eqnarray}
where $B_{\mu}$ is the St\"uckelberg field, $\lambda_B,\bar{\lambda}_B$ are respectively
the left- and right-handed St\"uckelberg gauginos, $D_B$ is the corresponding $D$-term
for the gauge supermultiplet of $B_\mu$ , $b$ is a complex scalar field, $\psi_{\bf b}$ is the
supersymmetric axion (axino) and $F_{\bf b}$ is the F-term of $\hat{{\bf b}}$.

After the integration over the Grassman variables the Lagrangean density is given by
\begin{eqnarray}
{\cal L}_{st}&=&2\left(\partial_{\mu}~\textrm{Im}~b+M_{st}B_{\mu} \right)^{2}
+i\psi_{{\bf b}}\sigma^{\mu}\partial_{\mu}\bar{\psi}_{{\bf b}}
+i\bar{\psi}_{{\bf b}}\bar{\sigma}^{\mu}\partial_{\mu}\psi_{{\bf b}}
+2F_{{\bf b}}F_{{\bf b}}^{\dagger}+4 M_{st}~\textrm{Re}~b~D_{B}\nonumber\\
&&-2\sqrt{2}M_{st}\left(\psi_{{\bf b}}\lambda_{B}+h.c. \right),
\end{eqnarray}
where the auxiliary fields $F_{\bf b}$ and $D_B$ will be defined in the next sections.

\subsection{The axion Lagrangean}

The axion Lagrangean contains the St\"uckelberg gauge-invariant terms introduced above and the Wess-Zumino 
interactions for the anomaly cancellation and it is given by
\begin{eqnarray}
{\cal L}_{axion}&=&\frac{1}{4}\int d^{4}\theta (\hat{{\bf b}}
+\hat{{\bf b}}^{\dagger}+2M_{st}\hat{B})^{2}-\frac{1}{2}\int d^{4}\theta\left\lbrace\left[
\frac{1}{2} b_{G} \,\textrm{Tr}({\cal G} {\cal G})\hat{{\bf b}}
+\frac{1}{2} b_{W} \,\textrm{Tr}(W W)\hat{{\bf b}}
\right.\right.
\nonumber\\
&&\left.\left.+b_{Y}\hat{{\bf b}}W^{Y}_{\alpha} W^{Y,\alpha}
+b_{B}\hat{{\bf b}}W^{B}_{\alpha}W^{B,\alpha}+b_{YB}\hat{{\bf b}}W^{Y}_{\alpha}W^{B,\alpha}\right]
\delta(\bar{\theta}^{2})
+h.c.\right\rbrace, \nonumber\\
\end{eqnarray}
where we have denoted with $\cal{G}$ the field-strength of $SU(3)_c$, 
with $W$ the supersymmetric field-strength of $SU(2)$, with
$W^{Y}$ and with $W^{B}$ the supersymmetric field-strength of $U(1)_{Y}$ and $U(1)_{B}$ respectively.
The factors in front of the WZ counterterms ($b_X$) are determined by the standard conditions of anomaly cancellation. The Lagrangean, in our case, contains extra WZ counterterms respect to \cite{Anastasopoulos:2008jt}, in particular we need to impose the cancellation of the mixed $B-SU(3)-SU(3)$ anomaly, which is now non-vanishing due to the charges of the two higgsinos in the model, which are not opposite. In the MSSM 
this cancellation is identical, due to the specific color charges of the fermions in each generation. 
This implies that in our case the effective action contains 
both a $b G G $ interaction of the axion with the gluons and 
a vertex involving the corresponding gauginos (gluinos).

Expanding the ${\cal L}_{axion}$ in the component fields  we obtain
\begin{eqnarray}
{\cal L}_{axion}&=&\frac{1}{2}\left( \partial_{\mu}\textrm{Im}\,b + M_{st}B_{\mu}\right)^{2}
+\frac{i}{4}\psi_{\bf b}\sigma^{\mu}\partial_{\mu}\bar{\psi_{\bf b}}
+\frac{i}{4}\bar{\psi_{\bf b}}\bar{\sigma}^{\mu}\partial_{\mu}\psi_{\bf b}
+\frac{1}{2}F_{\bf b}F_{\bf b}^{\dagger}-\frac{M_{st}}{\sqrt{2}}(\psi_{\bf b}\lambda_{B}+h.c.)
\nonumber\\
&&-\frac{1}{16}b_{G}\,\epsilon^{\mu\nu\rho\sigma} G^a_{\mu\nu} G^{a}_{\rho\sigma}\textrm{Im}\,b
-\frac{1}{16}b_{W}\,\epsilon^{\mu\nu\rho\sigma}W^{i}_{\mu\nu}W^{i}_{\rho\sigma}\textrm{Im}\,b
-\frac{1}{4}b_{Y}\epsilon^{\mu\nu\rho\sigma}F^{Y}_{\mu\nu}F^{Y}_{\rho\sigma}\textrm{Im}\,b
\nonumber\\
&&-\frac{1}{4}b_{B}\epsilon^{\mu\nu\rho\sigma}F^B_{\mu\nu}F^B_{\rho\sigma}\textrm{Im}\,b
-\frac{1}{4}b_{YB}\epsilon^{\mu\nu\rho\sigma}F^{Y}_{\mu\nu}F^B_{\rho\sigma}\textrm{Im}\,b
\nonumber\\
&&+\frac{1}{4}b_{W}[\frac{1}{4}\textrm{Im}\,b \lambda_{W^i}\sigma^{\mu}D_{\mu}\bar{\lambda}_{W^i}
-\frac{i}{4\sqrt{2}}\psi_{\bf b} \lambda_{W^i}\sigma^{\mu}\bar{\sigma}^{\nu}W^{i}_{\mu\nu}
+\frac{1}{4}F_{\bf b}\lambda_{W^i}\lambda_{W^i}+\frac{1}{2\sqrt{2}}\psi_{\bf b}\lambda_{W^i}D^{i}
+h.c.]
\nonumber\\
&&+\frac{1}{4}b_{G}[\frac{1}{4}\textrm{Im}\,b~\lambda_{g^a}\sigma^{\mu}D_{\mu}\bar{\lambda}_{g^a}
-\frac{i}{4\sqrt{2}}\psi_{\bf b}~\lambda_{g^a}\sigma^{\mu}\bar{\sigma}^{\nu} G^{a}_{\mu\nu}
+\frac{1}{4}F_{\bf b}~\lambda_{g^a}~\lambda_{g^a}+\frac{1}{2\sqrt{2}}\psi_{\bf b}~\lambda_{g^a}D^{a}
+h.c.]
\nonumber\\
&&+b_{Y}[\textrm{Im}\,b \lambda_{Y}\sigma^{\mu}D_{\mu}\bar{\lambda}_{Y}
-\frac{i}{2\sqrt{2}}\psi_{\bf b} \lambda_{Y}\sigma^{\mu}\bar{\sigma}^{\nu}F^{Y}_{\mu\nu}
+\frac{1}{2}F_{\bf b}\lambda_{Y}\lambda_{Y}+\frac{1}{\sqrt{2}}\psi_{\bf b}\lambda_{Y}D_{Y}
+h.c.]
\nonumber\\
&&+b_{B}[\textrm{Im}\,b \lambda_{B}\sigma^{\mu}D_{\mu}\bar{\lambda}_{B}
-\frac{i}{2\sqrt{2}}\psi_{\bf b} \lambda_{B}\sigma^{\mu}\bar{\sigma}^{\nu}F^B_{\mu\nu}
+\frac{1}{2}F_{\bf b}\lambda_{B}\lambda_{B}+\frac{1}{\sqrt{2}}\psi_{\bf b}\lambda_{B}D_{B}
+h.c.]
\nonumber\\
&&+b_{YB}[(\textrm{Im}\,b\lambda_{Y}\sigma^{\mu}\partial_{\mu}\bar{\lambda}_{B}
+\frac{1}{2}F_{\bf b}\lambda_{Y}\lambda_{B}+\frac{1}{\sqrt{2}}\psi_{\bf b}\lambda_{Y}D_{B}
-\frac{i}{2\sqrt{2}}\lambda_{Y}\sigma^{\mu}\bar{\sigma}^{\nu}F^B_{\mu\nu}\psi_{\bf b})
\nonumber\\
&&+(Y\leftrightarrow B)+h.c.],
\label{Lassionecomp}
\end{eqnarray}
with the  $F$ and $D$ terms given by
\ba
&&F_{\bf b}=-(b_{G}\bar{\lambda}_{g^a}\bar{\lambda}_{g^a}+b_{W}\bar{\lambda}_{W^i}\bar{\lambda}_{W^i}
+b_{Y}\bar{\lambda}_{Y}\bar{\lambda}_{Y}+b_{B}\bar{\lambda_{B}}\bar{\lambda}_{B}
+b_{YB}\bar{\lambda}_{Y}\bar{\lambda}_{B}),
\nonumber\\
&&D_B=-[\frac{g_B}{2\sqrt{2}}(B_{L}\tilde{L}^{\dagger}\tilde{L}+B_{R}\tilde{R}^{\dagger}\tilde{R}
+B_{Q}\tilde{Q}^{\dagger}\tilde{Q}+B_{U}\tilde{U}_{R}^{\dagger}\tilde{U}_{R}
+B_{D}\tilde{D}_{R}^{\dagger}\tilde{D}_{R}
\nonumber\\
&&\hspace{1cm}+B_{H_{1}}H_{1}^{\dagger}H_{1}
+B_{H_{2}}H_{2}^{\dagger}H_{2}+B_{S}S^{\dagger}S)+\frac{1}{2}\psi_{\bf b}(b_{B}\lambda_{B}
+b_{YB}\lambda_{Y})],
\nonumber\\
&&D_Y=-[\frac{g_Y}{2\sqrt{2}}(\tilde{L}^{\dagger}\tilde{L}-2\tilde{R}^{\dagger}\tilde{R}
-\frac{1}{3}\tilde{Q}^{\dagger}\tilde{Q}+\frac{4}{3}\tilde{U}_{R}^{\dagger}\tilde{U}_{R}
-\frac{2}{3}\tilde{D}_{R}^{\dagger}\tilde{D}_{R}+H_{1}^{\dagger}H_{1}-H_{2}^{\dagger}H_{2})
\nonumber\\
&&\hspace{1cm}+\frac{1}{2}\psi_{\bf b}(b_{Y}\lambda_{Y}+b_{YB}\lambda_{B})]
\nonumber\\
&&D^i=-\frac{1}{2}[g_{2}(\tilde{L}^{\dagger}\tau^{i}\tilde{L}
+\tilde{Q}^{\dagger}\tau^{i}\tilde{Q}+H_{1}^{\dagger}\tau^{i}H_{1}
+H_{2}^{\dagger}\tau^{i}H_{2})+\frac{b_W}{\sqrt{2}}\psi_{\bf b}\lambda_{W^i}]
\nonumber\\
&&D^a=-\frac{1}{2}[g_{s}(\tilde{Q}^{\dagger}T^{a}\tilde{Q} + \tilde{U}_R^{\dagger}T^{a}\tilde{U}_R
+ \tilde{D}_R^{\dagger}T^{a}\tilde{D}_R)+\frac{b_G}{\sqrt{2}}\psi_{\bf b}\lambda_{g^a}],
\label{Dterms}
\ea
in which we have terms coming both from ${\cal L}_{axion}$ and from the USSM Lagrangean 
that can be found in the appendix.

\subsection{The kinetic mixing}

In these type of supersymmetric models the extra $U(1)_B$ sector can
mix with $U(1)_Y$ in different ways. In particular, in the context of $USSM-A$,  
the kinetic mixing is treated as in the NMSSM with 
the inclusion of an anomalous $U(1)_B$ symmetry and the extra 
singlet $\hat{S}$ is charged under $B$.

The lagrangean for the gauge fields is modified by
introducing a mixing term $B-Y$ proportional to a small 
parameter $\sin{a}$ 
\ba
{\cal L}_{mixing}=-\frac{1}{4}\int{d^{4}\theta\,\,
2\sin{a} \, W^{Y \alpha} W^{B}_{\alpha} \,\,\delta^{2}(\bar{\theta})+h.c.}
\ea
where $\sin{a}$ represents the mixing between the two abelian structures $U(1)_{Y}$ and $U(1)_{B}$.
In the same way, the gauge mass terms lagrangean in the presence of kinetic mixing 
is modified by the inclusion of a term proportional to the mass parameter 
$M_{Y B}$ as follows
\ba
&&{\cal L}_{GMTmix}=\frac{1}{2}\int d^{4} \theta \left[ M_{Y B} W^{Y\alpha} W^{B}_{\alpha}
+h.c.\right] \delta^{4}(\theta,\bar{\theta}).
\ea
Furthermore, the $USSM-A$ is affected by another source of kinetic mixing 
coming from the mixed counterterm proportional to $b_{Y B}$ in the expression of
${\cal L}_{axion}$. Expanding this expression in component fields we observe that 
the multiplet $\hat{\bf b}$ contains the complex scalar field $b$ whose 
real part can be $\textrm{Re}\, b\neq 0$ and it generates a kinetic mixing 
proportional to $\propto b_{YB}\, \textrm{Re}b\, g_Y\, g_B $, where the coefficient
$b_{YB}$ fixed by the anomaly cancellation procedure, goes like 
the inverse of the St\"uckelberg mass and can be neglected in this first analysis
(see Ref. \cite{Anastasopoulos:2008jt}). 
In our formulation we assume $\sin{a}=0$ for simplicity  
and we will give a more detailed analysis of the 
kinetic mixing in the context of the USSM-A 
in a forthcoming paper \cite{preparation}.

\subsection{The Fayet-Iliopoulos terms}

To be as more general as possible, in theories with $U(1)$s gauge superfields
we should add to the lagrangean the following Fayet-Iliopoulos (FI) term
\ba
{\cal L}_{FI}=\xi_Y D_Y + \xi_B D_B.
\ea
which is allowed by symmetry reasons. Here $\xi_Y,\xi_B$ are two coefficients, while 
$D_Y$ and $D_B$ are the D-terms corresponding to the $U(1)_Y$ and $U(1)_B$ symmetry respectively.
In our analysis we omit these contributions even if a quadratically divergent  
FI always appears in a field theory at one loop \cite{Fischler:1981zk}.
The reason resides in the fact that, in the low-energy lagrangean there should be a counterterm,
which compensates precisely both the divergent and the finite part of the one loop
contributions (see Ref. \cite{Anastasopoulos:2008jt}). We are also omitting the terms 
responsible for the cancellation of gravitational anomalies.
A more comprehensive description will be given in \cite{preparation}.\\ 

Some of the notations used in our analysis are recalled in the appendix, here we just mention that the scalars of the model are denoted, as usual,  by a tilde ( $\tilde{\,}$ ).
It is convenient to combine the axion sector and the $F$ and $D$ terms extracted from the other sectors of the total Lagrangean of the model. This combination is in general defined to be the auxiliary Lagrangean, or ${\cal L}_{aux}$, 
which is given by
\ba
{\cal L}_{aux}&=&-y_{e}^{2}H_{1}^{\dagger}H_{1}\tilde{R}\tilde{R}^{\dagger}
-y_{u}^{2}H_{2}^{\dagger}H_{2}\tilde{U}_{R}\tilde{U}_{R}^{\dagger}
-y_{d}^{2}H_{1}^{\dagger}H_{1}\tilde{D}_{R}\tilde{D}_{R}^{\dagger}
-\vert\lambda H_{1}\cdot H_{2}\vert^{2}
\nonumber\\
&&-\vert\lambda S\vert^{2}(H_{2}^{\dagger}H_{2}+H_{1}^{\dagger}H_{1})
-y_{d}^{2}\tilde{D}_{R}^{\dagger}\tilde{D}_{R}\tilde{Q}^{\dagger}\tilde{Q}
-y_{e}^{2}\tilde{L}^{\dagger}\tilde{L}\tilde{R}\tilde{R}^{\dagger}
\nonumber\\
&&-y_{u}^{2}\tilde{U}_{R}\tilde{U}_{R}^{\dagger}\tilde{Q}^{\dagger}\tilde{Q}
-\lambda y_{u}\left(S\tilde{Q}^{\dagger}H_{1}\tilde{U}_{R}^{\dagger}+h.c.\right)
-\lambda y_{d}\left(S\tilde{Q}^{\dagger}H_{2}\tilde{D}_{R}^{\dagger}+h.c.\right)
\nonumber\\
&&\lambda y_{e}\left(S\tilde{L}^{\dagger}H_{2}\tilde{R}^{\dagger}+h.c.\right)
-y_{d}y_{u}\left(\tilde{U}_R^{\dagger}H_{2}^{\dagger}H_{1}\tilde{D}_{R}
+h.c.\right)
-y_{e}y_{d}\left(\tilde{D}_{R}^{\dagger}\tilde{Q}^{\dagger}\tilde{L}\tilde{R}
+h.c.\right)
\nonumber\\
&&+\vert(b_{G}\lambda_{g^a}\lambda_{g^a}+b_{W}\lambda_{W^i}\lambda_{W^i}
+b_{Y}\lambda_{Y}\lambda_{Y}+b_{B}\lambda_{B}\lambda_{B} + b_{YB}\lambda_{Y}\lambda_{B})\vert^2
\nonumber\\
&&-\frac{1}{2}[g_{s}(\tilde{Q}^{\dagger}T^{a}\tilde{Q} + \tilde{U}_R^{\dagger}T^{a}\tilde{U}_R
+ \tilde{D}_R^{\dagger}T^{a}\tilde{D}_R)+\frac{b_G}{\sqrt{2}}\psi_{\bf b}\lambda_{g^a}]^{2}
\nonumber\\
&&-\frac{1}{2}[g_{2}(\tilde{L}^{\dagger}\tau^{i}\tilde{L}
+\tilde{Q}^{\dagger}\tau^{i}\tilde{Q}+H_{1}^{\dagger}\tau^{i}H_{1}
+H_{2}^{\dagger}\tau^{i}H_{2})+\frac{b_W}{\sqrt{2}}\psi_{\bf b}\lambda_{W^i}]^{2}
\nonumber\\
&&-[\frac{g_Y}{2\sqrt{2}}(\tilde{L}^{\dagger}\tilde{L}-2\tilde{R}^{\dagger}\tilde{R}
-\frac{1}{3}\tilde{Q}^{\dagger}\tilde{Q}+\frac{4}{3}\tilde{U}_{R}^{\dagger}\tilde{U}_{R}
-\frac{2}{3}\tilde{D}_{R}^{\dagger}\tilde{D}_{R}+H_{1}^{\dagger}H_{1}-H_{2}^{\dagger}H_{2})
\nonumber\\
&&+\frac{1}{2}\psi_{\bf b}(b_{Y}\lambda_{Y}+b_{YB}\lambda_{B})]^{2}
-[\frac{g_B}{2\sqrt{2}}(B_{L}\tilde{L}^{\dagger}\tilde{L}+B_{R}\tilde{R}^{\dagger}\tilde{R}
+B_{Q}\tilde{Q}^{\dagger}\tilde{Q}+B_{U}\tilde{U}_{R}^{\dagger}\tilde{U}_{R}
\nonumber\\
&&+B_{D}\tilde{D}_{R}^{\dagger}\tilde{D}_{R}+B_{H_{1}}H_{1}^{\dagger}H_{1}
+B_{H_{2}}H_{2}^{\dagger}H_{2}+B_{S}S^{\dagger}S)+\frac{1}{2}\psi_{\bf b}(b_{B}\lambda_{B}
+b_{YB}\lambda_{Y})]^{2}
\nonumber\\
&&+\frac{1}{2}[\psi_{\bf b}\psi_{\bf b}(b_{G}^{2}\lambda_{g^a}\lambda_{g^a}
+b_{W}^{2}\lambda_{W^i}\lambda_{W^i}
+(b_{Y}^{2}+b_{YB}^{2})\lambda_{Y}\lambda_{Y}
\nonumber\\
&&+b_{B}^{2}\lambda_{B}\lambda_{B}+(b_{Y}+b_{B})b_{YB}\lambda_{Y}\lambda_{B}
+b_{B}b_{YB}\lambda_{B}\lambda_{Y})+h.c.]
\ea
where the expressions of the $D$ terms are now determined by Eq. (\ref{Dterms}).

\section{Goldstones of the potential and of the massive gauge bosons }
The identification of the goldstone modes of the model requires a combined analysis of the potential and of the 
bilinear mixing terms $Z_i \partial G_{Z_i}$ for all the broken (massive) gauge bosons. Naturally, the expansion near the vacuum is consistent if the stability conditions of the potential near the expansion point are satisfied. The neutral goldstone modes corresponding to the 
physical neutral gauge bosons after the breaking 
are part of the CP-odd sector together with other physical components, 
spanning together the entire CP-odd space. 
In general, in this sector, the potential contains a set of ``flat directions", which appear 
as goldstone modes of the matrix of its second derivatives. These goldstone modes do not necessarily coincide with the goldstone modes ($G_{Z'}$) identified from the bilinear mixings. This turns out to be the case if the St\"uckelberg decouples from the scalar potential while it gives mass to one of the anomalous gauge bosons. To clarify this point it is convenient to move back to the non-supersymmetric case.

The allowed structure of the potential involves $b$-independent ($V$) 
and $b$- dependent ($V'$) terms, just on the basis of 
the symmetries of the Lagrangean, given by 

\beq
V=\sum_{a=1,2}\Bigl(  \mu_a^2  H_a^{\dagger} H_a 
+ \l_{aa} (H_a^{\dagger} H_a)^2\Bigr)
-2\l_{12}(H_1^{\dagger} H_1)(H_2^{\dagger} H_2)
+2{\l^\prime_{12}} |H_2^T\tau_2 H_1|^2 \label{PQ},
\eeq
and
\begin{eqnarray}
V^\prime&=&\lambda_0\, H_2^{\dagger}H_1 e^{-i\sum_I(q_2^I-q_1^I)\frac{b_I}{M_I}}
+\l_1 \left(H_2^{\dagger}H_1 e^{-i\sum_I(q_2^I-q_1^I)\frac{b_I}{M_I}}\right)^2  
\nonumber\\
&&+\l_2 \left(H_2^{\dagger}H_2\right)
H_2^{\dagger}H_1 e^{-i\sum_I(q_2^I-q_1^I)\frac{b_I}{M_I}}
+\l_3 \left(H_1^{\dagger}H_1\right)
H_2^{\dagger}H_1 e^{-i\sum_I(q_2^I-q_1^I)\frac{b_I}{M_I}} + c.c.
\label{PQbreak}
\end{eqnarray}
respectively, where the sum over $I$ is a sum over the St\"uckelberg axions of the (several) anomalous $U(1)$'s.
 In the supersymmetric case this second contribution is, in general, not allowed, although it might appear after supersymmetry breaking. This second term or ``phase-dependent term" is directly responsible for Higgs-axion mixing and for producing a massive axion. The interesting point is that in the supersymmetric case (with $b$ a real field), even if $V'$ is not allowed, we may still, under some particular conditions, end up with a physical axion in the spectrum, as we are now going to elaborate.
  
As we have mentioned, the identification of the goldstones of the 
theory is necessarily done using the kinetic term of the scalars, 
including the St\"uckelberg, which in this case takes the form
\beq
|{\cal D}_\mu H_1|^2 + |{\cal D }_\mu H_2|^2 +
\frac{1}{2}(\partial_\mu b+ M_{st} B_\mu)^2\label{quadform}.
\label{KE}
\eeq
The expansion of this equation near the stable vacuum gives the usual bilinear mixings characterizing the derivative couplings of the physical massive gauge bosons to the corresponding goldstones; 
rather straightforwardly one obtains the combination 
\beq
M_Z Z^\mu \partial_\mu G_Z + M_{Z'}{Z'}^\mu \partial_\mu G_{Z'} + ...
\label{KEM}
\eeq
with $G_Z$ and $G_{Z'}$ being the true goldstone modes of the theory. Notice, if not obvious, that while 
$G_Z$ is just expressed as a linear combination of the two CP-odd components of the Higgs, $G_{Z'}$ on the other hand takes a contribution also from $b$, due to the St\"uckelberg mass term. Therefore, one of the special features of the combination of the Higgs and St\"uckelberg mechanisms is that in some cases the potential of the model - $V$ is an example of this situation, since it does not not include a $b$ field - is not sufficient to identify all the goldstone modes.  Clearly, if both $V$ and $V'$ are present, then $G_Z$ and $G_{Z'}$ can be extracted from the total potential and coincide with the goldstone modes extracted from the bilinear mixings of (\ref{KE}) and (\ref{KEM}). In this case the physical axion turns out to be massive. We recall that the quadratic part of the CP-odd potential takes the general form
\beqa
V_{CP-odd}&=&
 \left(\textrm{Im} H_1^0, \textrm{Im} H_2^0, b \right){\cal N}\left(\begin{array}{c}
\textrm{Im} H_1^0\\
\textrm{Im} H_2^0\\
b\\
\end{array}\right)
\eeqa
for a suitable ${\cal N}$ matrix whose explicit expression is important but not necessary for our dicussion. In the case 
of the MSSM the structure of the potential coincides with that of $V$ and one identifies only one physical CP-odd Higgs (called $A^0$ in the MSSM) which will not have an axion-like coupling, as can be verified by also a simple counting of the degrees of freedom before and after EWSB. 
In this case the orthogonal transformation that diagonalizes the CP-odd scalar sector takes the form
\ba
\pmatrix{\textrm{Im} H_2^0\cr \textrm{Im} H_1^0}=\; O\;
\pmatrix{A^0 \cr G^0}
\label{rotunit2}
\ea
and involves the physical (massive) CP-odd Higgs $A^0$ and a golstone mode $G^0$. 
The above discussion goes through in a similar way also for the anomalous $U(1)$ extension of the MSSM discussed in \cite{Anastasopoulos:2008jt}. For the case of a potential such as $V_{CP-odd}=V + V'$ instead, there is indeed a mixing between the components of the Higgs and $b$ and the diagonalization of the quadratic part of the potential gives
\beq
\pmatrix{\textrm{Im} H_2^0\cr \textrm{Im} H_1^0\cr b}=\; O_3\;
\pmatrix{\chi \cr G_1^0\cr G_2^0}
\label{rotunit}
\eeq
with $O_3$ being an orthogonal matrix. We have denoted the physical field by $\chi$ and the  NG-bosons
by $G_{1,2}^0$. 
In this case it is rather obvious that $\chi$ acquires an axion-like coupling, inherited from $b$. In other words $b$ has an expansion in terms of $\chi, G^0_1$ and $G^0_2$ or, equivalently, in terms of $\chi, G_Z$ and $G_{Z'}$, where $G_Z$ and $G_{Z'}$ are identified by Eq. (\ref{KEM}). The decomposition is clearly gauge dependent. 
One important comment concerns the nature of the $b F \tilde{F}$ interactions in this case.

In the unitary gauge the only axion-like couplings left involve the physical component of $b$, denoted by $\chi$, called ``the axi-Higgs", which gives typical $\chi F\tilde{F}$ interactions. As we have mentioned above, in the absence of $V'$, $b$ decouples from the rest of the Higgs sector in ${\mathcal N}$. In this case in the unitary gauge all the anomalous couplings can be removed, and the theory goes back again to its original anomalous form, with the old Lagrangean now replaced by an ordinary {\em massive} (and possibly anomalous) Yang-Mills theory. It is rather obvious that the truly new element in these types of actions shows up when a physical axion-like particle is induced in the spectrum. In the 
absence of this, the $bF\tilde{F}$ has dubious meaning, since this term does not cancel the anomaly, as emphasized by Preskill long ago \cite{Preskill:1990fr}. Rather, it allows a better power-counting of the modified (anomalous) action. A justification of this point of view comes from the fact that an anomalous (and massive) Yang-Mills theory can be given a typical St\"uckelberg form and a $b F\tilde{F}$ interaction by a field-enlarging transformation \cite{Coriano:2007fw}.

For this reason the only satisfactory potentials are those that either allow $b$ to be part of the scalar sector (such as for $V + V'$) or, alternatively, when they allow, under certain conditions that we are going to discuss next, a mixing between the CP-odd Higgs components and the St\"uckelberg. 

With these motivations in mind, we move to the case of the new superpotential.

\section{Scalar mass terms, the scalar potential and the mass of the gauge bosons}

Let's now move to a discussion of the other sectors of the theory, starting from the scalar one.
The Lagrangean for the scalar mass terms is given by
\ba
{\cal L}_{SMT}&=&-M^{2}_{L}\tilde{L}^{\dagger}\tilde{L}
-m^{2}_{R}\tilde{R}^{\dagger}\tilde{R}-M^{2}_{Q}\tilde{Q}^{\dagger}\tilde{Q}
-m^{2}_{U_{R}}\tilde{U}_{R}^{\dagger}\tilde{U}_{R}
-m^{2}_{D_{R}}\tilde{D}_{R}^{\dagger}\tilde{D}_{R}
-m_{1}^{2}H_{1}^{\dagger}H_{1}\nonumber\\
&&-m_{2}^{2}H_{2}^{\dagger}H_{2}-m_{S}^{2}S^{\dagger}{S}
-(a_{\lambda}S H_{1}\cdot H_{2}+h.c.)-(a_{e}H_{1}\cdot\tilde{L}\tilde{R}+h.c.)\nonumber\\
&&-(a_{d}H_{1}\cdot\tilde{Q}\tilde{D}_{R}+h.c.)
-(a_{u}H_{2}\cdot\tilde{Q}\tilde{U}_{R}+h.c.),
\ea
where $M_{L},M_{Q},m_R,m_{U_R},m_{D_R},m_1,m_2,m_S$ are 
the mass parameters for the explicit supersymmetry breaking, while
$a_e,a_{\lambda},a_u,a_d$ are coefficients with mass dimension one.

The computation of the Lagrangean containing the soft-breaking 
terms Lagrangean is, as usual, split into 
the scalar and gaugino mass terms  
\ba
{\cal L}_{Soft}={\cal L}_{SMT}+{\cal L}_{GMT}
+\frac{1}{2}M_{\bf b}\left(\psi_{\bf b}\psi_{\bf b}+\bar{\psi}_{\bf b}\bar{\psi}_{\bf b}\right),
\ea
where $M_{\bf b}$ is a mass parameter for the axino $\psi_{\bf b}$. 
The gaugino mass terms given by
\ba
&&{\cal L}_{GMT}=
-\frac{1}{2}M_G\left(\lambda_{g^{a}}\lambda_{g^{a}}
+\bar{\lambda}_{g^{a}}\bar{\lambda}_{g^{a}}\right)
-\frac{1}{2}M_w\left(\lambda_{W^{i}}\lambda_{W^{i}}
+\bar{\lambda}_{W^{i}}\bar{\lambda}_{W^{i}}\right)
\nonumber\\
&&\hspace{1.5cm}-\frac{1}{2}M_Y\left(\lambda_Y\lambda_Y+\bar{\lambda}_Y\bar{\lambda}_Y\right)
-\frac{1}{2}M_B\left(\lambda_{B}\lambda_{B}+\bar{\lambda}_B\bar{\lambda}_B\right),
\ea
where $\lambda_{g^a},\bar{\lambda}_{g^a}$ are respectively the 
left- and right-handed gauginos of the $SU(3)$ sector, $\lambda_{W^i},\bar{\lambda}_{W^i}$ are 
the left- and right-handed gauginos of the $SU(2)$ sector and $\lambda_Y,\bar{\lambda}_Y$
are the chiral gauginos of $U(1)_Y$.
The $M_G,M_w,M_Y,M_B$ mass terms are the SUSY breaking parameters for $SU(3)$, $SU(2)_W$, $U(1)_Y$
and $U(1)_B$ respectively.
Once we have imposed the equations of motion for the $F$-terms the on-shell Lagrangean is given by
\ba
{\cal L}_{aux-F}&=&-y_{e}^{2}H_{1}^{\dagger}H_{1}\tilde{R}\tilde{R}^{\dagger}
-y_{u}^{2}H_{2}^{\dagger}H_{2}\tilde{U}_{R}\tilde{U}_{R}^{\dagger}
-y_{d}^{2}H_{1}^{\dagger}H_{1}\tilde{D}_{R}\tilde{D}_{R}^{\dagger}
-\vert\lambda H_{1}\cdot H_{2}\vert^{2}
\nonumber\\
&&-\vert\lambda S\vert^{2}(H_{2}^{\dagger}H_{2}+H_{1}^{\dagger}H_{1})
-y_{d}^{2}\tilde{D}_{R}^{\dagger}\tilde{D}_{R}\tilde{Q}^{\dagger}\tilde{Q}
-y_{e}^{2}\tilde{L}^{\dagger}\tilde{L}\tilde{R}\tilde{R}^{\dagger}
\nonumber\\
&&-y_{u}^{2}\tilde{U}_{R}\tilde{U}_{R}^{\dagger}\tilde{Q}^{\dagger}\tilde{Q}
-\lambda y_{u}\left(S\tilde{Q}^{\dagger}H_{1}\tilde{U}_{R}^{\dagger}+h.c.\right)
-\lambda y_{d}\left(S\tilde{Q}^{\dagger}H_{2}\tilde{D}_{R}^{\dagger}+h.c.\right)
\nonumber\\
&&\lambda y_{e}\left(S\tilde{L}^{\dagger}H_{2}\tilde{R}^{\dagger}+h.c.\right)
-y_{d}y_{u}\left(\tilde{U}_R^{\dagger}H_{2}^{\dagger}H_{1}\tilde{D}_{R}
+h.c.\right)
-y_{e}y_{d}\left(\tilde{D}_{R}^{\dagger}\tilde{Q}^{\dagger}\tilde{L}\tilde{R}
+h.c.\right),
\nonumber\\
\ea
where the coefficients $y_{e},y_{u},y_{d}$ come from 
the Yukawa couplings of the superpotential,
while the $D$ terms are 
\ba
{\cal L}_{aux-D}&=&-\frac{g_2^{2}}{2}(\tilde{L}^{\dagger}\tau^{i}\tilde{L}
+\tilde{Q}^{\dagger}\tau^{i}\tilde{Q}+H_{1}^{\dagger}\tau^{i}H_{1}
+H_{2}^{\dagger}\tau^{i}H_{2})^2
-\frac{g_s^{2}}{2}(\tilde{Q}^{\dagger}T^{a}\tilde{Q} 
+ \tilde{U_R}^{\dagger}T^{a}\tilde{U_R}+ \tilde{D_R}^{\dagger}T^{a}\tilde{D_R})^2
\nonumber\\
&&-\frac{g_Y^{2}}{8}(\tilde{L}^{\dagger}\tilde{L}-2\tilde{R}^{\dagger}\tilde{R}
-\frac{1}{3}\tilde{Q}^{\dagger}\tilde{Q}
+\frac{4}{3}\tilde{U}_{R}^{\dagger}\tilde{U}_{R}
-\frac{2}{3}\tilde{D}_{R}^{\dagger}\tilde{D}_{R}
+H_{1}^{\dagger}H_{1}-H_{2}^{\dagger}H_{2})^{2}
\nonumber\\
&&-\frac{g_B^{2}}{8}(B_{L}\tilde{L}^{\dagger}\tilde{L}
+B_{R}\tilde{R}^{\dagger}\tilde{R}+B_{Q}\tilde{Q}^{\dagger}\tilde{Q}
+B_{U}\tilde{U}_{R}^{\dagger}\tilde{U}_{R}+B_{D}\tilde{D}_{R}^{\dagger}\tilde{D}_{R}
\nonumber\\
&&+B_{H_{1}}H_{1}^{\dagger}H_{1}
+B_{H_{2}}H_{2}^{\dagger}H_{2}+B_{S}S^{\dagger}S)^{2},
\label{LausUnMSSM}
\ea
where $B_L,B_R$ are the charges of the leptons chiral superfields,
$B_Q,B_U,B_D$ are the charges of the left and right chiral superfields
of the quark sector and $B_{H_1},B_{H_2},B_S$ are the charges of the two Higgs
doublet and of the extra singlet respectively.

\subsection{The scalar potential}

The study of EWSB in the case of these 
models proceeds similarly to the USSM \cite{Cvetic:1997ky}. 

The scalar potential is given by
\begin{eqnarray}
V&=&\vert\lambda H_{1}\cdot H_{2}\vert ^{2}
+\vert\lambda S\vert ^{2}(\vert H_{1}\vert^{2}+\vert H_{2}\vert^{2})
+\frac{1}{8}(g_2^{2}+g_Y^{2})(H_{1}^{\dagger}H_{1}-H_{2}^{\dagger}H_{2})^{2}\nonumber\\
&&+\frac{g_B^{2}}{8}(B_{H_{1}}H_{1}^{\dagger}H_{1}
+B_{H_{2}}H_{2}^{\dagger}H_{2}+B_{S}S^{\dagger}S)^{2}
+\frac{g_2^{2}}{2}\vert H_{1}^{\dagger}H_{2}\vert^{2}
+m_{1}^{2}\vert H_{1}\vert^{2}+m_{2}^{2}\vert H_{2}\vert^{2}\nonumber\\
&&+m_{S}^{2}\vert S\vert^{2}+(a_{\lambda}SH_{1}\cdot H_{2}+h.c.).
\end{eqnarray}

We introduce the following basis 
\ba
H_{1}=\frac{1}{\sqrt{2}}\left(
\begin{array}{c}
\textrm{Re}H_1^{0} + i~\textrm{Im}H_1^{0}\\
\textrm{Re}H_1^{-} + i~ \textrm{Im}H_1^{-}
\end{array}
\right),\hspace{0.5cm}
H_{2}=\frac{1}{\sqrt{2}}\left(
\begin{array}{c}
\textrm{Re}H_2^{+} + i ~\textrm{Im}H_2^{+}\\
\textrm{Re}H_2^{0} + i ~\textrm{Im}H_2^{0}
\end{array}
\right),\hspace{0.5cm} S=\frac{1}{\sqrt{2}}(\textrm{Re}S + i ~\textrm{Im}S),
\nonumber\\
\ea
where in correspondence of the minimum value of the potential
we use the following parametrization for the Higgs fields
\ba
\left\langle H_{1}\right\rangle=
\frac{1}{\sqrt{2}}\left(
\begin{array}{c}
v_{1}\\
0
\end{array}
\right),\hspace{1cm}
\left\langle H_{2}\right\rangle=\frac{1}{\sqrt{2}}\left(
\begin{array}{c}
0\\
v_{2}
\end{array}
\right),\hspace{1cm} \left\langle S\right\rangle=\frac{v_{S}}{\sqrt{2}}.
\ea
As usual, we require the existence of a stable vacuum imposing the conditions
\begin{eqnarray}
&&m_{1}^{2}v_{1}+\frac{1}{2}\lambda^{2}v_{1}(v_{2}^{2}+v_{S}^{2})+\frac{1}{\sqrt{2}}a_{\lambda}v_{2}v_{S}
-\frac{1}{8}v_{1}(v_{2}^{2}-v_{1}^{2})g^{2}\nonumber\\
&&\hspace{1cm}+\frac{1}{8}g_B^{2}B_{H_{1}}v_{1}(B_{H_{2}}v_{2}^{2}
+B_{H_{1}}v_{1}^{2}+B_{S}v_{S}^{2})=0,
\\\nonumber\\
&&m_{2}^{2}v_{2}+\frac{1}{2}\lambda^{2}v_{2}(v_{1}^{2}+v_{S}^{2})
+\frac{1}{\sqrt{2}}a_{\lambda}v_{1}v_{S}+\frac{1}{8}v_{2}(v_{2}^{2}
-v_{1}^{2})g^{2}\nonumber\\
&&\hspace{1cm}+\frac{1}{8}g_B^{2}B_{H_{2}}v_{2}(B_{H_{2}}v_{2}^{2}
+B_{H_{1}}v_{1}^{2}+B_{S}v_{S}^{2})=0,
\\\nonumber\\
&&\frac{1}{\sqrt{2}}a_{\lambda}v_{1}v_{2}+m_{S}^{2}v_{S}+\frac{1}{2}\lambda^{2}v_{S}v^2+\frac{1}{8}g_B^{2}B_{S}v_{S}(B_{H_{2}}v_{2}^{2}
+B_{H_{1}}v_{1}^{2}+B_{S}v_{S}^{2})=0,
\end{eqnarray}
where again $a_\lambda$ is a mass parameter of the model.

\subsection{Mass of the gauge bosons}
 
 The Lagrangean that describes the contributions to the mass of the gauge bosons
is given by
\ba
{\cal L}_{\textsc{q}}=\vert{\mathcal D}_{\mu}H_{1}\vert^{2}
+\vert{\mathcal D}_{\mu}H_{2}\vert^{2}
+\vert{\mathcal D}_{\mu}S\vert^{2}
+\frac{1}{2}\left(\partial_{\mu}\textrm{Im}~b + M_{st} B_{\mu}\right)^2
\ea
and involves, beside the two higgses, the SM bosonic singlet of $\hat{S}$, the bosonic component of 
the St\"uckelberg axion, $b$, and the St\"uckelberg mass $M_{st}$.
Collecting the quadratic terms we obtain the contributions to the gauge boson masses which are given by
\begin{eqnarray}
&&{\cal L}_{GM} = \frac{g_{2}^{2}}{4}(v_{1}^{2} + v_{2}^{2})W^{+\mu}W^{-}_{\mu}
+\frac{g_2^{2}}{8}(v_{1}^{2}+v_{2}^{2})W^{3\mu}W^{3}_{\mu}
-\frac{g_2 g_{Y}}{4}(v_{1}^{2}+v_{2}^{2})W^{3\mu}A^{Y}_{\mu}
\nonumber\\
&&\hspace{1cm}+\frac{g_{Y}^{2}}{8}(v_{1}^{2}+v_{2}^{2})A^{Y\mu}A^{Y}_{\mu}
+\frac{g_2 g_{B}}{4}(B_{H_{1}}v_{1}^{2}-B_{H_{2}}v_{2}^{2})W^{3}_{\mu}B^{\mu}
-\frac{g_{Y}g_{B}}{4}(B_{H_{1}}v_{1}^{2}-B_{H_{2}}v_{2}^{2})A^{Y}_{\mu}B^{\mu}
\nonumber\\
&&\hspace{1cm}
+\frac{g_{B}^{2}}{8}(B_{H_{1}}^{2}v_{1}^{2}
+B_{H_{2}}^{2}v_{2}^{2} +B_{S}^{2}v_{S}^{2})B_{\mu}B^{\mu}
+\frac{1}{2}M_{st}^2 B^{\mu}B_{\mu}.
\end{eqnarray}
Using the interaction basis of the gauge field components 
$(W^{3}_{\mu},A^{Y}_{\mu},B_{\mu})$ we obtain the corresponding mass matrix, which is given by
\ba
{\mathcal M}^2_{gauge}=\left(
\begin{array}{ccc}
\frac{g_2^{2}}{8} v^2 &
-\frac{g_2 g_{Y}}{8}v^2 &
\frac{g_2}{8} x_B\\
\\
-\frac{g_{2} g_{Y}}{8}v^2 &
\frac{g_{Y}^{2}}{8}v^2 &
\frac{g_{Y}}{8}x_B\\
\\
\frac{g_{2}}{8}x_B &
\frac{g_{Y}}{8}x_B &
\frac{N_{BB}}{8}+ \frac{M_{st}^{2}}{2}
\end{array}
\right),
\ea
where
\ba
x_B=g_{B}(v_{1}^{2}B_{H_{1}}-v_{2}^{2}B_{H_{2}}),
\hspace{0.5cm}
N_{BB}=g_{B}^{2}(B_{H_{1}}^{2}v_{1}^{2}
+B_{H_{2}}^{2}v_{2}^{2} +B_{S}^{2}v_{S}^{2}),
\hspace{0.5cm}
v^2=v_1^2+v_2^2.
\ea
Performing the diagonalization we obtain the rotation matrix  
\ba
\begin{small}
O^{A}_{susy}=\left(
\begin{array}{ccc}
\frac{g_Y}{g}&
\frac{g_2}{g}&
0\\\\
-\frac{g_2(f_{1}+\sqrt{f_{1}^{2}+4x_{B}^{2}g^{2}})}
{g\sqrt{2\left[4g^{2}x_{B}^{2}+f_{1}^2+ f_1\,\sqrt{f_{1}^{2}+4 g^{2}x_{B}^{2}}\right]}}&
\frac{g_Y(f_{1}+\sqrt{f_{1}^{2}+4x_{B}^{2}g^{2}})}
{g\sqrt{2\left[4g^{2}x_{B}^{2}+f_{1}^2+f_1\,\sqrt{f_{1}^{2}+4 g^{2}x_{B}^{2}}\right]}}&
\frac{g x_B \sqrt{2}}
{\sqrt{\left[4g^{2}\,x_{B}^{2}+ f_{1}^2+f_1\sqrt{f_{1}^{2}+4 g^{2}x_{B}^{2}}\right]}}\\
\\
-\frac{g_2(f_{1}-\sqrt{f_{1}^{2}+4x_{B}^{2}g^{2}})}
{g\sqrt{2\left[4g^{2}x_{B}^{2}+f_{1}^2- f_1\,\sqrt{f_{1}^{2}+4 g^{2}x_{B}^{2}}\right]}}&
\frac{g_Y(f_{1}-\sqrt{f_{1}^{2}+4x_{B}^{2}g^{2}})}
{g\sqrt{2\left[4g^{2}x_{B}^{2}+f_{1}^2-f_1\,\sqrt{f_{1}^{2}+4 g^{2}x_{B}^{2}}\right]}}&
\frac{g x_B \sqrt{2}}
{\sqrt{\left[4g^{2}\,x_{B}^{2}+ f_{1}^2-f_1\sqrt{f_{1}^{2}+4 g^{2}x_{B}^{2}}\right]}}
\end{array}
\right),
\end{small}
\nonumber\\
\ea
which acts on the interaction basis as
\ba
\left(
\begin{array}{c}
A^{\gamma}\\
Z\\
Z'
\end{array}
\right)=
O^{A}_{susy}\left(
\begin{array}{c}
W^{3}\\
A^{Y}\\
B
\end{array}
\right),
\ea
and where we have defined $g=\sqrt{g_Y^2+g_2^2}$ and $f_1=4 M_{st}^2 - g^2 v^2 + N_{BB}$.

We obtain one null eigenvalue corresponding to the photon, while the masses of the 
physical $Z$ and $Z'$ are given by
\begin{eqnarray}
M^{2}_{Z}&=& \frac{1}{8}\left(4 M_{st}^2+g^2\,v^2 +N_{BB}
-\sqrt{(4 M_{st}^2-g^2\,v^2+N_{BB})^2+4 g^2 x_B^2}\right)        
\nonumber\\
M^{2}_{Z'}&=& \frac{1}{8}\left(4 M_{st}^2+g^2\,v^2 +N_{BB}
+\sqrt{(4 M_{st}^2-g^2\,v^2+N_{BB})^2+4 g^2 x_B^2}\right).
\end{eqnarray}
Compared to the non-supersymmetric case \cite{Coriano:2005js}, the corrections to the masses of the gauge bosons involve also $v_S$, which is implicitly contained in $N_{BB}$.

\subsection{The charged and the CP-even sectors of the scalar potential}

The description of the charged sector of the model is performed using the standard basis 
$(\textrm{Re}H_2^{+},\textrm{Re}H_1^{-})$. We obtain the following mass matrix
\ba
{\mathcal M}^{2}_{c}=
\left(
\begin{array}{cc}
\frac{1}{2}(\frac{1}{2}g^{2}-\lambda^{2})v_{1}^{2}+a_{\lambda}\frac{v_{1}v_{S}}{\sqrt{2} v_{2}} & 
-\frac{1}{2}(\frac{1}{2}g^{2}-\lambda^{2})v_{1}v_{2}+a_{\lambda}\frac{v_{S}}{\sqrt{2}}\\
\\
-\frac{1}{2}(\frac{1}{2}g^{2} - \lambda^{2})v_{1}v_{2}+a_{\lambda}\frac{v_{S}}{\sqrt{2}}&
\frac{1}{2}(\frac{1}{2} g^{2} - \lambda^{2})v_{2}^{2}+a_{\lambda}\frac{v_{2}v_{S}}{\sqrt{2}v_{1}}
\end{array}
\right).
\ea
The same mass matrix is obtained in the basis $(-\textrm{Im}H_2^{+},\textrm{Im}H_1^{-})$.
We have one zero eigenvalue corresponding to a charged Goldstone boson and
a mass eigenvalue corresponding to the charged Higgs
\ba
m^{2}_{H^\pm}=\left(\frac{v_{1}}{v_{2}}+\frac{v_{2}}{v_{1}}\right)
\left(\frac{1}{4}g^{2}v_{1}v_{2}-\frac{1}{2}\lambda^{2}v_{1}v_{2}
+a_{\lambda}\frac{v_{S}}{\sqrt{2}}\right).
\ea
In the analysis of the CP-even sector we use the basis $(\textrm{Re}H_1^{0},\textrm{Re}H_2^{0},\textrm{Re}S)$. We obtain the matrix 
elements
\begin{eqnarray}
({\mathcal M}^{2}_{ev})_{11}&=&\frac{1}{4}\left(g_B^{2}B_{H_{1}}^{2}+g_Y^{2}+g_2^{2}\right)v_{1}^{2}
-a_{\lambda}\frac{v_{2}v_{S}}{\sqrt{2} v_{1}}
\nonumber\\
({\mathcal M}^{2}_{ev})_{12}&=&\left(\frac{g_B^{2}}{4}B_{H_{1}}B_{H_{2}}+\lambda^{2}
-\frac{g_2^{2}+g_Y^{2}}{4}\right)v_{1}v_{2}+a_{\lambda}\frac{v_{S}}{\sqrt{2}}
\nonumber\\
({\mathcal M}^{2}_{ev})_{13}&=&a_{\lambda}\frac{v_{2}}{\sqrt{2}}+\left(\frac{g_B^{2}}{4}B_{H_{1}}B_{S}
+\lambda^{2}\right)v_{1}v_{S}
\nonumber\\
({\mathcal M}^{2}_{ev})_{22}&=&\frac{1}{4}\left(g_B^{2}B_{H_{1}}^{2}+g_Y^{2}+g_2^{2}\right)v_{2}^{2}
-a_{\lambda}\frac{v_{2}v_{S}}{\sqrt{2} v_{1}}
\nonumber\\
({\mathcal M}^{2}_{ev})_{23}&=&a_{\lambda}\frac{v_{1}}{\sqrt{2}}+\left(\frac{g_B^{2}}{4} B_{H_{2}}B_{S}
+\lambda^{2}\right)v_{2}v_{S}
\nonumber\\
({\mathcal M}^{2}_{ev})_{33}&=&-a_{\lambda}\frac{v_{1}v_{2}}{\sqrt{2}v_{S}}
+\frac{1}{4}g_B^{2}B_{S}^{2}v_{S}^{2}\nonumber
\end{eqnarray}
with the other terms obtained by symmetry $({\mathcal M}_{12}={\mathcal M}_{21}, \textrm{etc}.)$.
The matrix has in general three massive eigenvalues corresponding to the three neutral Higgs particles
$(H^{0}_{1},H^{0}_{2},H^{0}_{3})$.

\subsection{The Neutral CP-odd sector and the axion}

The key sector that is responsible for the presence 
of a physical axion is the CP-odd one. Choosing the basis given by 
the components $(\textrm{Im}S,\textrm{Im}H_1^{0},\textrm{Im}H_2^{0})$, 
our superpotential with an extra singlet gives 
the mixing matrix
\ba
{\mathcal M}^{2}_{odd}=
\frac{a_{\lambda}}{\sqrt{2}}\left(
\begin{array}{ccc}
\frac{v_{1}v_{2}}{v_{S}}& v_{2} & v_{1}\\
v_{2} & \frac{v_{2}v_{S}}{v_{1}} & v_{S}\\
v_{1} & v_{S} & \frac{v_{1}v_{S}}{v_{2}}
\end{array}
\right).
\ea
Diagonalizing this mass matrix we can identify the orthogonal transformation $O^{odd}$ from the interaction to the mass eigenstates which is given by
\ba
\left(
\begin{array}{c}
\textrm{Im}S \\
\textrm{Im}H_1^{0}\\
\textrm{Im}H_2^{0}
\end{array}
\right)=O^{odd}\left(
\begin{array}{c}
G_1^{0}\\
G_2^{0}\\
H^{0}_{4}
\end{array}
\right).
\ea
A simple analysis gives two null eigenvalues, corresponding to two neutral goldstone bosons,
and one physical state, which is identified with a massive neutral Higgs boson
\ba
m^{2}_{H_{4}^{0}}=\frac{a_{\lambda}}{\sqrt{2}}\left(\frac{v_{1}v_{2}}{v_{S}}
+\frac{v_{1}v_{S}}{v_{2}}+\frac{v_{S}v_{2}}{v_{1}}\right).
\ea
From the diagonalization procedure we obtain 
\ba
O^{odd}=\left(
\begin{array}{ccc}
-\frac{v_S}{\sqrt{v_S^2+v_2^2} } & -\frac{v_S}{\sqrt{v_S^2+v_1^2} } &\frac{v_1 v_2}{\sqrt{v_1^2 v_2^2 +v^2 v_S^2} } \\
 0& \frac{v_1}{\sqrt{v_S^2+v_1^2} }& \frac{v_2 v_S}{\sqrt{v_1^2 v_2^2 + v^2 v_S^2} } \\
 \frac{v_2}{\sqrt{v_S^2+v_2^2} } & 0 & \frac{v_1 v_S}{\sqrt{v_1^2 v_2^2 +v^2 v_S^2} }
\end{array}
\right)
\ea
and the states are given by 
\ba
&&G_1^{0}=\frac{v_2 \textrm{Im}\,H_2^0-v_S \textrm{Im}\,S}{\sqrt{v_2^2+v_S^2}},
\nonumber\\
&&G_2^{0}=\frac{v_1 \textrm{Im}\,H_1^0 - v_S \textrm{Im}\,S}{\sqrt{v_1^2+v_S^2}},
\nonumber\\
&&H_{4}^{0}=\frac{v_1 v_2\textrm{Im}\,S +v_S v_2 \textrm{Im}\,H_1^0 
+v_1 v_S\textrm{Im}\,H_2^0}{\sqrt{v_1^2 v_2^2 + v_S^2 v^2}},
\ea
where $G_1^{0}$ and $G_2^{0}$ are two Goldstone modes, while $H_{4}^{0}$ is the physical
Higgs. 

Having identified the goldstones of the potential in the CP-odd sector, the parallel identification of the goldstones of the massive gauge bosons after EWSB is performed by an analysis of the bilinear mixings. In fact, from the Lagrangean density we can extract the following derivative coupling terms
\ba
{\cal L}_{DC}&=&\frac{1}{2} g_{2}W^{3}_{\mu}\partial^{\mu}G_Y - \frac{1}{2}g_{Y}A^{Y}_{\mu}\partial^{\mu} G_Y
+ \frac{1}{2}g_B B_{\mu}\partial^{\mu}G_B
\ea
where we have defined
\ba
&&G_Y=(v_{1}\textrm{Im}~H_1^0-v_{2}\textrm{Im}~H_2^0)
\nonumber\\
&&G_B=(B_{H_{1}}v_{1}\textrm{Im}~H_1^0+B_{H_{2}}v_{2}
\textrm{Im}~H_2^0+B_{S}v_{S}\textrm{Im}~S)+\frac{2 M_{st}}{g_B}~\textrm{Im}~b
\ea
which can be rotated onto the basis $(A^{\gamma}_{\mu},Z_{\mu},Z'_{\mu})$ using the $O^{A}_{susy}$ matrix
\ba
&&W^3_{\mu}=O^{A}_{W\g}A^{\g}_{\mu}+O^{A}_{WZ}Z_{\mu}+O^{A}_{WZ'}Z'_{\mu}
\nonumber\\
&&A^{Y}_{\mu}=O^{A}_{Y\g}A^{\g}_{\mu}+O^{A}_{YZ}Z_{\mu}+O^{A}_{YZ'}Z'_{\mu}
\nonumber\\
&&B_{\mu}=O^{A}_{B\g}A^{\g}_{\mu}+O^{A}_{BZ}Z_{\mu}+O^{A}_{BZ'}Z'_{\mu}
\ea
to obtain the expression for ${\cal L}_{DC}$ in terms of physical states
\ba
{\cal L}_{DC}&=& M_{Z} Z_{\mu} \partial^{\mu} G_Z + M_{Z'} Z'_{\mu} \partial^{\mu} G_{Z'}.
\ea
The two goldstone modes corresponding to the physical massive gauge bosons  are given by
\ba
&&M_Z G_Z=-A \left[\frac{v_1}{2 x_B} \left(f_1+\sqrt{f_1^2 + 4 g^2 x_B^2}\right) 
- v_1 g_B B_{H_1}\right]\textrm{Im}\,H^{0}_1
\nonumber\\
&&\hspace{1cm}+A \left[\frac{v_2}{2 x_B} \left(f_1+\sqrt{f_1^2 + 4 g^2 x_B^2}\right) 
+ v_2 g_B B_{H_2}\right] \textrm{Im}\,H^{0}_2
\nonumber\\
&&\hspace{1cm}
+B_S g_B v_S\,A\, \textrm{Im}\,S \,+\, 2 M_{st}\,A\, \textrm{Im}\,b
\nonumber\\\\
&&M_{Z'} G_{Z'}=
A' \left[\frac{v_1}{2 x_B} \left(\sqrt{f_1^2 + 4 g^2 x_B^2} -f_1\right) 
+ v_1 g_B B_{H_1}\right]\textrm{Im}\,H^{0}_1
\nonumber\\
&&\hspace{1cm}-A' \left[\frac{v_2}{2 x_B} \left(\sqrt{f_1^2 + 4 g^2 x_B^2} - f_1\right) 
- v_2 g_B B_{H_2}\right] \textrm{Im}\,H^{0}_2
\nonumber\\
&&\hspace{1cm}
+B_S g_B v_S\,A'\, \textrm{Im}\,S \,+\, 2 M_{st}\,A'\, \textrm{Im}\,b
\ea
where we have defined the following coefficients
\ba
A=\sqrt{\frac{1}{8}-\frac{f_1}{8\sqrt{f_1^2 +4g^2 x_B^2}}}
&&A'=\sqrt{\frac{1}{8}+\frac{f_1}{8\sqrt{f_1^2 +4g^2 x_B^2}}}.
\ea
It is simple to observe that $G_Z$ and $G_{Z'}$ are orthonormal.
At this point, a simple counting of the physical degrees of freedom before and after EWSB can give us a hint on the properties of this model.

Before EWSB we have ten degrees of freedom: two for $A^{Y}_{\mu}$, two for $W^3_{\mu}$, three for $B_{\mu}$,
two for the Higgs fields $\textrm{Im}H^{0}_1$ and $\textrm{Im}H^{0}_2$ and one for 
the singlet $\textrm{Im}~S$.
After the breaking, we are left with two polarization states for the physical photon,
three degrees of freedom for the $Z$ and the $Z'$ respectively,
one neutral Higgs state $H^{0}_4$ and one physical state which we are going to identify as the axi-Higgs.
Therefore we can build this new physical state requiring
its orthogonality with respect to the basis $\left\{H^{0}_4,G_{Z},{G}_{Z'}\right\}$
where $H^{0}_4$, identified as the physical direction of the potential, clearly belongs to the CP-odd sector.
We start from the following linear combination
\ba
\chi=b_{1}\textrm{Im}~H_1^0+b_{2}\textrm{Im}~H_2^0+b_{3}\textrm{Im}~S+b_{4}\textrm{Im}~b
\ea
and we determine the coefficients $b_1,\dots,b_4$ by the following 
constraints
\begin{eqnarray}
&&Y_1=b_{3}v_{1}v_{2}+b_{2}v_{1}v_{S}+b_{1}v_{2}v_{S}=0,
\nonumber\\
&&Y_2=4 b_4 M_{st}x_B +2 b_3 B_S v_S g_B x_B- b_1 v_1 (f_1-2 B_{H_1}g_B x_B+\sqrt{f_1^2 + 4 g^2 x_B^2} )
\nonumber\\
&&\hspace{1cm}
+b_2 v_2 (f_1 + 2 B_{H_2} g_B x_B + \sqrt{f_1^2 + 4 g^2 x_B^2} )=0
\nonumber\\
&&Y_3= 4 b_4 M_{st} x_B +2 b_3 B_S v_S g_B x_B+ b_2 v_2 (f_1+2 B_{H_2}g_B x_B-\sqrt{f_1^2 + 4 g^2 x_B^2} )
\nonumber\\
&&\hspace{1cm}
+b_1 v_1 (-f_1 + 2 B_{H_1}g_B x_B+\sqrt{f_1^2 + 4 g^2 x_B^2} )=0,
\end{eqnarray}
which give 
\begin{eqnarray}
b_{1}&=&b_{4}\frac{2 M_{st}}{g_B B_{S}}\frac{v_{1}v_{2}^{2}}{(v_{1}^{2}v_{2}^{2}
+v^{2}v_{S}^{2})}
\nonumber\\
b_{2}&=&b_{4}\frac{M_{st}}{4g_BB_{S}}\frac{v_{1}^{2}v_{2}}{(v_{1}^{2}v_{2}^{2}
+v^2 v_{S}^{2})}
\nonumber\\
b_{3}&=&-b_{4}\frac{M_{st}}{4g_BB_{S}}\frac{v^2 v_{S}}{(v_{1}^{2}v_{2}^{2}+v^{2}v_{S}^{2})},
\nonumber
\end{eqnarray}
where the coefficient $b_4$ is constrained by the normalization of the eigenstates. The physical axion will be given by
\ba
&&\chi=
\frac{1}{N_{\chi}}\left[2 M_{st}v_1 v_2^2\, \textrm{Im}H^0_1 + 2 M_{st}v_1^2 v_2\, \textrm{Im}H^0_2 
-2 M_{st}v^2 v_S \,\textrm{Im}\,S + B_S\,g_B (v^2 v_S^2 + v_1^2 v_2^2)\textrm{Im}\,b\right]
\nonumber\\
&&N_{\chi}=\sqrt{4 M_{st}^2 v^2 (v^2 v_S^2 + v_1^2 v_2^2)+B_S^2 g_B^2 (v^2 v_S^2 + v_1^2 v_2^2)^2}
\ea
where the new identified state has a nonvanishing projection over the St\"uckelberg field. Re-expressing 
$\textrm{Im}\, b$ in terms of $\chi$ and the goldstone modes of the massive gauge bosons, we discover that  the axion-like interactions (Wess-Zumino terms) mediated by the St\"uckelberg field can be rotated over $\chi$, giving trilinear vertices of the form $\chi F_I\wedge F_J$, where $I$ and $J$ denote the physical gauge bosons.
 
The rotation matrix $O^{\chi}_{susy}$ that rotates the physical components and the goldstones in the CP-odd sector takes the form
\ba
\left(
\begin{array}{c}
H_{4}^{0}\\
G_Z\\
G_Z'\\
\chi
\end{array}
\right)=
(O^{\chi}_{susy})
\left(
\begin{array}{c}
\textrm{Im}~H_1^0\\
\textrm{Im}~H_2^0\\
\textrm{Im}~S\\
\textrm{Im}~b
\end{array}
\right),
\ea
where all the entries are defined in Appendix B.

\subsubsection{The $B_S=0$ case: no physical axions}

In the case $B_{S}=0$, corresponding to a singlet of the entire gauge symmetry, we can proceed in the same way, obtaining, however, a different result compared to the previous case. In this case the general structure of the 
scalar potential can be modified by introducing linear or cubic terms in $\hat{S}$,
corresponding to the same structure of the nMSSM or of the NMSSM, with an additional $U(1)_{B}$ symmetry.
Adding a linear term we obtain
\footnote{At this stage we do not consider a cubic term in $\hat{S}$ in order to avoid 
the problem related to the formation of cosmological domain walls (see
\cite{Balazs:2007pf}, \cite{Panagiotakopoulos:1998yw}, \cite{Abel:1995wk}),
though even in this case one has two Higgs bosons and one Goldstone mode 
in the CP-odd sector.}
\begin{eqnarray}
V&=&\vert\lambda H_{1}\cdot H_{2} +\frac{m_{12}^2}{\lambda}\vert ^{2}
+\vert\lambda S\vert ^{2}(\vert H_{1}\vert^{2}+\vert H_{2}\vert^{2})
+\frac{1}{8}(g_2^{2}+g_Y^{2})(H_{1}^{\dagger}H_{1}-H_{2}^{\dagger}H_{2})^{2}\nonumber\\
&&+\frac{g_B^{2}}{8}B_{H_{1}}^2(H_{1}^{\dagger}H_{1}-H_{2}^{\dagger}H_{2})^2
+\frac{g_2^{2}}{2}\vert H_{1}^{\dagger}H_{2}\vert^{2}
+m_{1}^{2}\vert H_{1}\vert^{2}+m_{2}^{2}\vert H_{2}\vert^{2}\nonumber\\
&&+m_{S}^{2}\vert S\vert^{2} + (a_{\lambda} S H_{1} \cdot H_{2} + t_S S + h.c.),
\end{eqnarray}
where we have introduced the mass parameter $m_{12}^2/\lambda$ - which is the 
coefficient of $\hat{S}$ in the nMSSM superpotential - and $t_S$ , which is 
the coefficient of $\hat{S}$ in the soft breaking Lagrangean and has mass dimension three.
Notice that we have used the condition $B_{H_{1}}=-B_{H_{2}}$.
In this case, in the basis $\{\textrm{Im}\,S,\textrm{Im}\,H^{0}_1,\textrm{Im}\,H^{0}_2\}$, 
the CP-odd mass matrix is given by
\ba
{\cal M}_{odd}^{2}=\left(
\begin{array}{ccc}
-t_{S}\frac{\sqrt{2}}{v_{S}}-a_{\lambda}\frac{v_{1}v_{2}}{\sqrt{2} v_{S}} & -a_{\lambda}\frac{v_{2}}{\sqrt{2}} & -a_{\lambda}\frac{v_{1}}{\sqrt{2}}\\
-a_{\lambda}\frac{v_{2}}{\sqrt{2}} & -\frac{v_{2}}{v_{1}}(m_{12}^{2}+a_{\lambda}\frac{v_{S}}{\sqrt{2}}) & -m_{12}^{2}-a_{\lambda}\frac{v_{S}}{\sqrt{2}}\\
-a_{\lambda}\frac{v_{1}}{\sqrt{2}} & -m_{12}^{2}-a_{\lambda}\frac{v_{S}}{\sqrt{2}} & -\frac{v_{1}}{v_{2}}(m_{12}^{2}+a_{\lambda}\frac{v_{S}}{\sqrt{2}})
\end{array}
\right)
\ea
This sector provides two physical Higgs states and one goldstone mode of the form
\footnote{The same goldstone mode can be obtained from the NMSSM scalar potential \cite{Miller:2003ay}.}
\beq
G^{0}_{nMSSM}=\frac{1}{\sqrt{v_1^2+v_2^2}}\left(v_1 \textrm{Im}\,H^{0}_1-v_2 \textrm{Im}\,H^{0}_2\right).
\eeq
The other goldstone mode is obtained from the derivative coupling of the St\"uckelberg term ($B^{\mu}\partial_{\mu}\textrm{Im}\,b$).

Thus, from the derivative couplings, once we have performed a rotation on the physical basis,
we obtain the two orthogonal Goldstone modes $G_Z,G_{Z'}$ corresponding to 
the $Z$ and the $Z'$ bosons, which are a linear combination of $\textrm{Im}\,b$ and 
of the Goldstone mode obtained from the CP-odd sector,
\ba
G_Z= \alpha_1 G^{0}_{nMSSM} + \alpha_2 \textrm{Im}\,b ,
&&
G_{Z'}= \alpha'_1 G^{0}_{nMSSM} + \alpha'_2 \textrm{Im}\,b,
\ea
where the coefficients $\alpha_1\dots,\alpha'_2$ are not given in an explicit form
for simplicity. 

In this case the number of degrees of freedom before the symmetry breaking
is again equal to ten. In fact we have two for $W_3$, three for $B$, two for $Y$ 
and finally $\textrm{Im}\,H^{0}_1$, $\textrm{Im}\,H^{0}_2$ and $\textrm{Im}\,b$. 
After EWSB we are left with three degrees of freedom for the $Z$, three for the $Z'$,
two for the photon and two neutral higgs states, which are physical.
Therefore we do not have Higgs-axion mixing.

\section{The sfermion sector}

Coming to the scalar fermion sector (sfermions), 
the Lagrangean in terms of component fields is given by
\begin{eqnarray}
{\cal L}_{sfer}^{MSSM}&=&-\lambda\, y_{e}[S^{\dagger} H_{2}^{\dagger}\tilde{L}\tilde{R}
+S\tilde{L}^{\dagger}H_{2}\tilde{R}^{\dagger}]
-\lambda\, y_{d}[S^{\dagger}H_{2}^{\dagger}\tilde{Q}\tilde{D}_{R}
+S\tilde{Q}^{\dagger}H_{2}\tilde{D}_{R}^{\dagger}]
\nonumber\\
&&-\lambda\, y_{u}[S^{\dagger}H_{1}^{\dagger}\tilde{Q}\tilde{U}_{R}
+S\tilde{Q}^{\dagger}H_{1}\tilde{U}_{R}^{\dagger}]
-y_{e}^{2}[H_{1}^{\dagger}H_{1}(\tilde{L}^{\dagger}\tilde{L}
+\tilde{R}^{\dagger}\tilde{R})-H_{1}^{\dagger}\tilde{L}(H_{1}^{\dagger}\tilde{L})^{\dagger}]
\nonumber\\
&&-y_{d}^{2}[H_{1}^{\dagger}H_{1}(\tilde{Q}^{\dagger}\tilde{Q}
+\tilde{D}_{R}^{\dagger}\tilde{D}_{R})-H_{1}^{\dagger}\tilde{Q}(H_{1}^{\dagger}\tilde{Q})^{\dagger}]
\nonumber\\
&&-
y_{u}^{2}[H_{2}^{\dagger}H_{2}(\tilde{Q}^{\dagger}\tilde{Q}
+\tilde{U}_{R}^{\dagger}\tilde{U}_{R})
-H_{2}^{\dagger}\tilde{Q}(H_{2}^{\dagger}\tilde{Q})^{\dagger}]
-M^{2}_{L}\tilde{L}^{\dagger}\tilde{L}-m^{2}_{R}\tilde{R}^{\dagger}\tilde{R}
\nonumber\\
&&-M^{2}_{Q}\tilde{Q}^{\dagger}\tilde{Q}
-m^{2}_{U_{R}}\tilde{U}_{R}^{\dagger}\tilde{U}_{R}
-m^{2}_{D_{R}}\tilde{D}_{R}^{\dagger}\tilde{D}_{R}-(a_{e}H_{1}\cdot\tilde{L}\tilde{R}+h.c.)
\nonumber\\
&&-(a_{d}H_{1}\cdot\tilde{Q}\tilde{D}_{R}+h.c.)
-(a_{u}H_{2}\cdot\tilde{Q}\tilde{U}_{R}+h.c.)
\nonumber\\
&&-\frac{g_2^{2}}{2}(\tilde{L}^{\dagger}\tau^{i}\tilde{L}
+\tilde{Q}^{\dagger}\tau^{i}\tilde{Q}
+H_{1}^{\dagger}\tau^{i}H_{1}+H_{2}^{\dagger}\tau^{i}H_{2})^2
\nonumber\\
&&-\frac{g_s^{2}}{2}(\tilde{Q}^{\dagger}T^{a}\tilde{Q} 
+ \tilde{U_R}^{\dagger}T^{a}\tilde{U_R}
+\tilde{D_R}^{\dagger}T^{a}\tilde{D_R})^2
\nonumber\\
&&-\frac{g_Y^{2}}{8}(\tilde{L}^{\dagger}\tilde{L}
-2\tilde{R}^{\dagger}\tilde{R}-\frac{1}{3}\tilde{Q}^{\dagger}\tilde{Q}
+\frac{4}{3}\tilde{U}_{R}^{\dagger}\tilde{U}_{R}
-\frac{2}{3}\tilde{D}_{R}^{\dagger}\tilde{D}_{R}
+H_{1}^{\dagger}H_{1}-H_{2}^{\dagger}H_{2})^{2}.
\nonumber\\
\end{eqnarray}
In the presence of an extra ${U(1)}_B$ an additional piece coming from the D-terms 
must be added to the sfermion Lagrangean and it is given by
\ba
&&{\cal L}^{{U(1)}_B}_{sfer}=-\frac{g_B^{2}}{8}(B_{L}\tilde{L}^{\dagger}
\tilde{L}+B_{R}\tilde{R}^{\dagger}\tilde{R}+B_{Q}\tilde{Q}^{\dagger}\tilde{Q}
+B_{U}\tilde{U}_{R}^{\dagger}\tilde{U}_{R}+B_{D}\tilde{D}_{R}^{\dagger}\tilde{D}_{R}
\nonumber\\
&&\hspace{2cm}+B_{H_{1}}H_{1}^{\dagger}H_{1}+B_{H_{2}}H_{2}^{\dagger}H_{2}+B_{S}S^{\dagger}S)^{2}.
\ea
After spontaneous symmetry breaking we get
\begin{eqnarray}
{\cal L}^{tot}_{sfer}&=&-\frac{1}{2}\lambda\,v_S y_{e}v_{2}[\tilde{L}^{2}\tilde{R}+\tilde{L}^{2\dagger}\tilde{R}^{\dagger}]
-\frac{1}{2}\lambda\,v_S y_{d}v_{2}[\tilde{Q}^{2}\tilde{D}_{R}+\tilde{Q}^{2\dagger}\tilde{D}_{R}^{\dagger}]
-\frac{1}{2}\lambda\,v_S y_{u}v_{1}[\tilde{Q}^{1}\tilde{U}_{R}+\tilde{Q}^{1\dagger}\tilde{U}_{R}^{\dagger}]
\nonumber\\
&&-\frac{1}{2}y_{e}^{2}v_{1}^{2}[\tilde{L}^{2\dagger}\tilde{L}^{2}+\tilde{R}^{\dagger}\tilde{R}]
-\frac{1}{2}y_{d}^{2}v_{1}^{2}[\tilde{Q}^{2\dagger}\tilde{Q}^{2}+\tilde{D}_{R}^{\dagger}\tilde{D}_{R}]
-\frac{1}{2}y_{u}^{2}v_{2}^{2}[\tilde{Q}^{1\dagger}\tilde{Q}^{1}+\tilde{U}_{R}^{\dagger}\tilde{U}_{R}]
\nonumber\\
&&-M^{2}_{L}\tilde{L}^{\dagger}\tilde{L}-m^{2}_{R}\tilde{R}^{\dagger}\tilde{R}
-M^{2}_{Q}\tilde{Q}^{\dagger}\tilde{Q}-m^{2}_{U_{R}}\tilde{U}_{R}^{\dagger}\tilde{U}_{R}-m^{2}_{D_{R}}\tilde{D}_{R}^{\dagger}\tilde{D}_{R}
\nonumber\\
&&-(a_{e}\frac{v_{1}}{\sqrt{2}}\tilde{L}^{2}\tilde{R}+h.c.)-(a_{d}\frac{v_{1}}{\sqrt{2}}\tilde{Q}^{2}\tilde{D}_{R}+h.c.)+(a_{u}\frac{v_{2}}{\sqrt{2}}\tilde{Q}_{1}\tilde{U}_{R}+h.c.)
\nonumber\\
&&-\frac{g_2^{2}}{8}(v_{1}^{2}-v_{2}^{2})(\tilde{L}^{1\dagger}\tilde{L}^{1}-\tilde{L}^{2\dagger}\tilde{L}^{2}+\tilde{Q}^{1\dagger}\tilde{Q}^{1}-\tilde{Q}^{2\dagger}\tilde{Q}^{2})
\nonumber\\
&&-\frac{g_Y^{2}}{8}(v_{1}^{2}-v_{2}^{2})(\tilde{L}^{\dagger}\tilde{L}-2\tilde{R}^{\dagger}\tilde{R}-\frac{1}{3}\tilde{Q}^{\dagger}\tilde{Q}+\frac{4}{3}\tilde{U}_{R}^{\dagger}\tilde{U}_{R}-\frac{2}{3}\tilde{D}_{R}^{\dagger}\tilde{D}_{R})
\nonumber\\
&&-\frac{g_B^{2}}{8}\left(B_{H_{1}}v_1^2+B_{H_{2}}v_2^2+B_{S}v_S^2\right)(B_{L}\tilde{L}^{\dagger}\tilde{L}+B_{R}\tilde{R}^{\dagger}\tilde{R} +B_{Q}\tilde{Q}^{\dagger}\tilde{Q}+B_{U}\tilde{U}_{R}^{\dagger}\tilde{U}_{R}
+B_{D}\tilde{D}_{R}^{\dagger}\tilde{D}_{R});
\nonumber\\
\end{eqnarray}
here and in what follows superscripts on $\tilde{L}$ and $\tilde{Q}$ specify the doublet components.\\
In the basis $(\tilde{L}^{2},\tilde{R}^{\dagger})$, the entries of the mass matrix are given by
\begin{eqnarray}
&&(M_{\tilde{L}^{2},\tilde{R}})_{11}=y_{e}^{2}\frac{1}{2}v_{1}^{2}+M^{2}_{L}
-\frac{1}{8}(g_2^{2}-g_Y^{2})(v_{1}^{2}-v_{2}^{2})
+\frac{g_B^{2}}{8}B_{L}(B_{H_{1}}v_{1}^{2}+B_{H_{2}}v_{2}^{2}+B_{S}v_{S}^{2}),
\nonumber\\
&&(M_{\tilde{L}^{2},\tilde{R}})_{12}=
(M_{\tilde{L}^{2},\tilde{R}})_{21}=\frac{1}{2}\lambda v_{S} y_{e}v_{2}+a_{e}\frac{v_{1}}{\sqrt{2}},
\nonumber\\
&&(M_{\tilde{L}^{2},\tilde{R}})_{22}=\frac{1}{2}y_{e}^{2}v_{1}^{2}+m^{2}_{R}
-\frac{1}{4}g_Y^{2}(v_{1}^{2}-v_{2}^{2})
+\frac{g_B^{2}}{8}B_{R}(B_{H_{1}}v_{1}^{2}+B_{H_{2}}v_{2}^{2}+B_{S}v_{S}^{2}).
\end{eqnarray}
The former matrix can be diagonalized through a rotation defined by 
\ba
\tan 2\theta_{\tilde{L}^{2},\tilde{R}}=\frac{(\lambda v_{S} y_{e}v_{2}+a_{e}\sqrt{2} v_{1})}
{m^{2}_{R}-M^{2}_{L}+\frac{1}{8}(g_2^{2}-3g_Y^{2})(v_{1}^{2}-v_{2}^{2})
+\frac{g_B^{2}}{8}(B_{R}-B_{L})(B_{H_{1}}v_{1}^{2}+B_{H_{2}}v_{2}^{2}+B_{S}v_{S}^{2})}.
\ea
The eigenvalues have very lengthy expressions and we will omit them for brevity.
The three eigenstates are given by
\begin{eqnarray}
\tilde{l}_{1}&=&\cos\theta_{\tilde{L}^{2},\tilde{R}}\tilde{L}^{2}+\sin\theta_{\tilde{L}^{2},\tilde{R}}\tilde{R}^{\dagger}\nonumber\\
\tilde{l}_{2}&=&-\sin\theta_{\tilde{L}^{2},\tilde{R}}\tilde{L}^{2}+\cos\theta_{\tilde{L}^{2},\tilde{R}}\tilde{R}^{\dagger}
\nonumber\\
\tilde{l}_{3}&=&\tilde{L}^{1}.
\end{eqnarray}
The mass of $\tilde{L}^{1}$ is given by
\ba
M^{2}_{\tilde{L}^{1}}=\frac{1}{8}(g_2^{2}+g_Y^{2})(v_{1}^{2}-v_{2}^{2})+\frac{g_B^{2}}{8}B_{L}(B_{H_{1}}v_{1}^{2}+B_{H_{2}}v_{2}^{2}+B_{S}v_{S}^{2}).
\ea
Using the two basis $(\tilde{Q}^{2},\tilde{D}_{R}^{\dagger})$ and 
$(\tilde{Q}^{1},\tilde{U}_{R}^{\dagger})$, the mass sector of the squarks can be written as
\begin{eqnarray}
{\cal L}_{squark}&=&-
\left(
\begin{array}{cc}
\tilde{Q}^{2\dagger}&\tilde{D}_{R}
\end{array}
\right)
M_{\tilde{Q}^{2},\tilde{D}_{R}}
\left(
\begin{array}{cc}
\tilde{Q}^{2}\\
\tilde{D}_{R}^{\dagger}
\end{array}
\right)-
\left(
\begin{array}{cc}
\tilde{Q}^{1\dagger}&\tilde{U}_{R}
\end{array}
\right)
M_{\tilde{Q}^{1},\tilde{U}_{R}}
\left(
\begin{array}{cc}
\tilde{Q}^{1}\\
\tilde{U}_{R}^{\dagger}
\end{array}
\right),
\end{eqnarray}
where the $M_{\tilde{Q}^{2},\tilde{D}_{R}}$ matrix is defined as
\begin{eqnarray}
&&(M_{\tilde{Q}^{2},\tilde{D}_{R}})_{11}=\frac{1}{2}y_{d}^{2}v_{1}^{2}+M^{2}_{Q}
-\frac{1}{8}(g_2^{2}+\frac{1}{3}g_Y^{2})(v_{1}^{2}-v_{2}^{2})
+\frac{g_B^{2}}{8}B_{Q}(B_{H_{1}}v_{1}^{2}+B_{H_{2}}v_{2}^{2}+B_{S}v_{S}^{2}),
\nonumber\\
&&(M_{\tilde{Q}^{2},\tilde{D}_{R}})_{12}=(M_{\tilde{Q}^{2},\tilde{D}_{R}})_{21}=
\frac{1}{2}\lambda v_{S} y_{d}v_{2}+a_{d}\frac{v_{1}}{\sqrt{2}},
\nonumber\\
&&(M_{\tilde{Q}^{2},\tilde{D}_{R}})_{22}=\frac{1}{2}y_{d}^{2}v_{1}^{2}+m^{2}_{D_{R}}
-\frac{1}{12}g_Y^{2}(v_{1}^{2}-v_{2}^{2})
+\frac{g_B^{2}}{8}B_{D_{R}}(B_{H_{1}}v_{1}^{2}+B_{H_{2}}v_{2}^{2}+B_{S}v_{S}^{2}),
\nonumber
\end{eqnarray}
while for the $M_{\tilde{Q}^{1},\tilde{U}_{R}}$ matrix we get
\begin{eqnarray}
&&(M_{\tilde{Q}^{1},\tilde{U}_{R}})_{11}=\frac{1}{2}y_{u}^{2}v_{2}^{2}
+M^{2}_{Q}+\frac{1}{8}(g_2^{2}-\frac{1}{3}g_Y^{2})(v_{1}^{2}-v_{2}^{2})
+\frac{g_B^{2}}{8}B_{Q}(B_{H_{1}}v_{1}^{2}+B_{H_{2}}v_{2}^{2}+B_{S}v_{S}^{2}),
\nonumber\\
&&(M_{\tilde{Q}^{1},\tilde{U}_{R}})_{12}=
(M_{\tilde{Q}^{1},\tilde{U}_{R}})_{21}=
\frac{1}{2}\lambda v_{S} y_{u}v_{1}-a_{u}\frac{v_{2}}{\sqrt{2}}
\nonumber\\
&&(M_{\tilde{Q}^{1},\tilde{U}_{R}})_{22}=\frac{1}{2}y_{u}^{2}v_{2}^{2}+m^{2}_{U_{R}}
+\frac{1}{6}g_Y^{2}(v_{1}^{2}-v_{2}^{2})
+\frac{g_B^{2}}{8}B_{U_{R}}(B_{H_{1}}v_{1}^{2} + B_{H_{2}}v_{2}^{2}+B_{S}v_{S}^{2}).
\end{eqnarray}
The $M_{\tilde{Q}^{2},\tilde{D}_{R}}$ matrix can be diagonalized using 
\begin{eqnarray}
\tilde{q}_{1}&=&\cos\theta_{\tilde{Q}^{2},\tilde{D}_{R}}\tilde{Q}^{2}
+\sin\theta_{\tilde{Q}^{2},\tilde{D}_{R}}\tilde{D}_{R}^{\dagger}
\nonumber\\
\tilde{q}_{2}&=&-\sin\theta_{\tilde{Q}^{2},\tilde{D}_{R}}\tilde{Q}^{2}
+\cos\theta_{\tilde{Q}^{2},\tilde{D}_{R}}\tilde{D}_{R}^{\dagger}
\nonumber,
\end{eqnarray}
where the $\theta_{\tilde{Q}^{2},\tilde{D}_{R}}$ angle is defined by 
\ba
\tan 2\theta_{\tilde{Q}^{2},\tilde{D}_{R}}=\frac{(\lambda v_{S} y_{d}v_{2}+a_{d}\sqrt{2}v_{1})}
{m^{2}_{D_{R}}-M^{2}_{Q} + \frac{1}{8}(g_2^{2}-\frac{1}{3}g_Y^{2})(v_{1}^{2}-v_{2}^{2})
+\frac{g_B^{2}}{8}(B_{D_{R}}-B_{Q})(B_{H_{1}}v_{1}^{2}
+B_{H_{2}}v_{2}^{2}+B_{S}v_{S}^{2})}.
\ea
Again, we omit the explicit expression of the eigenvalues since they are quite lengthy. 
The $M_{\tilde{Q}^{1},\tilde{U}_{R}}$ matrix can be diagonalized by the following choice
\begin{eqnarray}
\tilde{q}_{3}&=&\cos\theta_{\tilde{Q}^{1},\tilde{U}_{R}}\tilde{Q}^{1}
+\sin\theta_{\tilde{Q}^{1},\tilde{U}_{R}}\tilde{U}_{R}^{\dagger}
\nonumber\\
\tilde{q}_{4}&=&-\sin\theta_{\tilde{Q}^{1},\tilde{U}_{R}}\tilde{Q}^{1}
+\cos\theta_{\tilde{Q}^{1},\tilde{U}_{R}}\tilde{U}_{R}^{\dagger}
\nonumber,
\end{eqnarray}
where $\theta_{\tilde{Q}^{1},\tilde{U}_{R}}$ is defined by 
\ba
\tan 2\theta_{\tilde{Q}^{1},\tilde{U}_{R}}=\frac{(\lambda v_{S} y_{u}\sqrt{2}v_{1}-a_{u}\sqrt{2}v_{2})}{m^{2}_{U_{R}}-M^{2}_{Q}
-\frac{1}{8}(g_2^{2}-\frac{5}{3}g_Y^{2})(v_{1}^{2}-v_{2}^{2})
+\frac{g_B^{2}}{8}(B_{U_{R}}-B_{Q})
(B_{H_{1}}v_{1}^{2}+B_{H_{2}}v_{2}^{2}+B_{S}v_{S}^{2})}.
\ea
Using the parameter values specified in the numerical analysis of the neutralino sector, typical values for 
sfermion masses are around a few TeV.

\section{Wess-Zumino counterterms and Chern-Simons interactions}

The cancellation of the gauge anomalies in these supersymmetric models are 
obtained by the introduction of axion counterterms. The supersymmetric form
of the corresponding Lagrangean introduces, beside the usual bosonic 
contributions of the form $b F\wedge F$ additional 
interactions between the axion and the gauginos and between the axino, the 
gauge fields and the corresponding gauginos. 
It is given by
\begin{eqnarray}
{\cal L}_{C}&=&-\int d^{4}\theta\left\lbrace\left[
\frac{1}{2} b_{G} \,\textrm{Tr}({\cal G} {\cal G})\hat{{\bf b}}
+\frac{1}{2} b_{W} \,\textrm{Tr}(W W)\hat{{\bf b}}
\right.\right.
\nonumber\\
&&\left.\left.
+b_{Y}\hat{{\bf b}}W^{Y}_{\alpha} W^{Y,\alpha}
+b_{B}\hat{{\bf b}}W^{B}_{\alpha}W^{B,\alpha}+b_{YB}\hat{{\bf b}}W^{Y}_{\alpha}W^{B,\alpha}\right]
\delta(\bar{\theta}^{2})
+h.c.\right\rbrace. \nonumber\\
\end{eqnarray}
whose general e
Expanding this expression in component fields using the WZ gauge we obtain
\ba
&&{\cal L}_{C}=-\frac{1}{8}\,b_G \,\epsilon^{\mu\nu\rho\sigma}
G^{a}_{\mu\nu}G^{a}_{\rho\sigma}\,\textrm{Im} b\,
-\frac{1}{8}b_W\,\epsilon^{\mu\nu\rho\sigma}
W^{i}_{\mu\nu}W^{i}_{\rho\sigma}\,\textrm{Im} b\,
\nonumber\\
&&-\frac{1}{4}b_Y\epsilon^{\mu\nu\rho\sigma}F^{Y}_{\mu\nu}F^{Y}_{\rho\sigma}\,\textrm{Im} b\,
-\frac{1}{4}b_B\epsilon^{\mu\nu\rho\sigma}F^{B}_{\mu\nu}F^{B}_{\rho\sigma}\,\textrm{Im} b\,
-\frac{1}{4}b_{YB}\epsilon^{\mu\nu\rho\sigma}F^{Y}_{\mu\nu}F^{B}_{\rho\sigma}\,\textrm{Im} b\,
\nonumber\\
&&+\,b_G[\,\textrm{Im} b\,
\frac{1}{2}(\lambda_{g^{a}}\sigma^{\mu}D_{\mu}\bar{\lambda}_{g^{a}})
-\frac{i}{2\sqrt{2}}\psi_{\bf b} \frac{1}{2}(\lambda_{g^{a}}
\sigma^{\mu}\bar{\sigma}^{\nu}G^{a}_{\mu\nu})
+\frac{1}{2}F_{\bf b}\frac{1}{2}(\lambda_{g^{a}}\lambda_{g^{a}})
\nonumber\\
&&+\frac{1}{\sqrt{2}}\psi_{\bf b}\frac{1}{2}(\lambda_{g^{a}}D_{G}^{a})+h.c.]
+b_W[\,\textrm{Im} b\, \frac{1}{2}(\lambda^{a}\sigma^{\mu}D_{\mu}\bar{\lambda}^{a})\
-\frac{i}{2\sqrt{2}}\psi_{\bf b} \frac{1}{2}(\lambda_{W^{i}}
\sigma^{\mu}\bar{\sigma}^{\nu}W^{i}_{\mu\nu})
\nonumber\\
&&+\frac{1}{2}F_{\bf b}\frac{1}{2}(\lambda_{W^{i}}\lambda_{W^{i}})
+\frac{1}{\sqrt{2}}\psi_{\bf b}\frac{1}{2}(\lambda_{W^{i}} D^{i})+h.c.]
+b_Y[\,\textrm{Im} b\, \lambda_Y\sigma^{\mu}D_{\mu}\bar{\lambda}_Y
-\frac{i}{2\sqrt{2}}\psi_{\bf b} \lambda_Y\sigma^{\mu}\bar{\sigma}^{\nu}F^{Y}_{\mu\nu}
\nonumber\\
&&+\frac{1}{2}F_{\bf b}\lambda_Y\lambda_Y
+\frac{1}{\sqrt{2}}\psi_{\bf b} \lambda_Y\,D_{Y}+h.c.]
+b_B[\,\textrm{Im} b\, \lambda_B\sigma^{\mu}D_{\mu}\bar{\lambda}_B
-\frac{i}{2\sqrt{2}}\psi_{\bf b} \lambda_B \sigma^{\mu}\bar{\sigma}^{\nu} F^{B}_{\mu\nu}
\nonumber\\
&&+\frac{1}{2}F_{\bf b}\lambda_B\lambda_B+\frac{1}{\sqrt{2}}
\psi_{\bf b}\lambda_B\,D_{B}+h.c.]
+b_{YB}[(\,\textrm{Im}b\,\lambda_Y\sigma^{\mu}\partial_{\mu}\bar{\lambda}_B
+\frac{1}{2}F_{\bf b}\lambda_Y \lambda_B
\nonumber\\
&&+\frac{1}{\sqrt{2}}\psi_{\bf b}
\lambda_Y\, D_{B} -\frac{i}{2\sqrt{2}}\lambda_Y
\sigma^{\mu}\bar{\sigma}^{\nu}F^B_{\mu\nu}\psi_{\bf b})+(Y\leftrightarrow B)+h.c.],
\nonumber\\
\label{Laxioncomp}
\ea
where we have additional contributions for the cancellation of 
the  $U(1)_B SU(3) SU(3)$ anomaly, which are typical of this model and 
are not present in previous similar formulations \cite{Anastasopoulos:2008jt}.

\subsection{The Chern-Simons Lagrangean}

As we have mentioned above, the Chern-Simons Lagrangean describes the freedom to re-distribute the anomaly in the trilinear gauge interactions of $AVV$ and $AAA$ type. In a bottom-up description of these models this freedom is equivalently formulated in terms of external Ward identities on the anomalous vertices. The corresponding Lagrangean is similar to the one given in \cite{Anastasopoulos:2008jt}, now with the addition of the 
gluonic terms. It takes the form 
\begin{eqnarray}
{\cal L}_{CS}&=&-\int d^{4}\theta\, \left\{ 
c_{1} \left[ (\hat{Y} D^{\alpha} \hat{B}-\hat{B} D^{\alpha}\hat{Y}) W_{\alpha}^B+h.c.\right]
\right.\nonumber\\
&&\left.-c_{2}\left[ (\hat{Y} D^{\alpha}\hat{B}-\hat{B} D^{\alpha}\hat{Y})W_{\alpha}^Y + h.c.\right]
\right.\nonumber\\
&&\left.-c_{3}\textrm{Tr}\left[(\hat{W} D^{\alpha}\hat{B}-\hat{B} D^{\alpha}\hat{W})W_{\alpha}
+\frac{1}{6}\hat{W}D^{\alpha}\hat{B} \bar{D}^{2}[D_{\alpha}\hat{W},\hat{W}]+h.c.\right]
\right.\nonumber\\
&&\left.-\,c_{4}\textrm{Tr}\left[(\hat{G} D^{\alpha}\hat{B}-\hat{B} D^{\alpha}\hat{G}){\cal G}_{\alpha}
+\frac{1}{6}\hat{G} D^{\alpha}\hat{B} \bar{D}^{2}[D_{\alpha}\hat{G},\hat{G}]+h.c.\right]
\right\}
\end{eqnarray} 
where the coefficients $c_1\dots c_4$ will be determined by the generalized Ward identities of the model.
Expanding this expression in terms of component fields we get
\begin{eqnarray}
{\cal L}_{CS}&=&
-c_{1}\epsilon^{\mu\nu\rho\sigma}B_{\mu} Y_{\nu} F_{\rho\sigma}^{B}
+c_{2}\epsilon^{\mu\nu\rho\sigma}B_{\mu} Y_{\nu} F_{\rho\sigma}^{Y}
+c_{3}\epsilon^{\mu\nu\rho\sigma}B_{\mu} \textrm{Tr}
\left(W_{\nu}F_{\rho\sigma} -\frac{i}{3}W_{\nu}[W_{\rho},W_{\sigma}]\right)
\nonumber\\
&&+c_{4}\epsilon^{\mu\nu\rho\sigma}B_{\mu}\textrm{Tr}
\left(G_{\nu}G_{\rho\sigma} -\frac{i}{3}G_{\nu}[G_{\rho},G_{\sigma}]\right)
-c_{1}\,(\lambda_B \sigma^{\mu}\bar{\lambda}_B A^{Y}_{\mu}
-\lambda_B\sigma^{\mu}\bar{\lambda}_{Y} B_{\mu}+h.c.)
\nonumber\\
&&+c_{2}\,(\lambda_Y\sigma^{\mu}\bar{\lambda}_Y B_{\mu}
-\lambda_Y \sigma^{\mu}\bar{\lambda}_B A^{Y}_{\mu}+h.c.)
+c_{3}\,\textrm{Tr}(\lambda_W\sigma^{\mu}\bar{\lambda_W}B_{\mu}-\lambda_W \sigma^{\mu}\bar{\lambda}_B W_{\mu}+h.c.)
\nonumber\\
&&+\,c_{4}\,\textrm{Tr}(\lambda_g\sigma^{\mu}\bar{\lambda}_{g} B_{\mu}
-\lambda_g \sigma^{\mu}\bar{\lambda}_B G_{\mu}+h.c.). 
\label{CScomponents}
\end{eqnarray}
The role of the Lagrangean is to redistribute the anomaly among the three anomalous 
vertices when the symmetry of the interaction is not enough to 
fix the partial contributions to the anomaly uniquely.

\section{Generalized broken Ward identities}

The anomaly cancellation mechanism for this supersymmetric model proceeds as in \cite{Coriano:2005js,Coriano:2007fw,Coriano:2007xg,Armillis:2007tb,
Coriano:2008pg,Armillis:2008bg}, where a detailed description of some physical cases can be found.
The resulting anomalies must be cancelled in the abelian sector $BBB,BYY,YBB$ and in the 
non-abelian $SU(2)$ and $SU(3)$ sectors.
If we start by using a parametrization of the one-loop trilinear gauge interactions 
with a symmetric distribution of the $AAA$ anomaly vertex ($\Delta_{AAA}$), in which we denote with
$-k_3=k_1+k_2$ the incoming momentum with the $\lambda$ index and with $k_1,k_2$ 
the outgoing momenta, with indices $\mu$ and $\nu$ respectively, 
we can introduce generalized Ward identities in the momentum space as defining conditions on the model. We obtain
\ba
&&k_{3,\lambda}{\cal A}_{BBB}\Delta_{AAA}^{\lambda\mu\nu}(k_3,k_1,k_2) 
-\frac{1}{4}b_B\,\varepsilon^{\mu\nu\alpha\beta}k_{1,\alpha}k_{2,\beta} - 2 m_f \Delta^{\mu\nu}_{BB}=0,
\ea
for the $BBB$ case, and analogous conditions in the other sectors. The expressions of $\Delta_{AAA}$,
$\Delta_{BB}$ and similar are given below; $m_f$ denotes the mass of the fermion in the anomaly loop.

\begin{figure}[ht]
{\centering \resizebox*{11cm}{!}{\rotatebox{0}
{\includegraphics{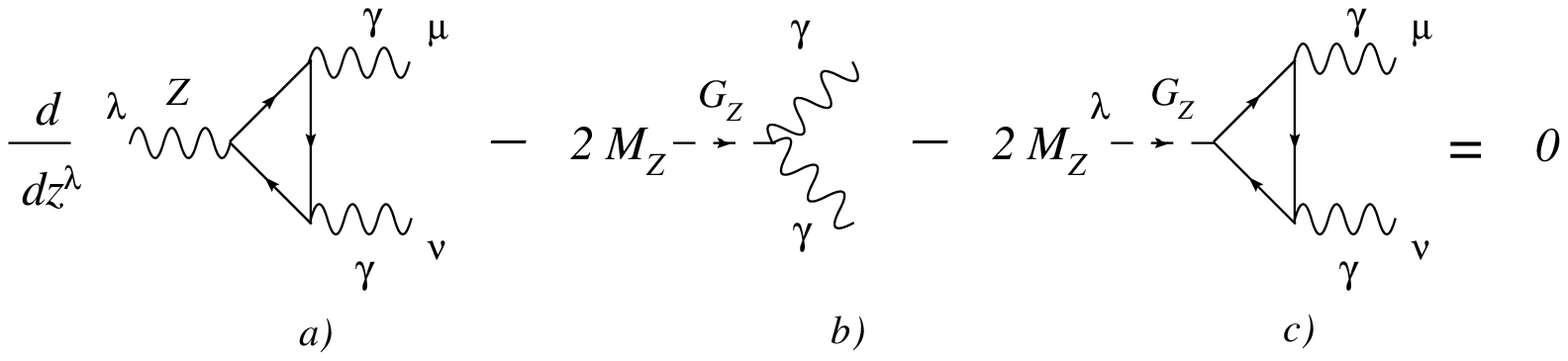}}}\par}
\caption{\small The generalized Ward identity for the $Z\gamma\gamma$ 
vertex in our anomalous model 
away from the chiral limit. The analogous STI for 
the SM case consists of only diagrams a) and c).}
\label{anomdiag3}
\end{figure}

Other two Ward identities are obtained 
by a cyclic permutation of the momenta. Also, notice that in this specific case we do not have 
Chern-Simons interactions in the defining condition. 
For a $BYY$ triangle we have
\ba
&&k_{3,\lambda}\left[{\cal A}_{BYY}\Delta_{AAA}^{\lambda\mu\nu}(k_3,k_1,k_2) 
-c_2 \varepsilon^{\lambda\mu\nu\alpha}(k_1-k_2)_{\alpha}\right]
-\frac{1}{4}b_Y\,\varepsilon^{\mu\nu\alpha\beta}k_{1,\alpha}k_{2,\beta}
- 2 m_f \Delta^{\mu\nu}_{YY}=0,
\nonumber\\
&&k_{1,\mu}\left[{\cal A}_{BYY}\Delta_{AAA}^{\lambda\mu\nu}(k_3,k_1,k_2) 
-c_2 \varepsilon^{\lambda\mu\nu\alpha}(k_1-k_2)_{\alpha}\right]- 2 m_f \Delta^{\lambda\nu}_{YY}=0,
\nonumber\\
&&k_{2,\nu}\left[{\cal A}_{BYY}\Delta_{AAA}^{\lambda\mu\nu}(k_3,k_1,k_2) 
-c_2 \varepsilon^{\lambda\mu\nu\alpha}(k_1-k_2)_{\alpha}\right]- 2 m_f \Delta^{\lambda\mu}_{YY}=0,
\ea
where the tensor structure of the triangles is given below.
For a $YBB$ triangle we have 
\ba
&&k_{3,\lambda}\left[{\cal A}_{YBB}\Delta_{AAA}^{\lambda\mu\nu}(k_3,k_1,k_2) 
-c_1 \varepsilon^{\lambda\mu\nu\alpha}(k_1-k_2)_{\alpha}\right]- 2 m_f \Delta^{\mu\nu}_{BB}=0,
\nonumber\\
&&k_{1,\mu}\left[{\cal A}_{YBB}\Delta_{AAA}^{\lambda\mu\nu}(k_3,k_1,k_2) 
-c_1 \varepsilon^{\lambda\mu\nu\alpha}(k_1-k_2)_{\alpha}\right]
-\frac{1}{4}b_{YB}\,\varepsilon^{\lambda\nu\alpha\beta}k_{2,\alpha}k_{3,\beta}- 2 m_f \Delta^{\lambda\nu}_{BB}=0,
\nonumber\\
&&k_{2,\nu}\left[{\cal A}_{YBB}\Delta_{AAA}^{\lambda\mu\nu}(k_3,k_1,k_2) 
-c_1 \varepsilon^{\lambda\mu\nu\alpha}(k_1-k_2)_{\alpha}\right]
-\frac{1}{4}b_{YB}\,\varepsilon^{\lambda\mu\alpha\beta}k_{3,\alpha}k_{1,\beta}- 2 m_f \Delta^{\lambda\mu}_{BB}=0,
\nonumber\\
\ea
where the coefficients $c_1,c_2$ are fixed by the BRST invariance under $U(1)_Y$.
The explicit form of the tensors $\Delta_{AAA}^{\lambda\mu\nu}$ and $\Delta^{\mu\nu}_{BB}$, 
in terms of Feynman integrals, are given by
\ba
&&\Delta_{AAA}^{\lambda\mu\nu}(m_f \neq 0)=\frac{1}{\pi^2}\int_0^1 dx\int_{0}^{1-x}dy\frac{1}{\Delta(m_f)}
\left\{
\right.\nonumber\\
&&\left.\hspace{2cm}
\varepsilon[k_1,\lambda,\mu,\nu]\left[-\frac{\Delta(m_f)-m_f^2}{3} + k_2\cdot k_2 y(y-1)- x y k_1\cdot k_2\right]
\right.\nonumber\\
&&\left.\hspace{2cm}
+\varepsilon[k_2,\lambda,\mu,\nu]\left[\frac{\Delta(m_f)-m_f^2}{3}- k_1\cdot k_1 x(x-1)+ x y k_1\cdot k_2\right]
\right.\nonumber\\
&&\left.\hspace{2cm}
+\varepsilon[k_1,k_2,\lambda,\nu](k_1^{\mu} x(x-1)- x y k_2^{\mu})
\right.\nonumber\\
&&\left.\hspace{2cm}
+\varepsilon[k_1,k_2,\lambda,\mu](k_2^{\nu} y(1-y)+ x y k_1^{\nu})
\right\}\,,
\ea
and
\ba
\Delta^{\mu\nu}_{BB}=-\frac{m_f}{3\pi^2}\varepsilon^{\mu\nu\alpha\beta}k_{1\alpha}k_{2\beta}
\int_0^1\int_0^{1-x}dx dy\frac{1}{\Delta(m_f)},
\ea
where $\Delta(m_f)=[m_f^2 + (y-1)y k_2^2 +(x-1)x k_1^2 -2 xyk_1\cdot k_2]$\,.
For $\Delta^{\mu\nu}_{YY}$ and $\Delta^{\mu\nu}_{YB}$ we obtain similar expressions. The same relations can be reformulated 
in the mass eigenstate basis in terms of the physical gauge bosons $Z$ and $Z^{\prime}$. The structure of the (generalized) Ward identity in this case is shown in Fig. \ref{anomdiag3}, written in configuration space,  where the first term corresponds to the anomaly, the second is the axion counterterm projected out on the goldstone $G_Z$, and the third diagram describes the mass corrections due to the coupling of the goldstone to the massive fermion in the loop. In the chiral limit, obviously, the third term is absent. 

The generalized Ward identities for the case $U(1)_B \,SU(2)\, SU(2)$ have similar expressions, while
the case $U(1)_B\, SU(3)\, SU(3)$ requires a further comment.
As a matter of fact, in this case the higgsinos do not circulate in the loop, but 
the $BGG$ triangle exhibits an anomaly when $B_S\neq 0$, (see Eq.(\ref{anomalies1})).
For the same reason we do not have a $BGG$ anomaly in the MLSOM \cite{Coriano:2005js} 
(Minimal Low Scale Orientifold Model) case 
when the Higgs charges under $U(1)_{B}$ are equal.

\section{$Z$ decay into four fermions: Chern-Simons interactions}

One interesting signature of trilinear anomalous vertices involving three anomalous 
gauge bosons can be investigated in the decay process of the $Z/Z'$ into four fermions by the mediation
of two extra anomalous currents. 
This kind of process is phenomenologically relevant since it is sensitive to the presence
of (at least) two or more extra anomalous $U(1)$.
As a matter of fact, in the MLSOM (non supersymmetric case) 
in the presence of an abelian symmetry given by $G_1=U(1)_{Y}\times U(1)_{B}$ 
where $B$ is anomalous, the off-shell effective vertex does not contain any Chern-Simons interaction
by construction. If we take, for instance, the triangle $\langle ZZ'Z'\rangle$, some of 
the relevant effective vertices coming from the interaction eigenstate basis which 
have an anomalous component are $\langle B B B\rangle$ and $\langle Y B B\rangle$. 
In the $BBB$ case the Chern-Simons interaction vanishes trivially, while in the $YBB$ case
the corresponding Chern-Simons counterterm must be ``absorbed'' in a redefinition of the triangle
in order to ensure the BRST invariance. Equivalently, the $YBB$ vertex does not allow a partial anomaly on the $Y$ leg, since there is no axion for $Y$.
An analysis of the anomalous trilinear interactions in the context of the MLSOM can be found in \cite{Armillis:2007tb}.

In the presence of multiple anomalous $U(1)$'s (such as $U(1)_Y\times U(1)_B \times U(1)_{B'}$) the situation is quite different.
The $Z$ decay into four fermions can be mediated by two different extra neutral currents 
and the off-shell vertex can be of the type $\langle ZZ'Z''\rangle$, while from the 
interaction eigenstate basis a contribution $B B'B'$ appears. A simple inspection
of the gauge invariance of this vertex shows that a Chern-Simons interaction can not be 
absorbed into a redefinition of the $B B'B'$ triangle. 

\begin{figure}[t]
\begin{center}
\includegraphics[scale=0.8,angle=0]{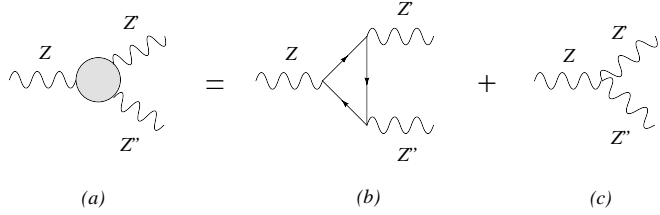}
\caption{\small Redefinition of the effective trilinear vertex including the Chern-Simons interactions.}
\label{4fermnol}
\end{center}
\end{figure}

\begin{figure}[t]
\begin{center}
\includegraphics[scale=0.8,angle=0]{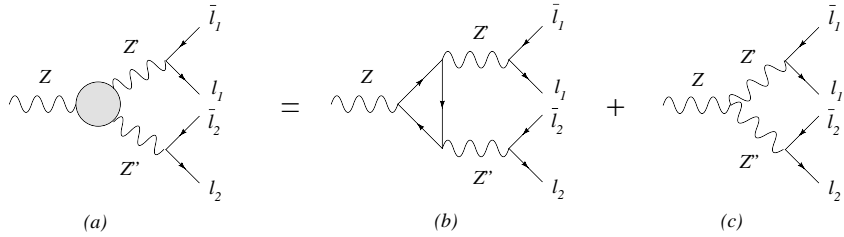}
\caption{\small Decay of the Z boson into 4 fermions plus the Chern-Simons contribution.}
\label{4fermionsblob}
\end{center}
\end{figure}

A symmetric distribution of the anomaly on the $B B'B'$ triangle, with outgoing momenta $k_1,k_2$
and incoming momentum $k$, fixes the Rosenberg parametrization as follows
\footnote{We have defined $a_n=\frac{i}{2\pi^2}$ and we use the notation 
$\varepsilon[k_1,k_2,\mu,\nu]=\varepsilon^{\alpha\beta\mu\nu}k_{1,\alpha}k_{2,\beta}$}
\ba
&&T^{\lambda\mu\nu}_{AAA}=(-A_5 k_1\cdot k_2 - A_6 k_2^2 -\frac{a_n}{3})\varepsilon[k_1,\lambda,\mu,\nu]
+(-A_4 k_1\cdot k_2 - A_3 k_1^2 +\frac{a_n}{3})\varepsilon[k_2,\lambda,\mu,\nu] 
\nonumber\\
&&+A_3 k_1^{\mu}\varepsilon[k_1,k_2,\lambda,\nu]+A_4 k_2^{\mu}\varepsilon[k_1,k_2,\lambda,\nu]
+A_5 k_1^{\nu}\varepsilon[k_1,k_2,\lambda,\mu]+A_6 k_2^{\nu}\varepsilon[k_1,k_2,\lambda,\mu],
\ea
thus, we have a partial anomaly equal to $\frac{a_n}{3}$ on each Lorentz index
\ba
&&k^{\lambda} T^{\lambda\mu\nu}_{AAA}=\frac{a_n}{3}\varepsilon[k_1,k_2,\mu,\nu]
\nonumber\\
&&k_1^{\mu} T^{\lambda\mu\nu}_{AAA}=\frac{a_n}{3}\varepsilon[k_1,k_2,\lambda,\nu]
\nonumber\\
&&k_2^{\nu} T^{\lambda\mu\nu}_{AAA}=-\frac{a_n}{3}\varepsilon[k_1,k_2,\lambda,\mu].
\ea
The generalized Chern-Simons interaction allowed by the presence of multiple anomalous $U(1)$s
can be formally written as 
\ba
V_{CS}^{\lambda\mu\nu}=a_n^{(1)}\varepsilon[\lambda,\mu,\nu,\alpha](k_1^{\alpha}-k_2^{\alpha})
+a_n^{(2)}\varepsilon[\lambda,\mu,\nu,\alpha](k_2^{\alpha}-k_3^{\alpha})
+a_n^{(3)}\varepsilon[\lambda,\mu,\nu,\alpha](k_3^{\alpha}-k_1^{\alpha})
\ea
where $k_3=-k$ and the coefficients $a_n^{(i)}$ $i=1,2,3$ depend on the model and satisfy
the relation $a_n^{(1)}+a_n^{(2)}+a_n^{(3)}=a_n$. 
Therefore, in the definition of the effective vertex the contributions coming from the Chern-Simons
interactions appear explicitly and spoil the symmetric distribution 
of the anomaly on $BB'B'$. Moreover, the cancellation of the anomaly is ensured by the presence of 
the WZ interactions, which are constrained by the BRST invariance of the model.
For example, the computation of the diagrams described in Figs. \ref{4fermnol} and \ref{4fermionsblob} gives
\ba
&&\bar{T}=\varepsilon^{\lambda}(k) \left(T^{\lambda\mu\nu}_{AAA} + V_{CS}^{\lambda\mu\nu}\right)
\left[\left(g^{\mu\mu'} -\frac{k_1^{\mu}k_1^{\mu'}}{M_{Z'}^2}\right)\frac{-i}{k_1^2-M_{Z'}^2}  
\bar{u}(q_1)\Gamma_{\mu'}v(q_2)            
\right.\nonumber\\
&&\hspace{2cm}\left.
\left(g^{\nu\nu'} -\frac{k_2^{\nu}k_2^{\nu'}}{M_{Z''}^2}\right)\frac{-i}{k_2^2-M_{Z''}^2}  
\bar{u}(q_3)\Gamma_{\nu'}v(q_4)
\right],
\ea
where we have indicated with $\Gamma_{\nu'}$ the generic Lorentz structure of the fermion coupling 
to the extra $Z'/Z''$.
For instance, the Chern-Simons contribution gives
\ba
&&\bar{T}_{CS}=\varepsilon^{\lambda}(k)\left[
a^{(1)}\varepsilon[\lambda,\mu,\nu,k_1-k_2]+a^{(2)}\varepsilon[\lambda,\mu,\nu,k_2-k_3]+a^{(3)}\varepsilon[\lambda,\mu,\nu,k_3-k_1] \right]\times
\nonumber\\
&&\hspace{2cm}\bar{u}(q_1)\Gamma^{\mu}v(q_2) 
\bar{u}(q_3)\Gamma^{\nu}v(q_4)
\frac{-1}{(k_1^2-M_{Z'}^2)(k_2^2-M_{Z''}^2)}.
\ea
The detection of these interactions is rather difficult 
experimentally, given the low production rates due to the large mass 
of the extra $Z^\prime$, currently bound to be larger than $900$ GeV.

\section{The Neutralino sector}

Moving to the neutralino sector, here the mass matrix is 
7-dimensional because of the presence of the axino,
the singlino and the $B$-ino in the spectrum. In the $B_S\neq 0$ case we obtain
\begin{eqnarray}
{\mathcal L}_{\tilde{\chi}^{0}}&=&
-\frac{1}{2}M_{w}\lambda_{W^3}\lambda_{W^3}
-\frac{1}{2}M_{Y}\lambda_{Y}\lambda_{Y}-\frac{1}{2}M_{B}\lambda_{B}\lambda_{B}
+\frac{iv_{1}}{\sqrt{2}}g_2\lambda_{W^3}\tilde{H}_{1}^{1}
-\frac{iv_{2}}{\sqrt{2}}g_2\lambda_{W^3}\tilde{H}_{2}^{2}
-\frac{iv_{1}}{\sqrt{2}}g_{Y}\lambda_{Y}\tilde{H}_{1}^{1}
\nonumber\\
&&+\frac{iv_{2}}{\sqrt{2}}g_Y\lambda_{Y}\tilde{H}_{2}^{2}
+\frac{iv_{1}}{\sqrt{2}}g_B B_{{H}_{1}}\lambda_{B}\tilde{H}_{1}^{1}
+\frac{iv_{2}}{\sqrt{2}}g_B B_{{H}_{2}}\lambda_{B}\tilde{H}_{2}^{2}
+\frac{iv_S}{\sqrt{2}}g_B B_{S}\lambda_{B}\tilde{S}
-\lambda~v_{S}\tilde{H}_{1}^{1}\tilde{H}_{2}^{2}
\nonumber\\
&&-\lambda~v_{1}\tilde{S}\tilde{H}_{2}^{2}
-\lambda~v_{2}\tilde{S}\tilde{H}_{1}^{1}
+\frac{M_{st}}{2\sqrt{2}}\psi_{\bf b}\lambda_{B}
-\frac{1}{2} M_{\bf b}\psi_{\bf b}\psi_{\bf b} + h.c.,
\end{eqnarray}
where $M_{w},M_{Y},M_{B},M_{\bf b}$ are mass 
parameters and the term $\lambda\,v_S/\sqrt{2}$ plays the role of the $\mu$-term;
notice that $\lambda$ is a dimensionless parameter.
We have indicated with $\lambda_{W^3}, \lambda_{Y}, \lambda_{B}$
the gauginos of $W^{3},A^{Y},B$ respectively and with $\psi_{\bf b}$
the SUSY particle associated to $b$.
The fields $\tilde{H}^{i}_1$ and $\tilde{H}^{i}_2$ ($i=1,2$) denote the supersymmetric
partners of the two Higgs doublets, while $\tilde{S}$ is the SUSY partner
of the extra singlet $S$.

In the basis
$(-i\lambda_{W^3},-i\lambda_{Y},-i\lambda_{B},\tilde{H_{1}^{1}},
\tilde{H_{2}^{2}},\tilde{S},-i\psi_{\bf b})$ the mass matrix takes the form
\ba
M_{\tilde{\chi}^{0}}=\left(
\begin{array}{ccccccc}
M_{w} & 0& 0& -\frac{v_{1}}{2} g_2 & \frac{v_{2}}{2}g_2 & 0 & 0\\
0 &M_Y& 0& \frac{v_{1}}{2}g_Y & -\frac{v_{2}}{2}g_Y & 0 &0\\
0 &0 &M_B& -\frac{v_{1}}{2}g_B B_{H_{1}} & -\frac{v_{2}}{2}g_B B_{H_{2}} & -\frac{v_{S}}{2}g_B B_{S} & -\frac{M_{st}}{\sqrt{2}}\\
-\frac{v_{1}}{2}g_2 & \frac{v_{1}}{2}g_Y & -\frac{v_{1}}{2}g_B B_{H_{1}} & 0 & -\lambda \frac{v_{S}}{\sqrt{2}} & -\lambda \frac{v_{2}}{\sqrt{2}} & 0\\
\frac{v_{2}}{2}g_2 & -\frac{v_{2}}{2}g_Y & -\frac{v_{2}}{2}g_B B_{H_{2}} & -\lambda \frac{v_{S}}{\sqrt{2}} & 0 & -\lambda \frac{v_{1}}{\sqrt{2}} & 0\\
0 & 0 & -\frac{v_{S}}{2}g_B B_{S} & -\lambda \frac{v_{2}}{\sqrt{2}} & -\lambda \frac{v_{1}}{\sqrt{2}} & 0 & 0\\
0 & 0 & -\frac{M_{st}}{\sqrt{2}} & 0 & 0 & 0 & M_{\bf b}
\end{array}
\right)
\ea
that will be analyzed numerically in a section below.

\subsection{A preliminary choice}

A preliminary choice \cite{Cvetic:1997ky} which allows to simplify the 
structure of the $7 \times 7$ neutralino matrix
is made by setting $M_{w}=M_Y=M_B=M_{\bf b}=\lambda=0$.
In these conditions the diagonalization is rather straightforward and we obtain
three null eigenvalues. 
The first corresponds to a physical pure-photino which is obtained from the rotation
\ba
&&\lambda_{\gamma}=\sin\theta_W \lambda_{W^3}+\cos\theta_W \lambda_{Y},
\nonumber\\
&&\lambda_{Z_{SM}}=\cos\theta_W \lambda_{W^3}-\sin\theta_W \lambda_{Y},
\ea
where $\lambda_{Z_{SM}}$ is an intermediate unphysical state.
The second state, corresponding to a null eigenvalue, 
is given by a mixture of Higgsino and axino states
\ba
\tilde{\chi}^{0}_2=\frac{M_{st}}{2 g_B v_1 B_S}
\tilde{H}^{1}_{1} + \frac{M_{st}}{2 g_B v_2 B_S}\tilde{H}^{2}_{2} + \psi_{\bf b},
\ea
while the third is a pure Higgsino state which corresponds 
to the SUSY partner of $H^{0}_{4}$ and it is given by the expression
\ba
\tilde{\chi}^{0}_3=\frac{v_S}{v_1}\tilde{H}^{1}_{1}+\frac{v_S}{v_2}\tilde{H}^{2}_{2}+\tilde{S}.
\ea
The other states corresponding to the non-zero eigenvalues are complicated combinations
of higgsinos, gauginos ($\lambda_{Z_{SM}},\lambda_{B}$) and the axino.

Notice that in our treatment we are considering for simplicity a real-valued neutralino matrix.
In the most general cases - for example in some CP-noninvariant 
theories -  these matrix elements are complex 
and they may contain phase factors which are physical and can not be eliminated by a redefinition of the fields.

\section{Supersymmetric interactions of the axion with the neutralinos}

In this section we proceed with a study of the basic tree-level 
interaction vertices involving the physical axion (axi-Higgs). 
Analyzing each sector of the whole Lagrangean we have different types 
of interactions involving the axi-Higgs.

\begin{figure}[t]
\begin{center}
\includegraphics[scale=0.5,angle=0]{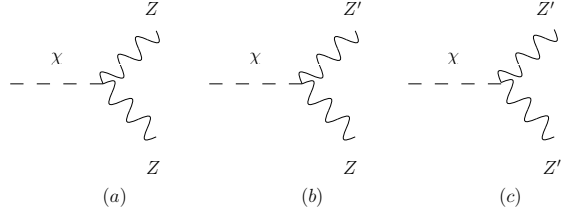}
\caption{\small Trilinear interactions between $\chi$ and the neutral currents}
\label{fig1}
\end{center}
\end{figure}

First of all, from the counterterm Lagrangean 
we have trilinear interactions obtained by rotating the WZ counterterms 
on the physical basis, which formally give terms of the type
\ba
{\cal L}_{\chi Z Z} = R_1\,\epsilon^{\mu\nu\rho\sigma} Z^{abel}_{\mu\nu} Z^{abel}_{\rho\sigma}\,\chi
+ R_2\,\epsilon^{\mu\nu\rho\sigma} Z'^{abel}_{\mu\nu} Z'^{abel}_{\rho\sigma}\,\chi 
+ R_3\,\epsilon^{\mu\nu\rho\sigma} Z^{abel}_{\mu\nu} Z'^{abel}_{\rho\sigma}\,\chi,
\ea
where for simplicity we have indicated with 
$R_1,R_2,R_3$ the coefficients which appear in front of each vertex. These include the rotation 
matrices, the coupling constants of the gauge groups and the coefficients 
coming from the anomaly cancellation procedure.
We omit their explicit expressions since they are not relevant for this discussion. Notice that in this case only the abelian part of field strengths contribute to the 
counterterms for the neutral currents and that $Z^{abel}_{\mu\nu}=\partial_{\mu}Z_{\nu}-\partial_{\nu}Z_{\mu}$. 
The interactions coming from these terms are shown in Fig.\ref{fig1}.

\begin{figure}[t]
\begin{center}
\includegraphics[scale=0.5,angle=0]{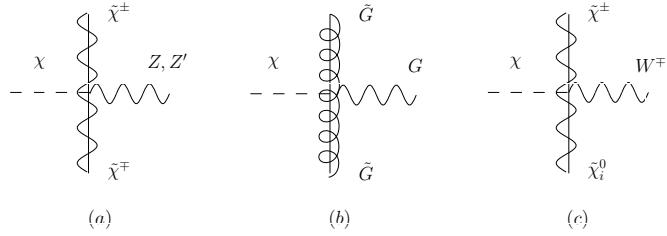}
\caption{\small Quadrilinear interactions involving $\chi$, charginos/gluinos/neutralinos and a gauge boson.}
\label{fig2}
\end{center}
\end{figure}

\begin{figure}[ht]
\begin{center}
\includegraphics[scale=0.5,angle=0]{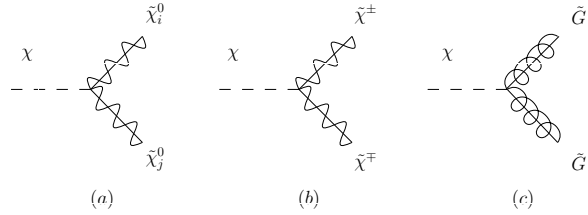}
\caption{\small Derivative trilinear interactions between $\chi$ and charginos/gluinos/neutralinos.}
\label{fig3}
\end{center}
\end{figure}

\begin{figure}[ht]
\begin{center}
\includegraphics[scale=0.5,angle=0]{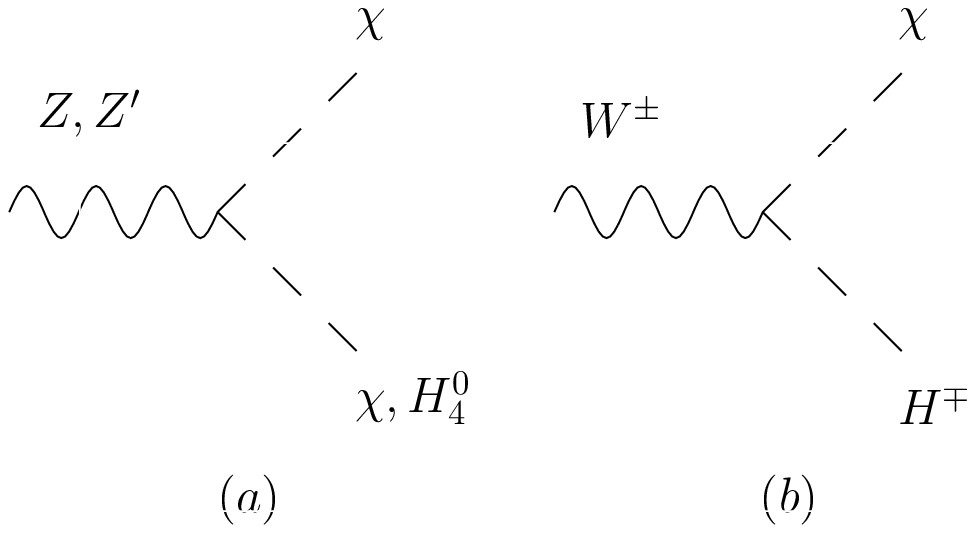}
\caption{\small Trilinear interactions between $\chi$, an Higgs boson and an electroweak gauge boson.}
\label{fig4}
\end{center}
\end{figure}

From the axion Lagrangean ${\cal L}_{axion}$ we obtain quadrilinear interactions 
between $\chi$, the neutralinos/gluinos/charginos, the neutral/charged gauge bosons 
and trilinear derivative interactions, illustrated in Fig.\ref{fig2}-\ref{fig3}. 
In fact, by a careful inspection of ${\cal L}_{axion}$ we find  
\ba
&&{\cal L}_{axion}^{\chi \tilde{\chi}\tilde{\chi} Gauge}= 
R^{Z}\,\chi\, \bar{\tilde{\chi}}^{\pm} \gamma^{\mu} \tilde{\chi}^{\mp}\, Z_{\mu}
+R^{G}\,\chi\, \bar{\tilde{G}} \gamma^{\mu} \tilde{G}\, G_{\mu}
+R^{W}\,\chi\, \bar{\tilde{\chi}}^{\pm} \Gamma^{\mu} \tilde{\chi}^{0}_i\, W^{\mp}_{\mu}
+\{Z\rightarrow Z'\}\,,
\ea
while the derivative trilinear interactions are given by  
\ba
{\cal L}_{axion}^{\chi \tilde{\chi} \tilde{G}}= 
R^{\chi i j}\,\chi\, \bar{\tilde{\chi}}^0_i \Gamma^{\mu}\partial_{\mu} \tilde{\chi}^0_j + 
R^{\chi \tilde{G} \tilde{G}}\,\chi\, \bar{\tilde{G}} \gamma^{\mu}\partial_{\mu} \tilde{G}
+R^{\chi \pm}\,\chi\, \bar{\tilde{\chi}}^{\pm} \Gamma^{\mu}\partial_{\mu} \tilde{\chi}^{\mp}\,,
\ea
where $\Gamma^{\mu}$ indicates that we can have vector or axial-vector interactions.
Trilinear interactions between one neutral current and two axion-like particles 
can be obtained from ${\cal L}_{\textsc{quad}}$ and have the form
\ba
{\cal L}_{\textsc{quad}}^{\chi H Z}= 
R^{\chi H^0_i Z}\, \chi \stackrel{\leftrightarrow}{\partial^{\mu}} H^0_i Z_{\mu}+
R^{\chi H^{\pm} W^{\mp}}\, \chi \stackrel{\leftrightarrow}{\partial^{\mu}} H^{\pm} W^{\mp}_{\mu}
+ \{Z\rightarrow Z'\};
\ea
to these terms correspond the interactions shown in Fig.\ref{fig4}; 
\begin{figure}[h]
\begin{center}
\includegraphics[scale=0.5,angle=0]{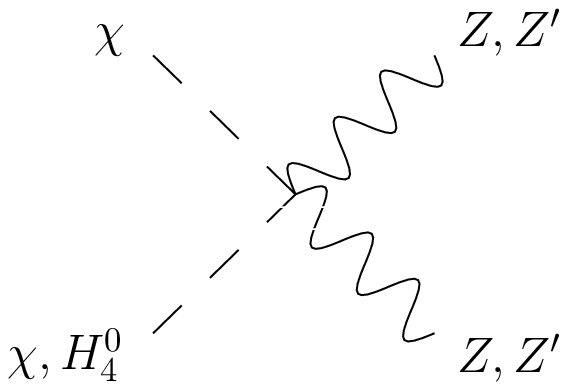}
\caption{\small Quadrilinear interaction involving $\chi$, two electroweak neutral gauge bosons and the CP-odd Higgs.}
\label{fig5}
\end{center}
\end{figure}
Analogously, the quadrilinear interactions between two axion like particles 
and two neutral gauge bosons are given by (see Fig. \ref{fig5})  
\ba
{\cal L}_{\textsc{quad}}^{\chi\chi Z Z}= R^{ZZ}_1\, \chi \chi Z_{\mu}Z^{\mu} 
+ R^{ZZ}_2\, \chi\, H^0_4 Z_{\mu}Z^{\mu} + R^{ZZ'}_1\, \chi\, \chi \,Z_{\mu}Z'^{\mu}
+ R^{ZZ'}_2\, \chi\, H^0_4 Z_{\mu}Z'^{\mu} + \{Z\rightarrow Z'\}
\nonumber\\
\ea
where, again, we have introduced the coefficients $R^{ZZ}_i,R^{ZZ}_j$ containing 
the rotation matrices and the couplings, for simplicity.

\begin{figure}[t]
\begin{center}
\includegraphics[scale=0.5,angle=0]{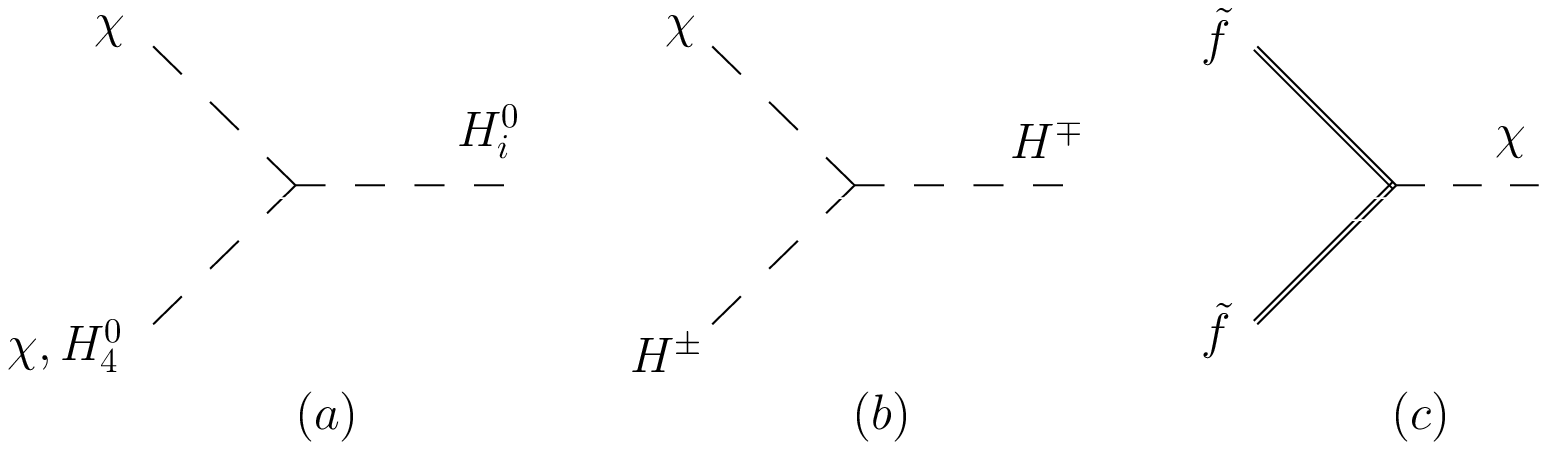}
\caption{\small Trilinear interaction involving $\chi$ and Higgs bosons/sfermions.}
\label{fig6}
\end{center}
\end{figure}

From the Lagrangean of the scalar mass terms ${\cal L}_{SMT}$ we obtain 
the following trilinear interactions involving the axi-Higgs, the Higgs
bosons coming from the scalar sector (CP-even, CP-odd, charged) and 
the sfermions
\ba
{\cal L}_{SMT}^{\chi\chi even-odd}= R^{\chi^2 i}\,\chi^2 \,H^{0}_i 
+ R^{\chi i}\,\chi\, H^{0}_4 H^{0}_i 
+R^{\chi \pm}\, \chi\, H^{\mp} H^{\pm} 
+ R^{\chi \tilde{f} \tilde{f}} \chi\, \tilde{f} \tilde{f},
\ea
where $H^{0}_i$ with $i=1,\dots 3$ indicates the physical Higgs states
coming from the CP-even sector (see Fig.\ref{fig6}).
\begin{figure}[ht]
\begin{center}
\includegraphics[scale=0.5,angle=0]{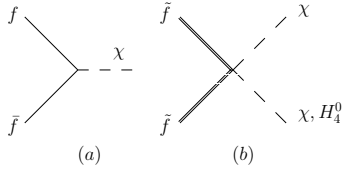}
\caption{\small Interactions obtained from ${\cal L}_{Yuk}^{\chi}$.}
\label{fig7}
\end{center}
\end{figure}
\begin{figure}[ht]
\begin{center}
\includegraphics[scale=0.5,angle=0]{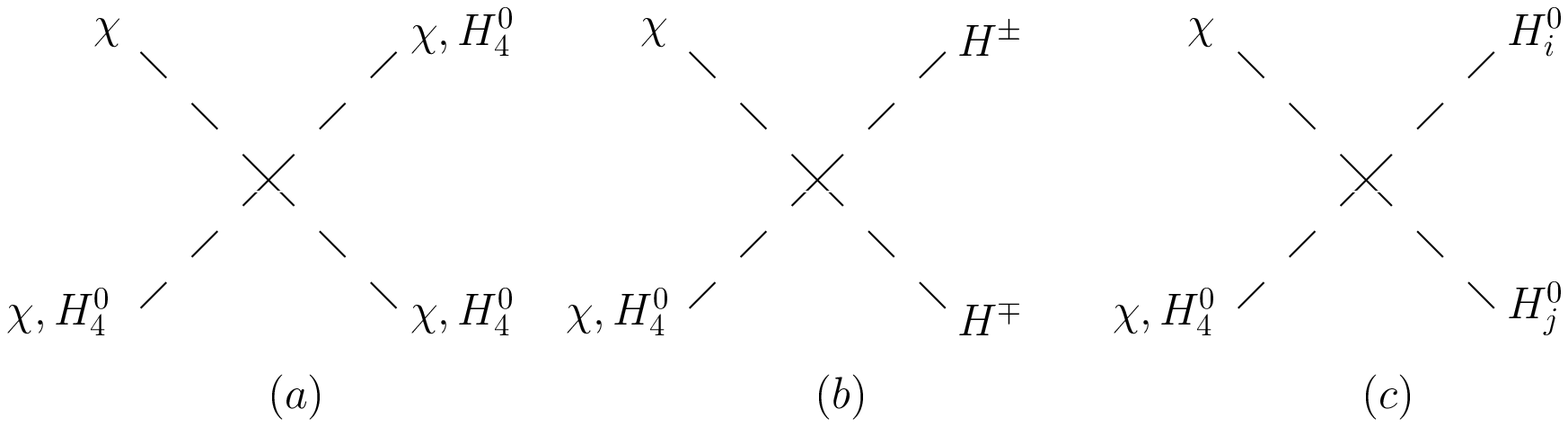}
\caption{\small Quadrilinear interactions involving $\chi$ and CP-odd/CP-even/charged Higgs.}
\label{fig8}
\end{center}
\end{figure}
We denote with ${\cal L_{W}}$ the on-shell Lagrangean coming 
from the superpotential, once that the $F$-terms have been removed, and containing 
all the Yukawa-type interactions
\ba
{\cal L_{W}}={\cal L}_{Yuk} + {\cal L}_{S} + {\cal L}_{Yuk-F}
\ea
where ${\cal L}_{Yuk}$ represents the Yukawa interactions that do not contain the extra singlet $S$
and are linear in $y_e,y_u,y_d$, while ${\cal L}_{S}$ indicates all the Yukawa interactions 
containing $S$. Finally, with ${\cal L}_{Yuk-F}$ we indicate those interactions that are quadratic
in $y_e,y_u,y_d$ and in $\lambda$. 
%and are related to the elimination of the $F$-terms. 
Then we have
\ba
&&{\cal L}_{Yuk}=y_{e}\epsilon^{ij}[-\tilde{H}_{1}^{i}L^{j}\tilde{R}
-\bar{\tilde{H}_{1}}^{i}\bar{L}^{j}\tilde{R}^{\dagger}
-H_{1}^{i}L^{j}\bar R-H_{1}^{i\dagger}\bar{L}^{j} R
-\bar R\tilde{H}_{1}^{i}\tilde{L}^{j}
-R\bar{\tilde{H}_{1}}^{i}\tilde{L}^{j\dagger}]
\nonumber\\
&&+y_{d}\epsilon^{ij}[-\tilde{H}_{1}^{i}Q^{j}\tilde{D}_{R}
-\bar{\tilde{H}}_{1}^{i}\bar{Q}^{j}\tilde{D}_{R}^{\dagger}
-H_{1}^{i}Q^{j}\bar D_R - H_{1}^{i\dagger}\bar{Q}^{j}D_{R}
-\bar D_R\tilde{H}_{1}^{i}\tilde{Q}^{j}
-D_{R}\bar{\tilde{H}}_{1}^{i}\tilde{Q}^{j\dagger}]
\nonumber\\
&&+y_{u}\epsilon^{ij}[
-\tilde{H}_{2}^{i}Q^{j}\tilde{U}_{R}
-\bar{\tilde{H}}_{2}^{i}\bar{Q}^{j}\tilde{U}_{R}^{\dagger}
-H_{2}^{i}Q^{j}\bar U_R - H_{2}^{i\dagger}\bar{Q}^{j}U_{R}
-\bar U_R \tilde{H}_{2}^{i}\tilde{Q}^{j}
-U_{R}\bar{\tilde{H}}_{2}^{i}\tilde{Q}^{j\dagger}],
\ea
\ba
&&{\cal L}_{S}=
-\lambda y_{e}[S^\dagger H_{2}^{\dagger}\tilde{L}\tilde{R}
+S\tilde{L}^{\dagger}H_{2}\tilde{R}^{\dagger}]
-\lambda y_{d}[S^\dagger H_{2}^{\dagger}\tilde{Q}\tilde{D}_{R}
+S\tilde{Q}^{\dagger}H_{2}\tilde{D}_{R}^{\dagger}]
\nonumber\\
&&-\lambda y_{u}[S^\dagger H_{1}^{\dagger}\tilde{Q}\tilde{U}_{R}
+S\tilde{Q}^{\dagger}H_{1}\tilde{U}_{R}^{\dagger}]
+\lambda\epsilon^{ij}[-S\tilde{H}_{1}^{i}\tilde{H}_{2}^{j}
-S^\dagger\bar{\tilde{H}_{1}}^{i}\bar{\tilde{H}_{2}}^{j}]
\nonumber\\
&&-|\lambda S|^2(H_2^{\dagger} H_2 +H_1^{\dagger}H_1)
\ea
and finally
\ba
&&{\cal L}_{Yuk-F}=-|\lambda H_1\cdot H_2|^2
-y_{e}^{2}[\tilde{L}^{\dagger}\tilde{L}\tilde{R}^{\dagger}\tilde{R}
+H_{1}^{\dagger}H_{1}(\tilde{L}^{\dagger}\tilde{L}+\tilde{R}^{\dagger}\tilde{R})
\nonumber\\
&&-H_{1}^{\dagger}\tilde{L}(H_{1}^{\dagger}\tilde{L})^{\dagger}]
-y_{d}^{2}[\tilde{Q}^{\dagger}\tilde{Q}\tilde{D}_{R}^{\dagger}\tilde{D}_{R}
+H_{1}^{\dagger}H_{1}(\tilde{Q}^{\dagger}\tilde{Q}+\tilde{D}_{R}^{\dagger}\tilde{D}_{R})
-H_{1}^{\dagger}\tilde{Q}(H_{1}^{\dagger}\tilde{Q})^{\dagger}]
\nonumber\\
&&-y_{u}^{2}[\tilde{Q}^{\dagger}\tilde{Q}\tilde{U}_{R}^{\dagger}\tilde{U}_{R}
+H_{2}^{\dagger}H_{2}(\tilde{Q}^{\dagger}\tilde{Q}+\tilde{U}_{R}^{\dagger}\tilde{U}_{R})
-H_{2}^{\dagger}\tilde{Q}(H_{2}^{\dagger}\tilde{Q})^{\dagger}]\,.
\nonumber\\
\ea
From the Yukawa mass terms contained in ${\cal L}_{Yuk}$ and in ${\cal L}_{S}$
we can isolate the pseudoscalar coupling of the axi-Higgs to the fermions
and a quadrilinear scalar interaction with the sfermions
\ba
{\cal L}_{Yuk-S}^{\chi}=R_{Yuk}^{\chi \bar{f} f} \,\bar{\psi}_f \gamma^5 \psi_f \, \chi  
+R_{S}^{\chi^2 \tilde{f} \tilde{f}} \chi\, \chi \,\tilde{f} \tilde{f}
+R_{S}^{\chi\, H^0_4\,\tilde{f} \tilde{f}} \chi\, H^0_4 \,\tilde{f} \tilde{f}
\ea
where we have indicated with $\psi_f$ the generic fermion and
with $\tilde{f}$ the generic sfermion (see Fig.\ref{fig7}).

Quadrilinear axionic self interactions can be obtained from
${\cal L}_{S}$ and from ${\cal L}_{Yuk-F}$
\ba
&&{\cal L}_{\cal W}^{\chi H^0_4}= R^{\chi^4} \chi^4
+ R^{\chi^3} \chi^3 H^0_4 + R^{\chi^2 \pm} \chi^2 H^{\pm} H^{\mp}
+ R^{\chi^2} \chi^2 (H^0_4)^2 +  R^{\chi} \chi (H^0_4)^3
\nonumber\\
&&+ R^{\chi^2 i j}\, \chi^2 H^{0}_i H^{0}_j
+ R^{\chi H^0_4 i j}\, \chi H^0_4 H^{0}_i H^{0}_j
+ R^{\chi H^0_4 \pm}\, \chi H^0_4 H^{\mp} H^{\pm}
\ea
and are listed in Fig.\ref{fig8}.

\section{Numerical Analysis}

In this section we present a numerical analysis of the neutralino sector.
We have performed the numerical diagonalization of the $7\times 7$ neutralino
matrix and we have studied the eigenvalues dependence with respect to
the free parameters of the model. Furthermore, since in this model the neutralino sector exhibits
an axino component due to the presence of St\"uckelberg interactions,
we have investigated, in the case of the lightest neutralino state,
its mixing with the other states.
In Tab. \ref{parameters} we have listed all the values of the
parameters that we have used in our analysis. In our analysis we have followed, in spirit, 
the approach of Kalinowski and collaborators in \cite{Kalinowski:2008iq}. In their paper 
the authors, who deal with the USSM, present two scenarios: 
in the first one they assume unified values for the gaugino mass terms and 
in a second scenario they consider with different values (arbitrary values). 
We refer to their analysis for further justifications and motivations of this choice.
\begin{figure}[bh]
\subfigure[\small Eigenvalues as a function of $M_{st}$]
{\includegraphics[
width=6cm,
angle=-90]{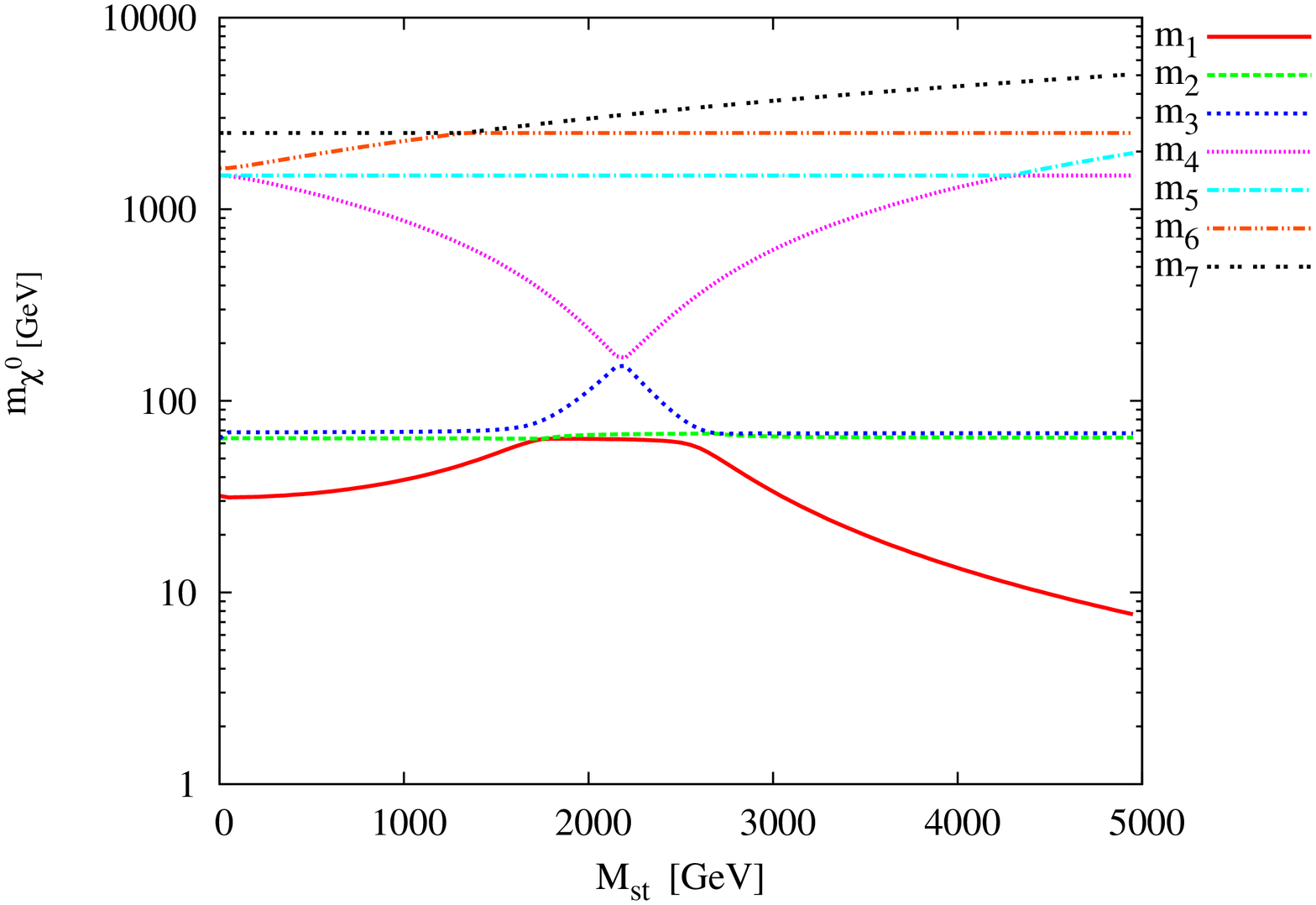}}
\subfigure[\small Squared components of the lightest neutralino]
{\includegraphics[
width=6cm,
angle=-90]{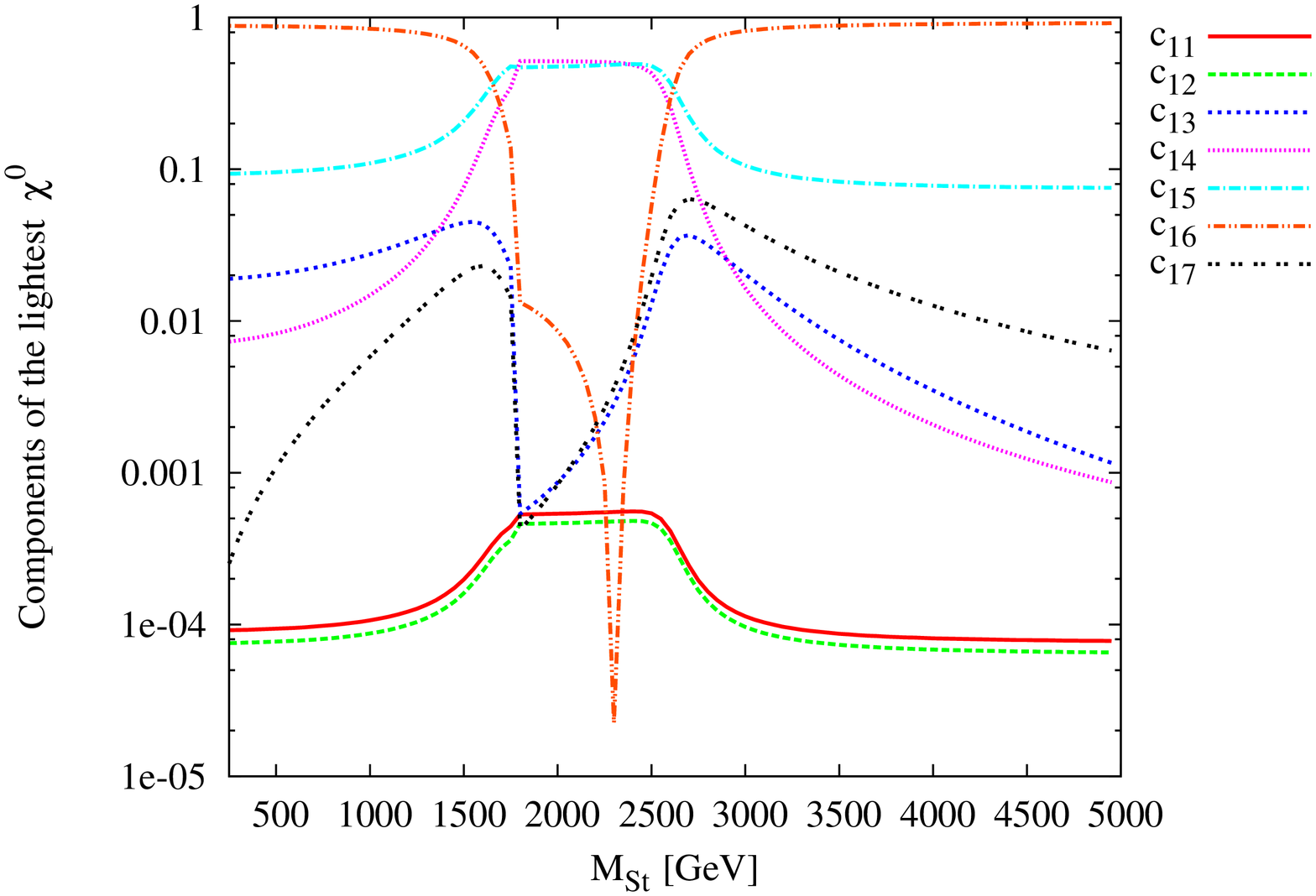}}
\caption{\small Study of the neutralino eigenvalues as
a function of St\"uckelberg mass $M_{st}$.}
\label{neutralino1}
\end{figure}
\begin{figure}[ht]
\subfigure[\small Eigenvalues as a function of $M_B$]
{\includegraphics[
width=6cm,
angle=-90]{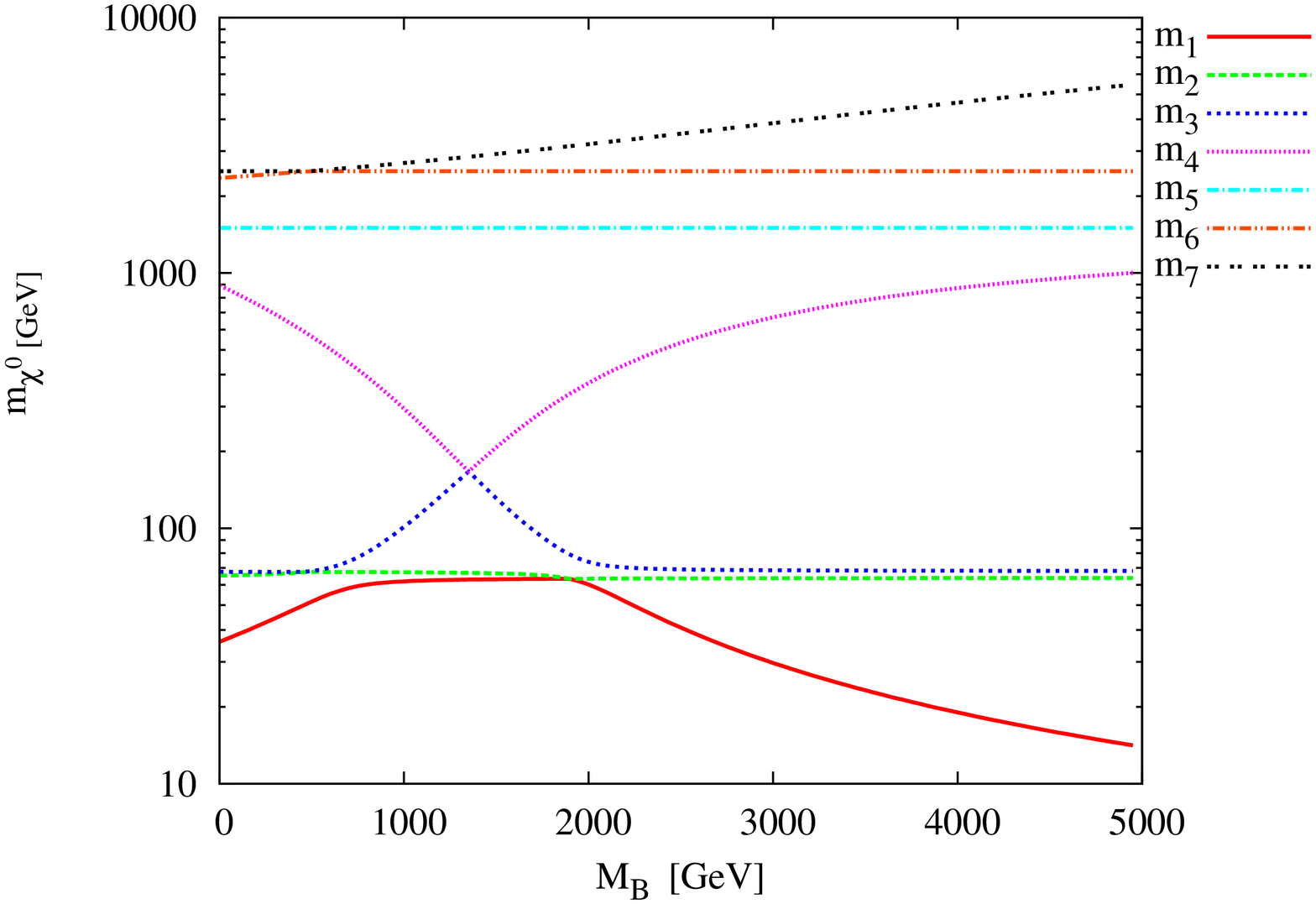}}
\subfigure[\small Squared components of the lightest neutralino]
{\includegraphics[
width=6cm,
angle=-90]{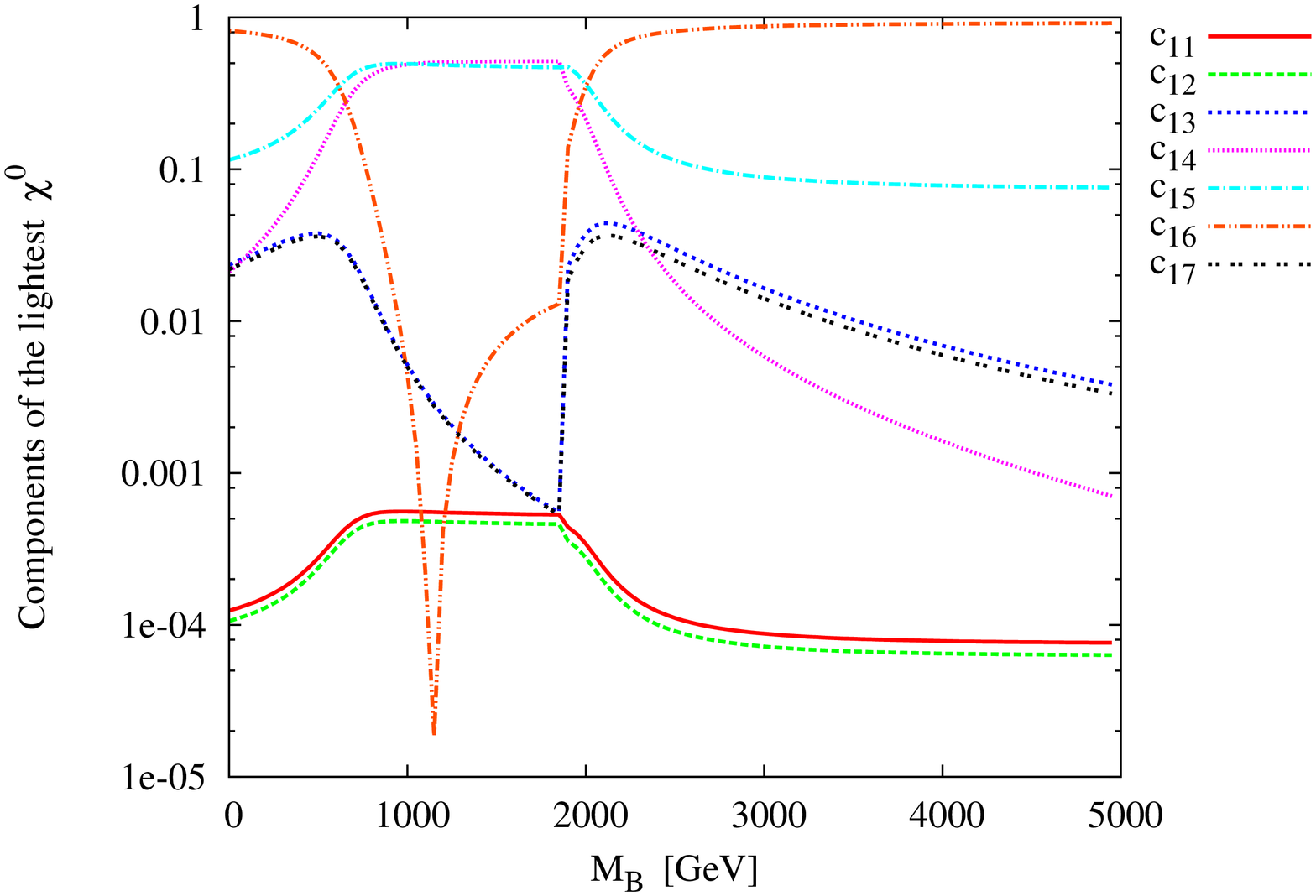}}
% \caption{\small The same as Fig.(\ref{neutralino1}) but as a function of $M_Y$, $M_w$ and $M_B$.}
% \label{neutralino2}
% \end{figure}
%  %
% \begin{figure}[h]
\subfigure[\small Eigenvalues as a function of $M_{\bf b}$]
{\includegraphics[
width=6cm,
angle=-90]{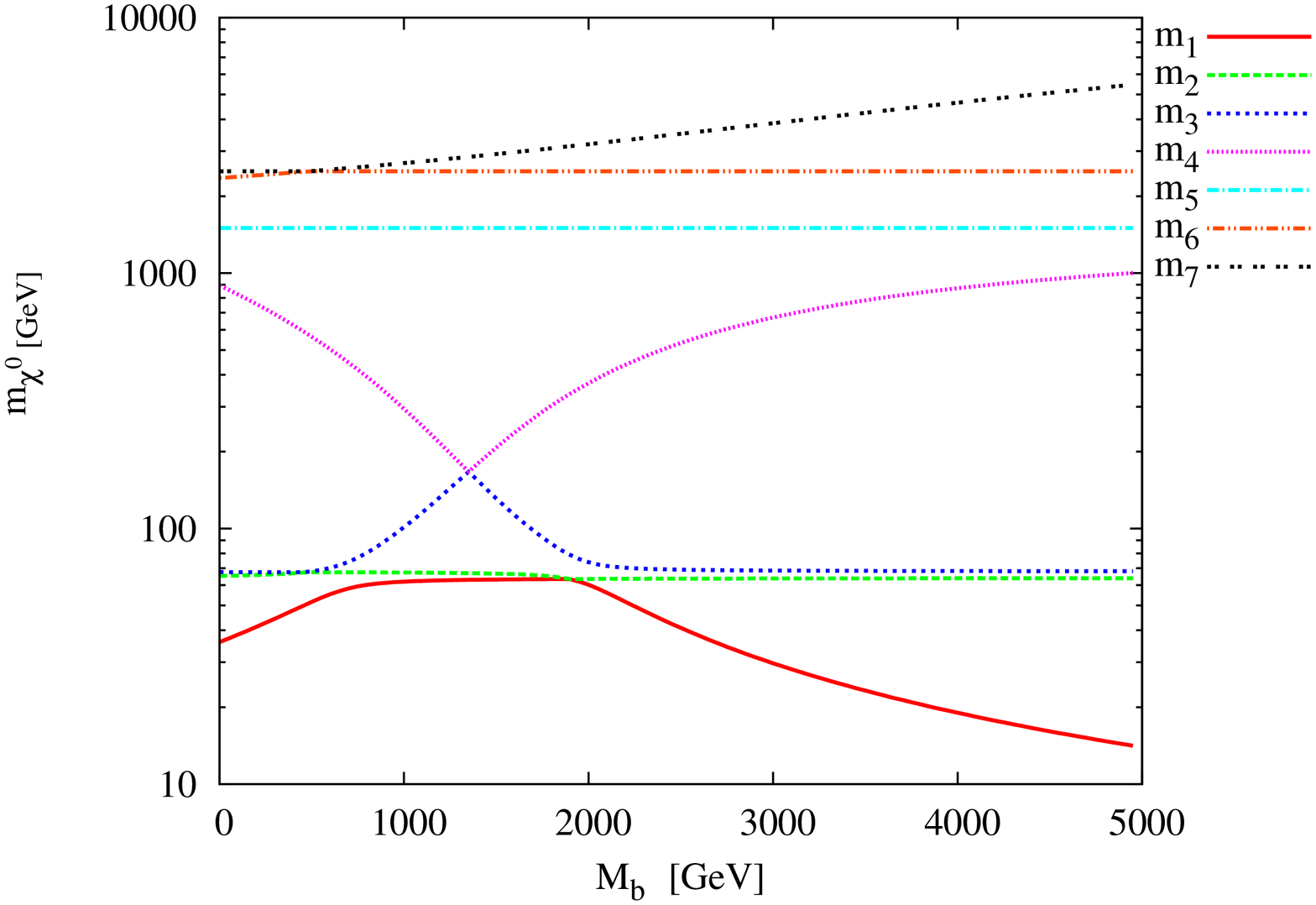}}
\subfigure[\small Squared components of the lightest neutralino]
{\includegraphics[
width=6cm,
angle=-90]{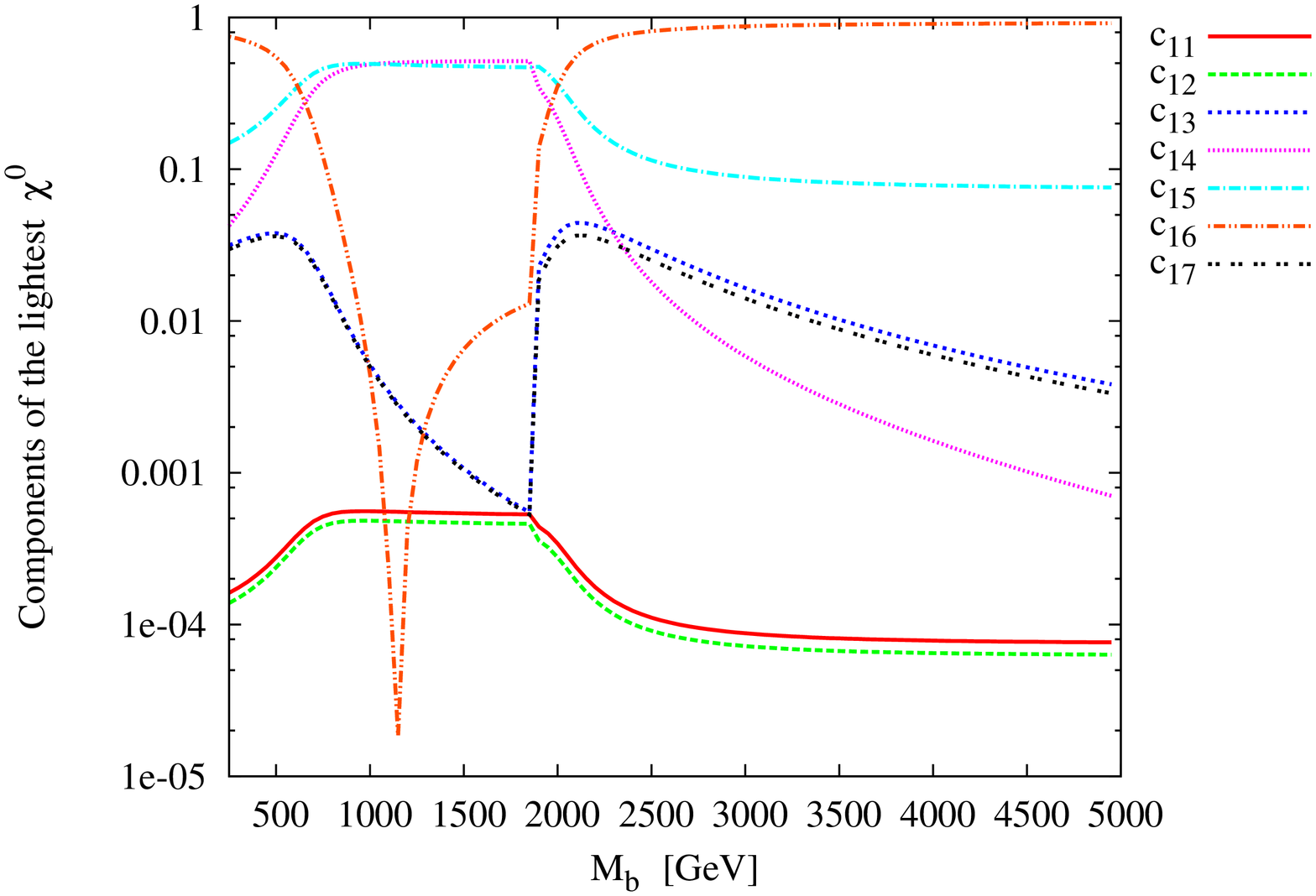}}
\caption{\small The same as Fig.(\ref{neutralino1}) but as a function of $M_B$ and $M_{\bf b}$.}
\label{neutralino2}
\end{figure}
\begin{figure}[ht]
\subfigure[\small Eigenvalues as a function of $M_Y$]
{\includegraphics[
width=6cm,
angle=-90]{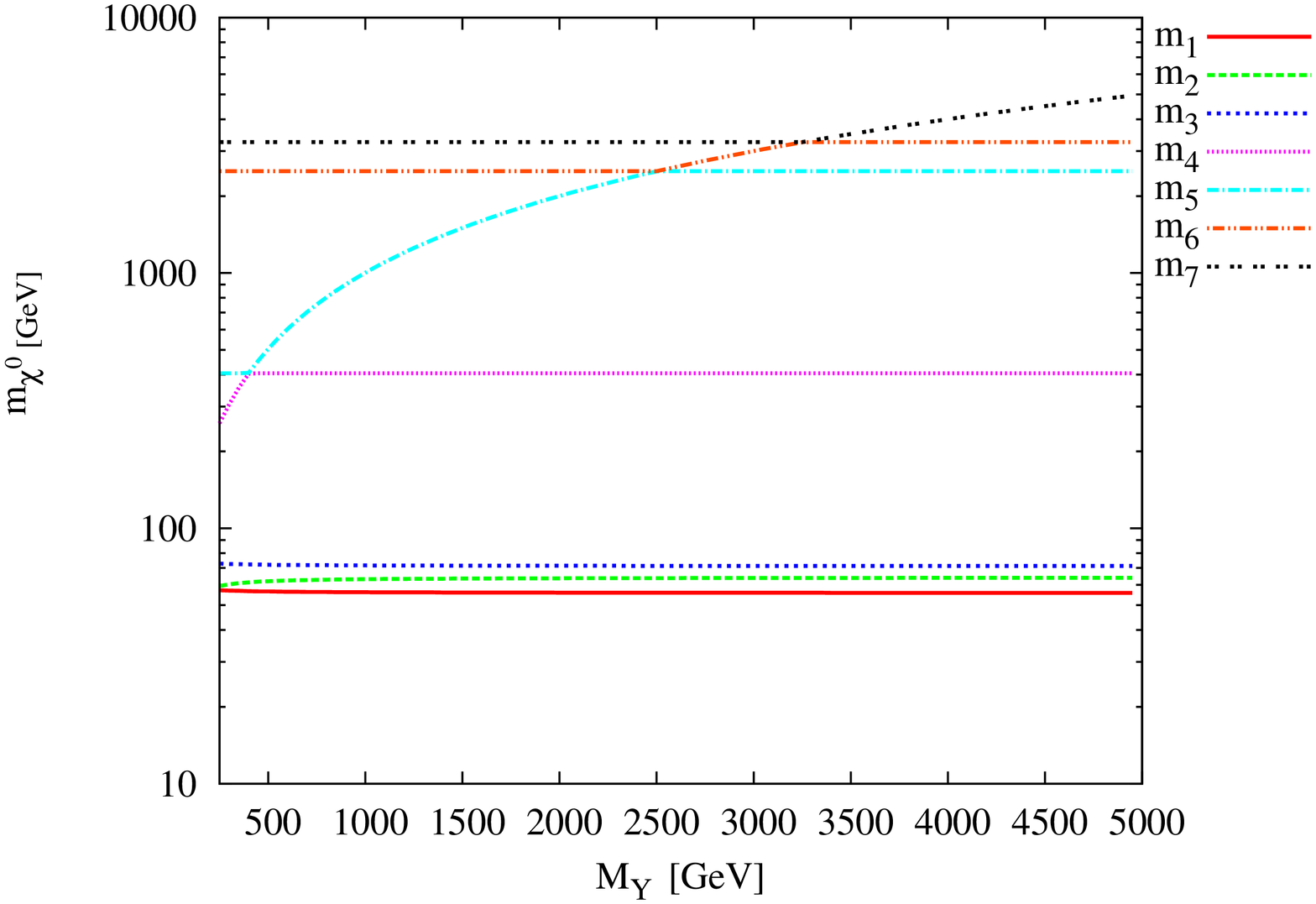}}
\subfigure[\small Squared components of the lightest neutralino]
{\includegraphics[
width=6cm,
angle=-90]{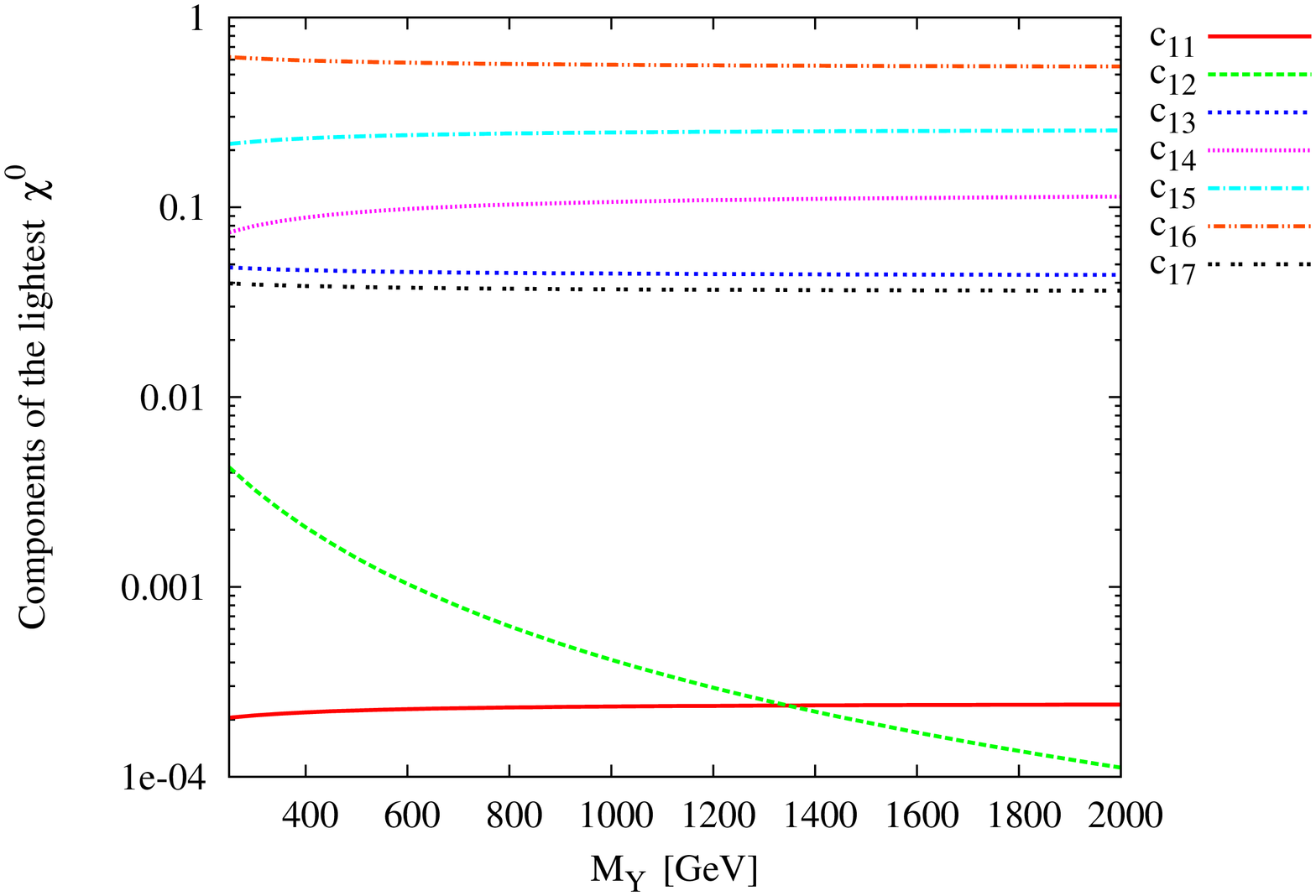}}
% \caption{\small The same as Fig.(\ref{neutralino1}) but as a function of the
% soft breaking parameter $M_Y$.}
% \label{neutralino2}
% \end{figure}
% %
% \begin{figure}[h]
\subfigure[\small Eigenvalues as a function of $M_w$]
{\includegraphics[
width=6cm,
angle=-90]{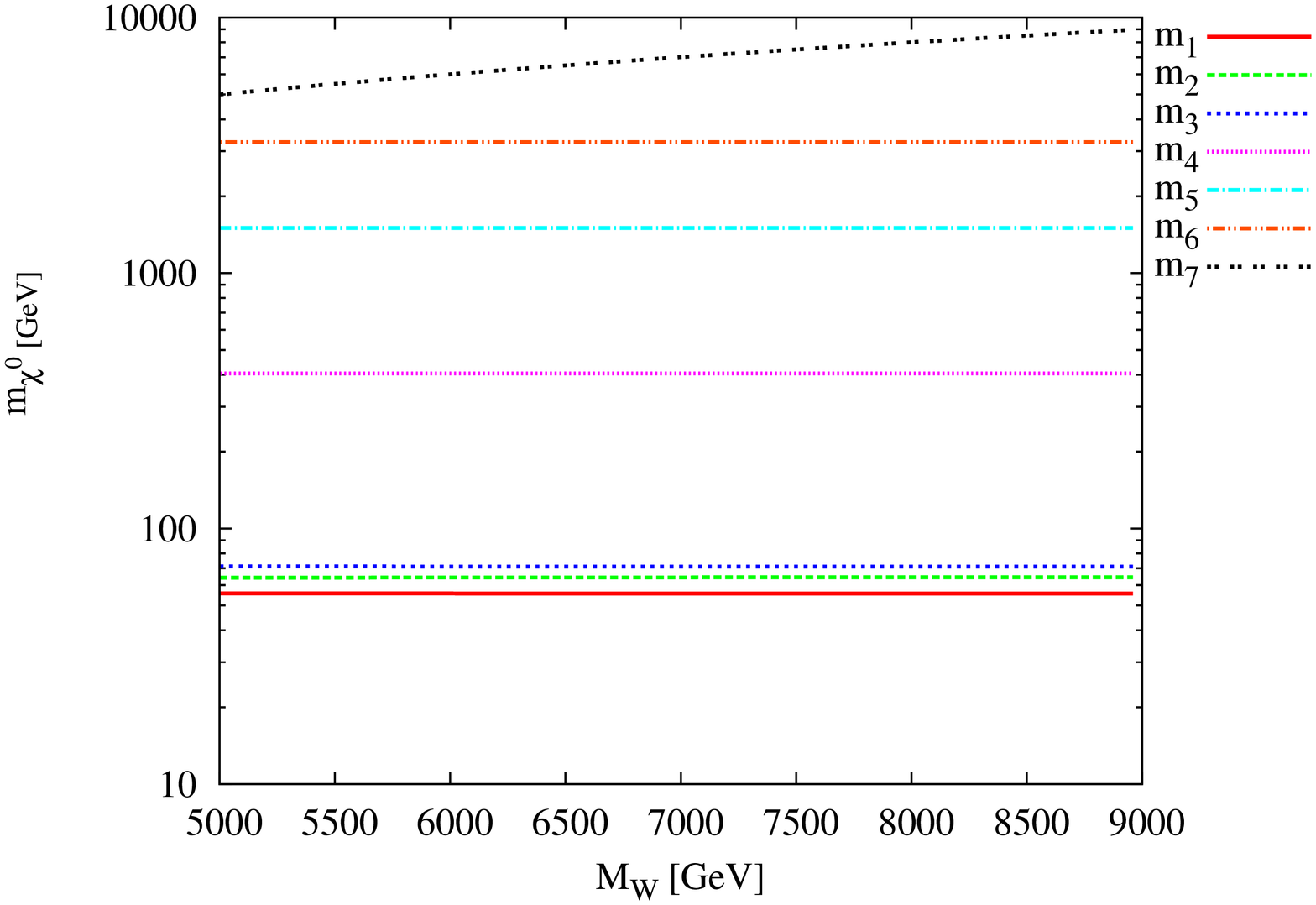}}
\subfigure[\small Squared components of the lightest neutralino]
{\includegraphics[
width=6cm,
angle=-90]{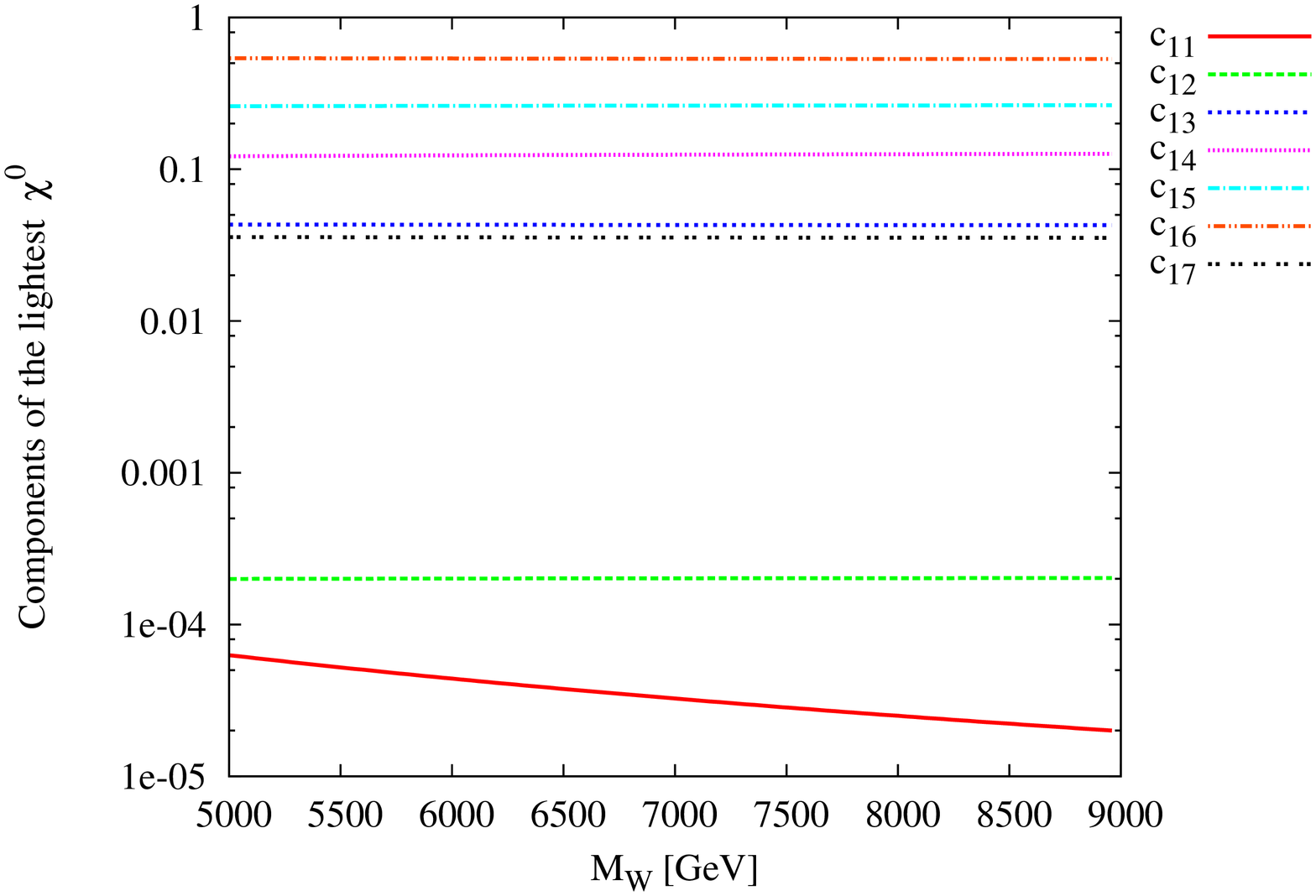}}
\caption{\small The same as Fig.(\ref{neutralino1}) but as a function of $M_Y$ and $M_w$.}
\label{neutralino3}
\end{figure}
\begin{figure}[ht]
\subfigure[\small Eigenvalues as a function of $g_B$]
{\includegraphics[
width=6cm,
angle=-90]{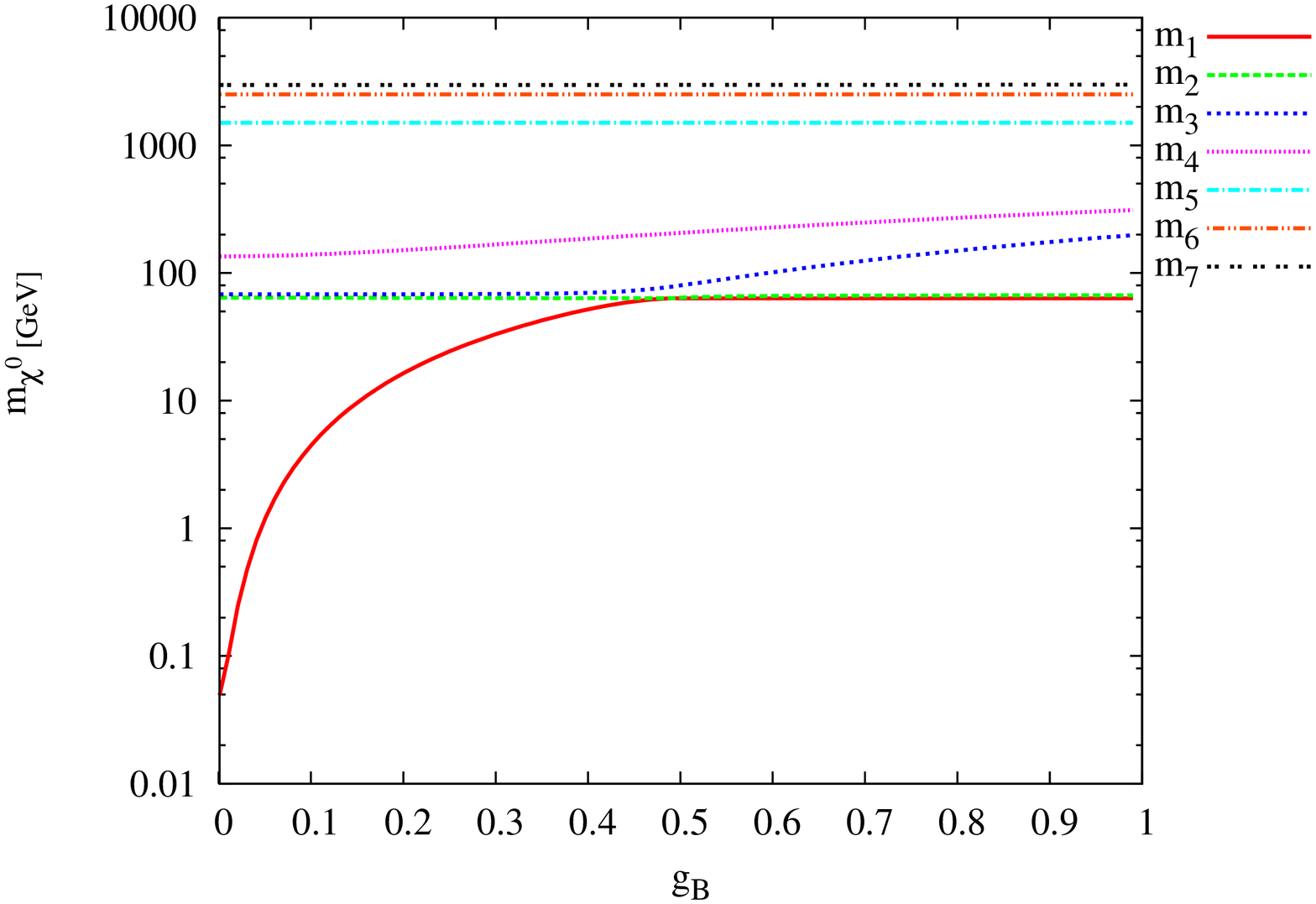}}
\subfigure[\small Squared components of the lightest neutralino]
{\includegraphics[
width=6cm,
angle=-90]{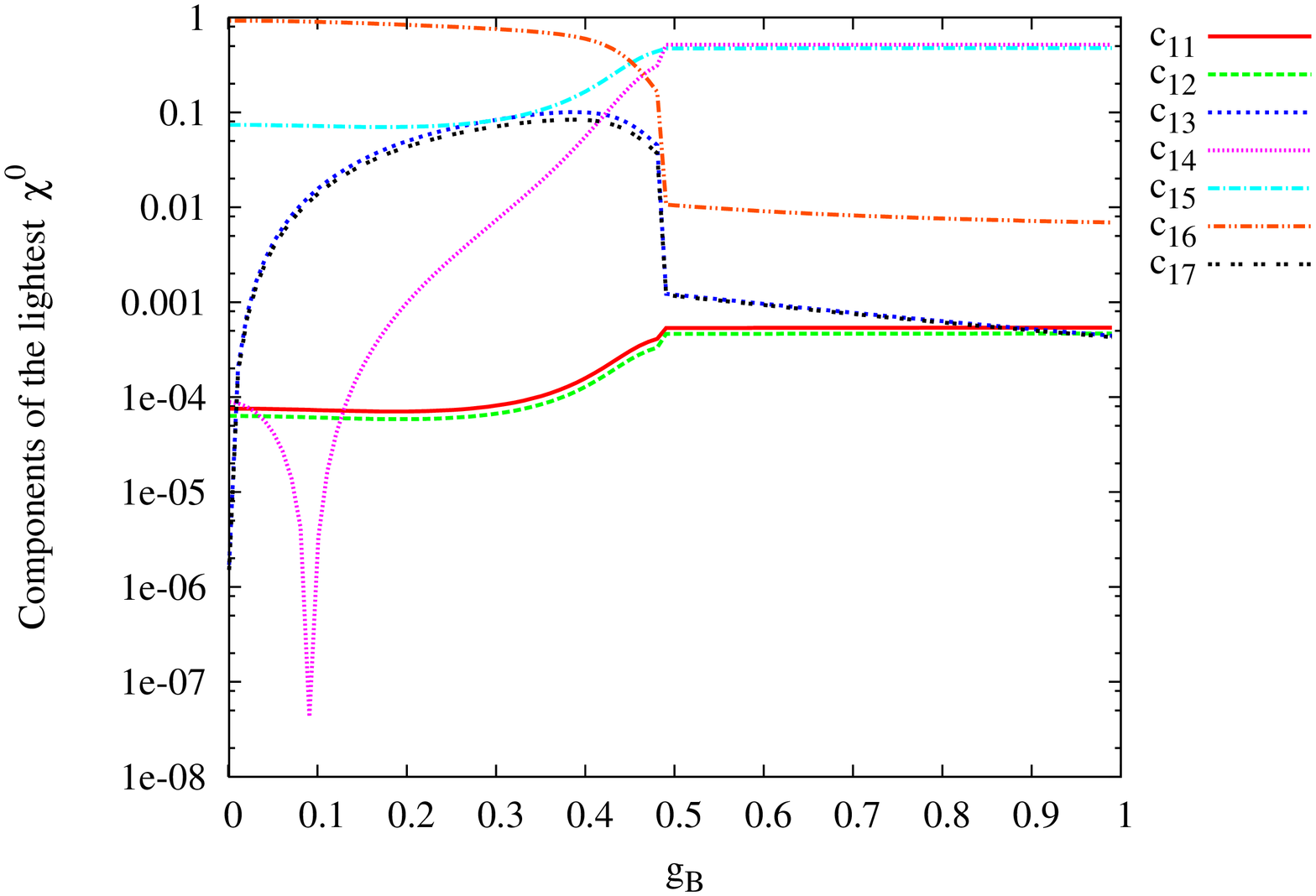}}
\subfigure[\small Eigenvalues as a function of $\tan{\beta}$]
{\includegraphics[
width=6cm,
angle=-90]{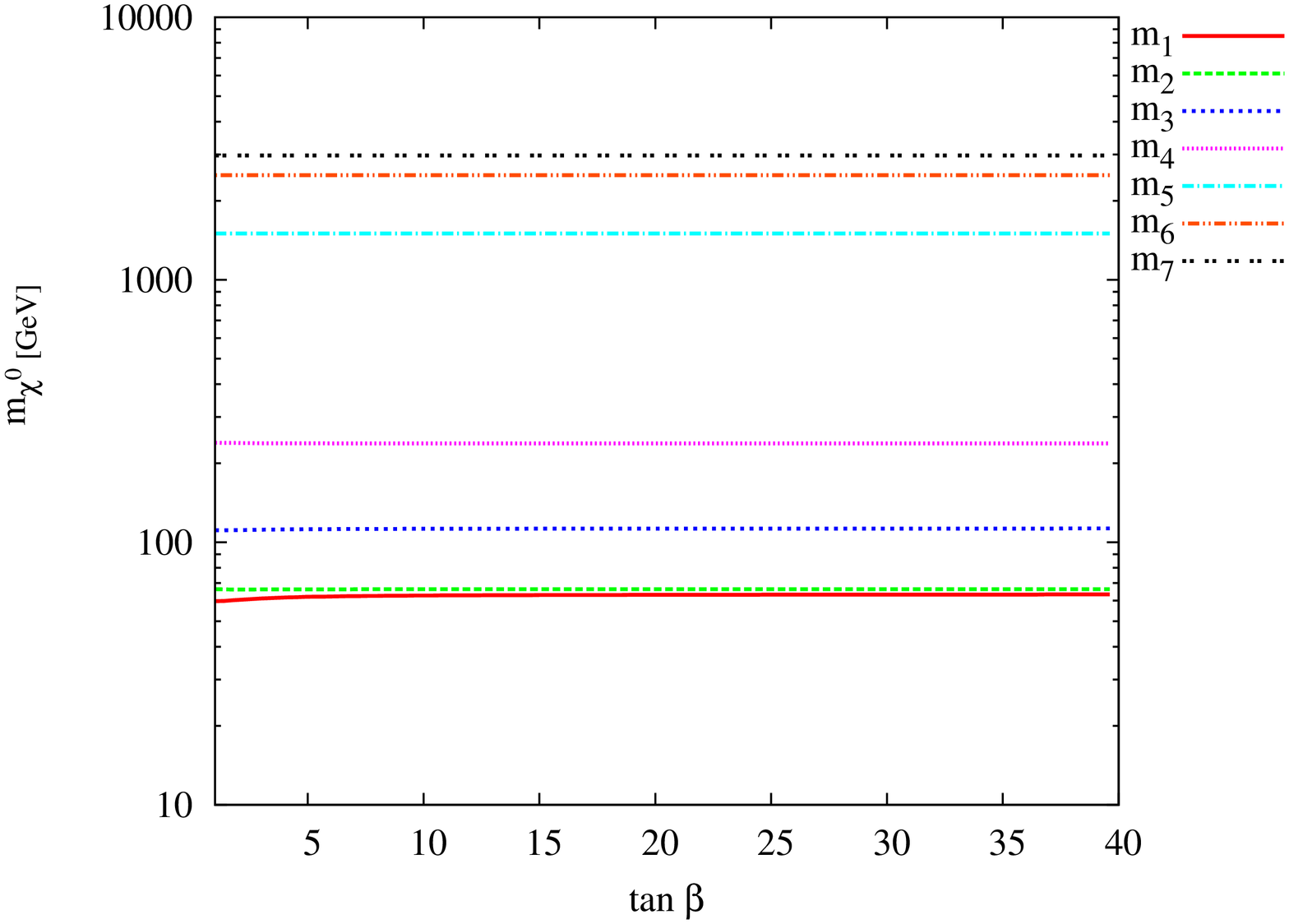}}
\subfigure[\small Squared components of the lightest neutralino]
{\includegraphics[
width=6cm,
angle=-90]{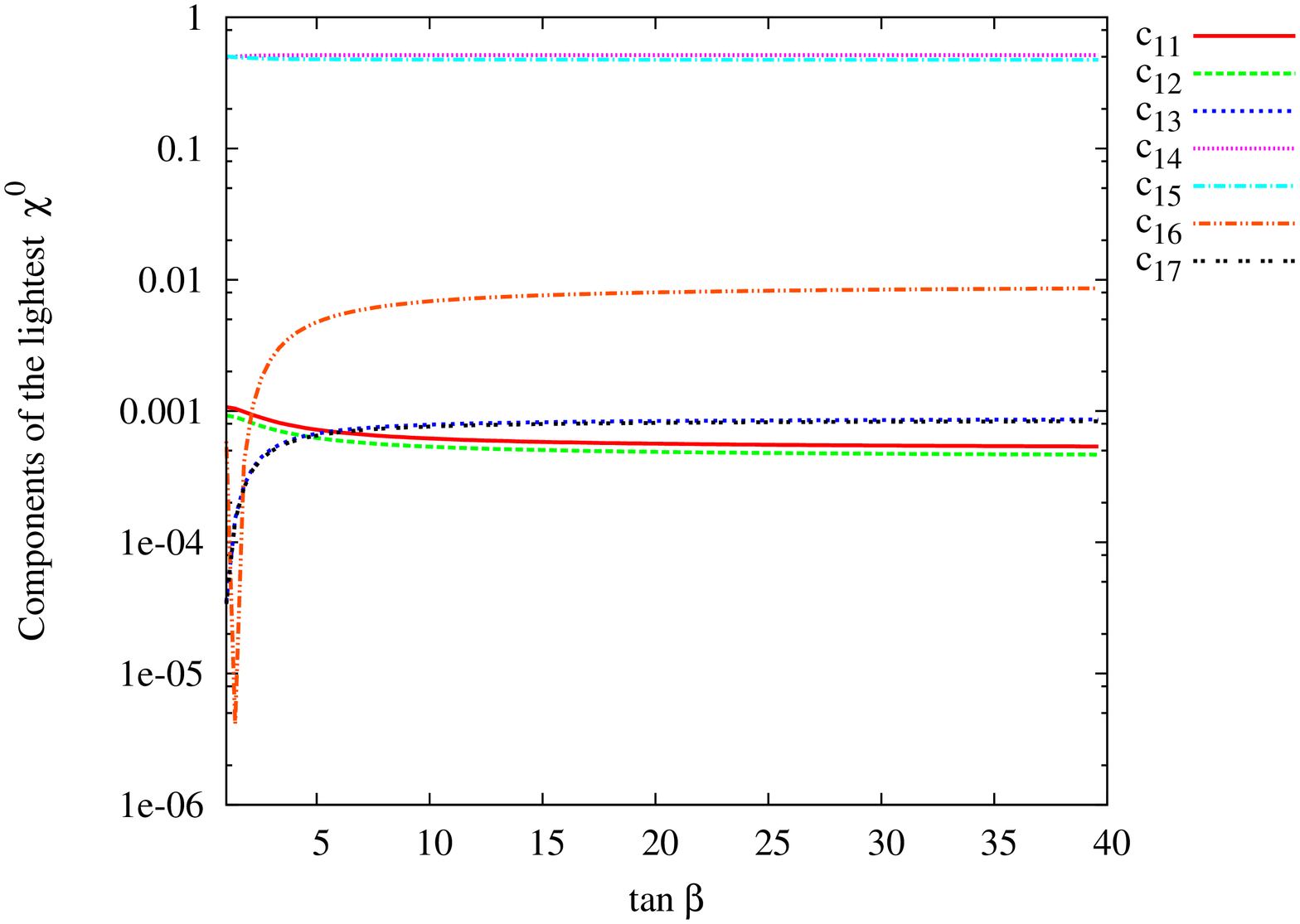}}
\subfigure[\small Eigenvalues as a function of $v_S$]
{\includegraphics[
width=6cm,
angle=-90]{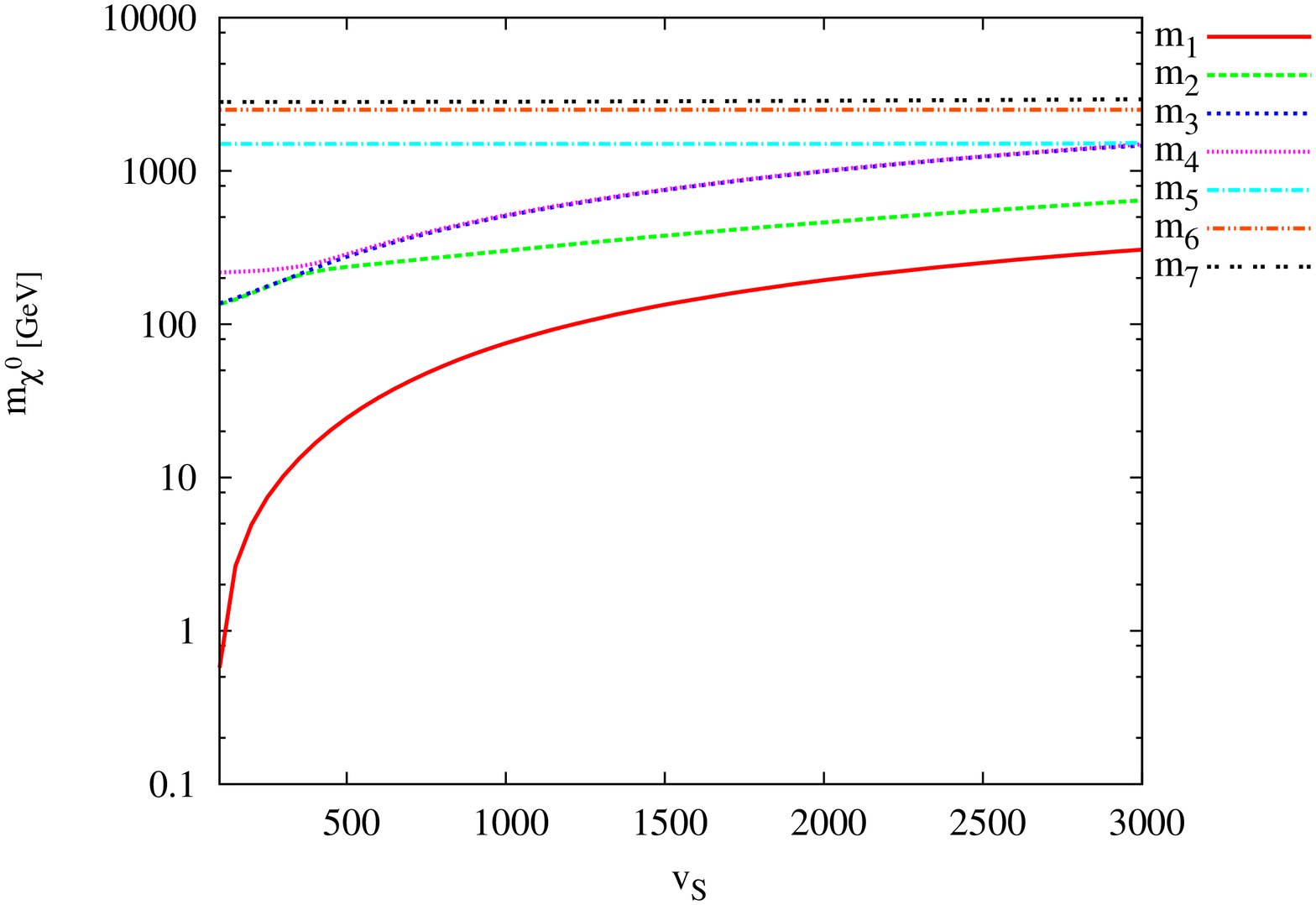}}
\subfigure[\small Squared components of the lightest neutralino]
{\includegraphics[
width=6cm,
angle=-90]{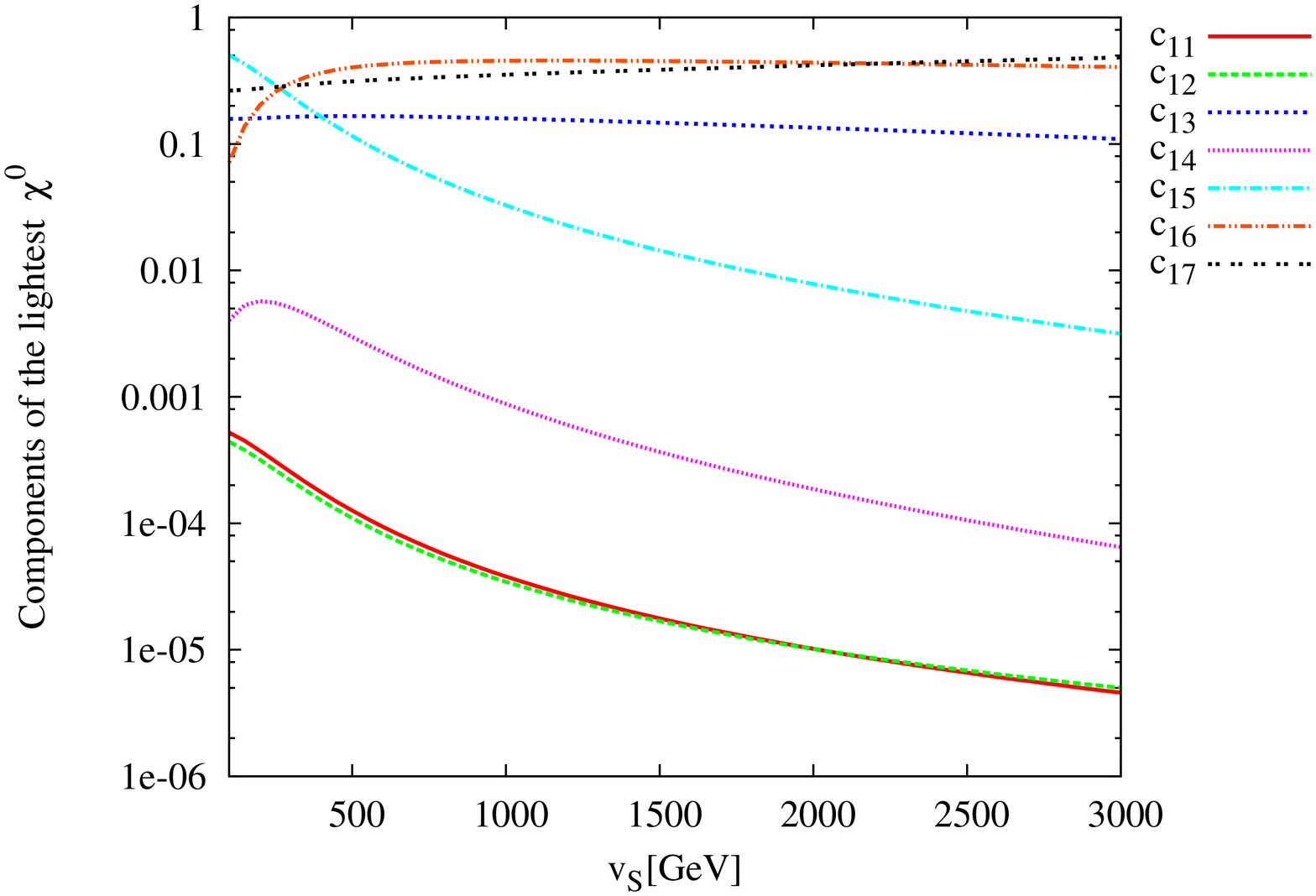}}
\caption{\small The same as Fig.(\ref{neutralino1}) but as a function of $g_B$, $\tan\beta$ and $v_S$.}
\label{neutralino4}
\end{figure}
We have chosen $\tan\beta\approx 40$ and we have constrained the value
of $v_1$ in order to be consistent with the value of the mass of the
$Z_0$ boson, while the value of the coupling constant $g_B$ is $0.65$.

The values $\lambda<0.7$ and $v_S$ around 1 TeV
are consistent with the MSSM value of the Higgs masses.

The charges $B_{H_1}$ and $B_{H_2}$ are free parameters because we have
only four equations coming from the gauge invariance of the superpotential
and eight charges to be constrained.
One possible choice is $B_{H_1}=-3/(2\sqrt{10})$ and $B_{H_2}=-1/(\sqrt{10})$,
which is obtained from the E$_6$SSM model \cite{Kalinowski:2008iq}.

In Figs.\ref{neutralino1}-\ref{neutralino4},
we plot on the left-hand side the numerical value of the neutralino masses
obtained from the diagonalization procedure as a function of the
mass parameters $M_{st},M_B,M_{\bf b},M_Y,Mw$ and of $g_B$ and $\tan\beta$.
On the right-hand side we plot the squared value of each component
of the lightest neutralino state in order to establish which
component is dominant, since every neutralino state appears as
a mixture of the axino, the singlino etc.
We can formally decompose the generic $i$-th neutralino state ($i=1,\dots,7$)
in the basis $\{-i\lambda_{W_3},-i\lambda_{Y},-i\lambda_{B},
\tilde{H}^1_1,\tilde{H}^2_2,\tilde{S},-i\psi_{\bf b} \}$
\ba
\tilde{\chi}^{0}_i=a_{i1}\,\lambda_{W_3}+ a_{i2}\,\lambda_{Y}+ a_{i3}\,\lambda_{B}
+ a_{i4}\,\tilde{H}^1_1 + a_{i5}\,\tilde{H}^2_2 + a_{i6}\,\tilde{S} + a_{i7}\,\psi_{\bf b}
\ea
and in the figures we indicate the square of each component as $c_{ij}=|a_{ij}|^2$,
where the lightest state corresponds to the $i=1$ choice.
From the left panel of Figs.\ref{neutralino1} and \ref{neutralino2}
we observe that the value of the mass
of the lightest neutralino state that is consistent with the current experimental bounds \cite{PDG}
is obtained approximately by varying the values of
$M_{st}$ in the interval $1.7 \div 2.5$ TeV, while $M_B$ and $M_{\bf b}$ in the interval $1 \div 2$ TeV.
In the right panel of Figs.\ref{neutralino1} and \ref{neutralino2}
it is interesting to observe that for these values of the soft breaking parameters
we have a tiny region beyond 1 TeV in which the axino and the B-ino components are
almost coincident, the two higgsinos are dominant, while the singlino is the most suppressed
component.
For values of $M_{st}, M_{B}, M_{\bf b}$ below $1$ TeV and beyond $2.5$ TeV, the lightest
neutralino is ``mostly'' singlino, while the $W$-ino and the $Y$-ino components are suppressed and
the eigenvalues appear to be non-degenerate apart
from the states $\tilde{\chi}^{0}_2-\tilde{\chi}^{0}_3$.
From the left-hand side of Fig. \ref{neutralino3} it is evident that all the eigenvalues
do not exhibit substantial variations with respect to $M_Y,M_{w}$ and the heaviest states are
non degenerate. In both cases (see Fig. \ref{neutralino3} (b), (d)), the singlino component
is the leading one.
A similar feature can be found in the USSM case \cite{Kalinowski:2008iq}, where
the singlino is always dominant with respect to the other components. 

Finally, in Fig. \ref{neutralino4} we have analyzed the dependence upon the coupling constant 
$g_B$, $\tan\beta$ and $v_S$. In the left-hand side (a) the mass value of the lightest state
starts to be greater than $50$ GeV
once $g_B>0.4$ and it is almost degenerate with $\tilde{\chi}^0_2$.

From the analysis of each component
in the right panel (b), for $g_B$ less than $0.5$ the main contribution comes from the singlino,
while the axino and the $B$-ino are almost degenerate and subdominant with respect to the 
$\tilde{H}^2_2$ contribution. 
When $g_B$ becomes greater than $0.5$ we have an inversion: the two Higgsinos are dominant
and almost equal, while the singlino is subleading and the
combination axino-$B$-ino is more suppressed.

As a consequence of our constraint on the vev $v_1$, the eigenvalues dependence on $\tan\beta$
is weak (see Fig. \ref{neutralino4} (c)), while we have a strong impact of low values
of $\tan\beta$ on the axino, $B$-ino and on the singlino components.
Even in this case, with the choice of the parameters that we have made
in Tab. \ref{parameters}, we can identify a small region in which the
contribution of the singlino is highly suppressed.

In the last scenario, represented in Fig. \ref{neutralino4} (c,d), it seems possible to have an axino dominated lightest neutralino. This is achieved with a larger value of the effective $\mu$- term (given by $\lambda v_S$) and a slightly lower one for the axino susy breaking parameter $M_{\bf b}$. 

Given these results, one important issue that one would like 
to address concerns the modifications implied by our model respect 
to standard scenarios of neutralino densities -for instance in the MSSM 
or in the nMSSM - which require a separate investigation of the (rather large) 
parameter space. We just remark that a related analysis \cite{Racioppi}, 
based on an anomalous version of the MSSM which shares various similarities 
with our model, shows that for an axino-dominated LSP (light supersimmetric particle) - 
in the range between 50 GeV - 2 TeV- with a mass gap around 1-5 \% between 
the LSP and the NLSP (next to lightest supersymmetric particle), the constraints 
from WMAP can be satisfied. The NLSP, in that model,  has components which are 
typical of the (non anomalous) MSSM, with a dominant gaugino and/or a gaugino-higgsino 
projection. In the presence of extra singlets and with a physical axion, which is our case, 
this scenario should be  modified even further, but we expect some similarities 
with these previous studies, especially in the neutralino sector, to hold. 
In a recent study of the axion in the MLSOM, for instance, the possibility 
of having the axion as a long lived particle require a very small mass for this 
particle ($\sim 10^{-4}$ eV) \cite{CG}. In the USSM-A the presence of an axion in the 
bosonic sector and of a neutralino in the fermionic sector as possible dark matter 
components raises the issue of the interplay between the two sectors. At the same time, 
in the fermionic neutral sector, the role of the
co-annihilation becomes crucial, especially in the presence of mass degeneracy, which 
modifies substantially the
neutralino relic densities already in this sector. We hope to return with a complete 
analysis of these points in the near future \cite{preparation}

\begin{table}[h]
\begin{center}
\begin{footnotesize}
\begin{tabular}{|c|c|c|c|c|c|c|c|c|c|}
\hline
      & $M_Y$ [TeV] & $M_w$ [TeV] & $M_B$ [TeV] & $M_{st}$ [TeV]& $M_{\bf b}$ [TeV]& $\lambda$ & $v_S$ [TeV]&
         $\tan\beta$ &  $g_B$\\
\hline
Fig. (\ref{neutralino1}) (a,b) &1.5 & 2.5 & 1.6 &  0$\div$ 5 & 1.5 & 0.1 & 0.9 & 40 & 0.65\\
\hline
Fig. (\ref{neutralino2}) (a,b) &1.5 & 2.5 & 0$\div$ 5 & 2 & 1.5 & 0.1 & 0.9 & 40 & 0.65\\
\hline
Fig. (\ref{neutralino2}) (c,d) &1.5 & 2.5 & 1.6 & 2 & 0$\div$ 5 & 0.1 & 0.9 & 40 & 0.65\\
\hline
Fig. (\ref{neutralino3}) (a,b) &0$\div$ 5 & 2.5 & 2.1 & 2 & 1.5 & 0.1 & 0.9 & 40 & 0.65\\
\hline
Fig. (\ref{neutralino3}) (c,d) &1.5 & 5$\div$ 9 & 2.1 & 2 & 1.5 & 0.1 & 0.9 & 40 & 0.65\\
\hline
Fig. (\ref{neutralino4}) (a,b) &1.5 & 2.5 & 1.6 &  2 & 1.5 & 0.1 & 0.9 & 40 & 0.1 $\div$ 1\\
\hline
Fig. (\ref{neutralino4}) (c,d) &1.5 & 2.5 & 1.6 &  2 & 1.5 & 0.1 & 0.9 & 1 $\div$ 40 & 0.65\\
\hline
Fig. (\ref{neutralino4}) (e,f) &1.5 & 2.5 & 1.6 &  2.1 & 1 & 0.7 & 0.1$\div$3 & 40 & 0.65\\
\hline
\end{tabular}
\end{footnotesize}
\end{center}
\caption{\small Parameters for the neutralino eigenvalues analysis
for the charge assignment $B_{H_1}=-3/(2\sqrt{10})$ and $B_{H_2}=-1/(\sqrt{10})$.}
\label{parameters}
\end{table}

\clearpage
\section{Unitarity bound of the model}
Being the theory an effective description of an anomalous Lagrangean  in which the presence of the axion is the low energy signature of a more complicated mechanism of cancellation which would eventually induce higher derivative terms in the effective action, it is necessary at this stage to comment about
the unitarity of this class of models. This point has been raised in \cite{Coriano:2008pg} and further developed in 
\cite{Armillis:2008bg}. One of the most natural contexts for discussing unitarity
is related to $2\rightarrow 2$ processes mediated by
BIM (Bouchiat - Iliopoulos - Meyer) amplitudes,
in particular those involving gluons and photons.
These processes exhibit an anomalous behavior when
the $gg\rightarrow \gamma\gamma$ amplitude
is mediated by the exchange in the $s$-channel of neutral gauge bosons
that couple to the fermion loops via axial-vector interactions.
\begin{figure}[h]
\begin{center}
%\subfigure[\small ]
{\includegraphics[width=8cm, angle=0]{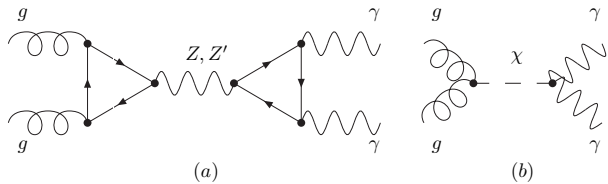}}
\caption{\small BIM amplitude for $gg\rightarrow\g\g$ plus the amplitude
obtained by the exchange of $\chi$.}
\label{unit}
\end{center}
\end{figure}
As shown in these previous analysis, this class of amplitudes, at partonic level,
violate the Froissart bound in the ultraviolet limit.
As a matter of fact, although the Wess-Zumino counterterms are introduced in the Lagrangean
as dimension-5 local operators to ensure the BRST invariance
of the effective action, their contributions to the amplitudes are not sufficient
to cancel the divergent behavior of the anomalous poles which affect the BIM amplitude shown in Fig. \ref{unit} (a).
In the supersymmetric generalization of the model that we have presented,
this issue of unitarity remains basically the same as for the non-supersymmetric case.

As we have discussed above, in the latter case the physical axion appears as a massive degree of freedom
in the CP-odd sector, due to the presence of a Peccei-Quinn breaking term in the
scalar potential. After EWSB
the St\"uckelberg axion $b$ is rotated directly on the physical axion $\chi$ and on the two goldstones $G_Z,G_{Z'}$. 
Therefore, if we choose the unitary gauge, the only diagram that we can draw in order to erase the bad high energy behaviour of 
Fig. \ref{unit} (a) is the second graph (b), where the same amplitude of (a) is mediated by the exchange of the massive axi-Higgs, $\chi$. One can show by a direct study of these two graphs that there is no cancellation of these two contributions at high energy  \cite{Coriano:2008pg}. The problem remains also in the case of the USSM-A model discussed here. We have again a unitarity bound in the supersymmetric case
since the only difference with respect to the non-supersymmetric case
is the contribution of extra fermions circulating in
the loops of the BIM amplitude, in particular the charginos.

\section{Conclusions and Perspectives} 

We have presented a generalization of the USSM in the presence 
of an anomalous $U(1)$ and of a physical axion in the 
CP-odd scalar sector of the theory, model that we call the USSM-A. 
This model, which is a direct generalization of a similar 
construction based on the potential of the MSSM \cite{Anastasopoulos:2008jt}, 
allows higgs-axion mixing. Both constructions are 
extensions of a non-supersymmetric formulation, studied previously 
\cite{Coriano:2005js} (the MLSOM) developed in the context of orientifold vacua of string theory.
In the case of the MLSOM, Higgs-axion mixing has been obtained by requiring that the 
anomalous gauge boson becomes massive by a combination of the Higgs and of the St\"uckelberg mechanisms, with an axion 
that is part of the scalar potential. 
Moving to the supersymmetric case, the generalization of this construction - obtained by using the MSSM superpotential with an extra anomalous $U(1)$ - is found to be characterized by an axino in the spectrum, which appears as a component of the neutralino sector, but not by an axion, since the St\"uckelberg field does not acquire an axion-like coupling and remains a goldstone mode.
The failure of the MSSM superpotential to provide such a mixing has to be attributed 
to the structure of the scalar potential of the model. Supersymmetry 
prohibits a term with a direct presence of the axion in the scalar potential, 
which otherwise would allow such a mixing. 

In our model the mixing occurs indirectly, but the CP-odd sector has to be non-minimal, with an extra singlet which is charged under the anomalous $U(1)$. This approach, as we have emphasized, is quite generic, since its essential working requirement, respect to the MSSM, is the enlargement of the CP-odd sector with one extra SM singlet. Given these minimal requirements, which can be easily satisfied in rather different string vacua, these 
low energy effective theories capture the essential physical implications of several high energy scenarios, either with a low scale string scale or a much higher scale, as in the heterotic case. Explicit formulations of superpotentials, such as those, for instance, derived from free fermionic models \cite{Faraggi:2006qa}, offer the natural ground where to apply the methodology discussed in this work. 

Anomalous $U(1)$'s are quite common in string theory but can also be generated, in the corresponding effective lagrangean, by the decoupling of heavy fermions (and gauge bosons) in grand unified scenarios \cite{CG}. It is then natural to ask what is left at low energy if such decoupling has indeed occurred at some higher scale and it reasonable to foresee that the axion is likely to play a fundamental role \cite{CG} in formulating the answer to this question. Clearly, there are corrections to the action discussed in this work,  which should be characterized by higher derivative contributions (of dimension larger than 5), i.e. beyond the typical Wess-Zumino terms. Arguments in favor of a possible generalization in this direction of the construction presented in this work have been discussed in previous works \cite{CI}
and especially in \cite{CG}; they are motivated by the fact that anomalies cannot be canceled with local counterterms. 

A related issue concerns the size of the mass of the extra $Z'$ in the various models. It is clear that if its decoupling occurs at the Planck scale, then the St\"uckelberg mass term takes approximately 
the value of that decoupling scale. This implies that the axion-like couplings induced at low energy are also heavily suppressed. Other interactions, however, in the non-supersymmetric case, have been found to remain sizeable \cite{CG}. 
 
A final comment concerns supersymmetry breaking, which may induce phase-dependent terms in the potential. As discussed in \cite{Coriano:2005js} for the MLSOM, the axion, in that specific case, gets a sizeable mass which can be as large as the electroweak scale. Similar considerations could remain true in the supersymmetric model that we have presented, although here we have analyzed - by a deliberate choice - the case of a light axion, since we consider this scenario more interesting phenomenologically.   In the presence of these phases the pseudoscalar, however, becomes massive. For instance, a mass region of few GeV's is certainly not excluded, as well as a scenario characterized by a very light axion ($\sim 10^{-4}$ eV), and both can be easily included within our analysis. In particular, for an axion in the GeV mass range, for instance, the interactions of this particle are rather similar to those of a light CP-odd Higgs boson, but now with extra interaction with the gauge fields, due to the anomaly, which are not allowed for the  rest of the CP-odd sector.

\centerline{\bf Acknowledgments}
We thank Nikos Irges, Antonio Racioppi and Elisa Manno for discussions. The work of C.C. was supported in part
by the European Union through the Marie Curie Research and Training Network 
``Universenet'' (MRTN-CT-2006-035863) and by The Interreg II Crete-Cyprus program.

\section{Appendix A: Notations}

In this appendix we specify our notations.

The covariant derivatives are given by
\begin{eqnarray}
\bar{D}_{\dot{A}}=-\bar{\partial}_{\dot{A}}-i\theta^{B}\sigma^{\mu}_{B\dot{A}}\partial_{\mu}\hspace{1cm} 
D_{A}=\partial_{A}+i\sigma^{\mu}_{A\dot{B}}\bar{\theta}^{\dot{B}}\partial_{\mu}.
\label{dersusycov}
\end{eqnarray}
The left/right chiral superfields in terms of field components are given in a generic form as follows
\begin{eqnarray}
\hat{\Phi}_L(x,\theta,\bar{\theta})&=&A(x)+i\theta\sigma^{\mu}\bar{\theta}\partial_{\mu}A(x)
-\frac{1}{4}\theta\theta\bar{\theta}\bar{\theta}\Box A(x)+\sqrt{2}\theta\psi(x)
\nonumber\\
&&+\frac{i}{\sqrt{2}}\theta\theta\sigma^{\mu}\bar{\theta}\partial_{\mu}\psi(x)+\theta\theta F(x),
\label{compchiraleleft}\\
\hat{\Phi}_R^{\dagger}(x,\theta,\bar{\theta})&=&A^{*}(x)
-i\theta\sigma^{\mu}\bar{\theta}\partial_{\mu}A^{*}(x)
-\frac{1}{4}\theta\theta\bar{\theta}\bar{\theta}\Box A^{*}(x)+\sqrt{2}\bar{\theta}\bar{\psi}(x)
\nonumber\\
&&-\frac{i}{\sqrt{2}}\bar{\theta}\bar{\theta}\theta\sigma^{\mu}\partial_{\mu}\bar{\psi}(x)
+\bar{\theta}\bar{\theta}F^{*}(x).
\end{eqnarray}
A generic scalar superfield $\hat{V}$ in the Wess-Zumino gauge is given by
\begin{eqnarray}
\hat{V}(x,\theta,\bar{\theta})&=&\theta\sigma^{\mu}\bar{\theta}
[V_{\mu}(x)-\partial_{\mu}B(x)]+\theta\theta\bar{\theta}\bar{\lambda}(x)
+\bar{\theta}\bar{\theta}\theta\lambda(x)+\theta\theta\bar{\theta}\bar{\theta}d(x)
\end{eqnarray}
where $B(x)$ is a generic real valued scalar field.
\begin{table}
\begin{center}
\begin{tabular}{|l||c|c|c|}
\hline
Superfield & Bosonic & Fermionic & Auxiliary \\
\hline
$\hat{\bf b}(x,\theta,\bar{\theta})$& $b(x)$ & $\psi_{\bf b}(x)$ & $F_{\bf b}(x)$ \\
$\hat{S}(x,\theta,\bar{\theta})$    & $S(x)$ & $\tilde{S}(x)$ & $F_{S}(x)$ \\
$\hat{L}(x,\theta,\bar{\theta})$    & $\tilde{L}(x)$ & $L(x)$ & $F_{L}(x)$\\
$\hat{R}(x,\theta,\bar{\theta})$    & $\tilde{R}(x)$ & $\bar{R}(x)$ & $F_{R}(x)$ \\
$\hat{Q}(x,\theta,\bar{\theta})$    & $\tilde{Q}(x)$ & $Q(x)$ & $F_{Q}(x)$ \\
$\hat{U}_{R}(x,\theta,\bar{\theta})$& $\tilde{U}_R(x)$ & $\bar{U}_R(x)$ & $F_{U_R}(x)$ \\
$\hat{D}_{R}(x,\theta,\bar{\theta})$& $\tilde{D}_R(x)$ & $\bar{D}_R(x)$ & $F_{D_R}(x)$ \\
$\hat{H}_{1}(x,\theta,\bar{\theta})$& $H_1(x)$ & $\tilde{H_1}(x)$ & $F_{H_1}(x)$ \\
$\hat{H}_{2}(x,\theta,\bar{\theta})$& $H_2(x)$ & $\tilde{H_2}(x)$ & $F_{H_2}(x)$ \\
$\hat{B}(x,\theta,\bar{\theta})$    & $B_{\mu}(x)$ & $\lambda_{B}(x),\bar{\lambda}_{B}(x)$ & $D_B(x)$ \\
$\hat{Y}(x,\theta,\bar{\theta})$    & $A^{Y}_{\mu}(x)$ & $\lambda_{Y}(x),\bar{\lambda}_{Y}(x)$ & $D_Y(x)$ \\
$\hat{W}^{i}(x,\theta,\bar{\theta})$& $W^{i}_{\mu}(x)$ & $\lambda_{W^{i}}(x),\bar{\lambda}_{W^{i}}(x)$ & $D_{W^{i}}(x)$ \\
$\hat{G}^{a}(x,\theta,\bar{\theta})$& $G^{a}_{\mu}(x)$ & $\lambda_{g^a}(x),\bar{\lambda}_{g^a}(x)$ & $D_{G^{a}}(x)$ \\
\hline
\end{tabular}
\end{center}
\label{tab2}
\caption{Superfields and their components.}
\end{table}
The generic expressions for the field-strengths are 
\begin{eqnarray}
W^{Y}_{\alpha}&=&-\frac{1}{4}\bar{D}\bar{D}D_{\alpha}\hat{Y},
\nonumber\\ 
W^{B}_{\alpha}&=&-\frac{1}{4}\bar{D}\bar{D}D_{\alpha}\hat{B},
\nonumber\\ 
W_{\alpha}&=&-\frac{1}{8 g_2}\bar{D}\bar{D}e^{-2g_2\hat{W}}D_{\alpha}e^{2g_2\hat{W}},
\nonumber\\ 
{\cal G}_{\alpha}&=&-\frac{1}{8 g_s}\bar{D}\bar{D}e^{-2g_s\hat{G}}D_{\alpha}e^{2 g_s\hat{G}}
\label{fieldstrength}
\end{eqnarray}
where we have used $\hat{W}=\tau^{i}\hat{W}^{i}$ with $\tau^{i}$ being the $SU(2)$ generators, 
while $\hat{G}=T^{a}\hat{G}^{a}$ with $T^{a}$ being the $SU(3)$ generators.
The non supersymmetric field-strength are defined as
\ba
F^{Y}_{\mu\nu}&=&\partial_{\mu}A^{Y}_{\nu}-\partial_{\nu}A^{Y}_{\mu},
\nonumber\\
F^{B}_{\mu\nu}&=&\partial_{\mu}B_{\nu}-\partial_{\nu}B_{\mu},
\nonumber\\
W_{\mu\nu}^{i}&=&\partial_{\mu}W^{i}_{\nu}-\partial_{\nu}W^{i}_{\mu}
-g_2\varepsilon^{ijk}W^{j}_{\mu}W^{k}_{\nu}
\nonumber\\
G_{\mu\nu}^{a}&=&\partial_{\mu}G^{a}_{\nu}-\partial_{\nu}G^{a}_{\mu}
-g_s f^{a b c}G^{b}_{\mu}G^{c}_{\nu}
\ea

\section*{Appendix B: The USSM Lagrangean}

For completeness we introduce in what follows 
the USSM Lagrangean that is a part of the total Lagrangean given by 
${\cal L}_{Tot}={\cal L}_{USSM}+{\cal L}_{axion}+{\cal L}_{CS}$.

\ba
{\cal L}_{USSM}={\cal L}_{lep}+{\cal L}_{quark}+{\cal L}_{Higgs}+{\cal L}_{gauge}+
{\cal L}_{SMT}+{\cal L}_{GMT}
\ea

\begin{eqnarray}
&&{\cal L}_{lep}=\int{d^{4}\theta\left[\hat{L}^{\dagger}e^{2g_2\hat{W}
+g_Y\hat{Y}+g_B\hat{B}}\hat{L}+\hat{R}^{\dagger}e^{2g_2\hat{W}
+g_Y\hat{Y}+g_B\hat{B}}\hat{R}\right]}\\
&&{\cal L}_{quark}=\int d^{4}\theta\left[\hat{Q}^{\dagger}e^{2g_s\hat{G}+2g_2\hat{W}
+g_Y\hat{Y}+g_B\hat{B}}\hat{Q}+\hat{U}_{R}^{\dagger}e^{2g_s\hat{G}
+g_Y\hat{Y}+g_B\hat{B}}\hat{U}_{R}+\hat{D}_{R}^{\dagger}e^{2g_s\hat{G}
+g_Y\hat{Y}+g_B\hat{B}}\hat{D}_{R}\right]
\nonumber\\
\\
&&{\cal L}_{Higgs}=\int{d^{4}\theta\left[\hat{H}_{1}^{\dagger}e^{2g_2\hat{W}+g_Y\hat{Y}
+g_B\hat{B}}\hat{H}_{1}+\hat{H}_{2}^{\dagger}e^{2g_2\hat{W}+g_Y\hat{Y}
+g_B\hat{B}}\hat{H}_{2}+\hat{S}^{\dagger}e^{g_B\hat{B}}\hat{S}
+{\cal W}\delta^{2}(\bar{\theta})+\bar{{\cal W}}\delta^{2}(\theta)\right]}\nonumber\\
\\
%{\cal L}_{S}&=&\int{d^{4}\theta\hat{S}^{\dagger}e^{g_B\hat{B}}\hat{S}}\\
&&{\cal L}_{gauge}=\frac{1}{4}\int{d^{4}\theta\left[{\cal G}^{\alpha}{\cal G}_{\alpha}+W^{\alpha}W_{\alpha}+W^{Y\alpha}W^{Y}_{\alpha}
+W^{B\alpha}W^{B}_{\alpha}\right]\delta^{2}(\bar{\theta})+h.c.}
\\\nonumber\\
&&{\cal L}_{SMT}=-\int d^{4}\theta\,\delta^{4}(\theta,\bar{\theta})\,[M^{2}_{L}\hat{L}^{\dagger}\hat{L}
+m^{2}_{R}\hat{R}^{\dagger}\hat{R}+M^{2}_{Q}\hat{Q}^{\dagger}\hat{Q}
+m^{2}_{U}\hat{U}_{R}^{\dagger}\hat{U}_{R}+m^{2}_{D}\hat{D}_{R}^{\dagger}\hat{D}_{R}
\nonumber\\
&&\hspace{2.5cm}+m_{1}^{2}\hat{H}_{1}^{\dagger}\hat{H}_{1}
+m_{2}^{2}\hat{H}_{2}^{\dagger}\hat{H}_{2}+m_{S}^{2}\hat{S}^{\dagger}\hat{S}
+(a_{\lambda}\hat{S}\hat{H}_{1}\cdot\hat{H}_{2}+h.c.)+(a_{e}\hat{H}_{1}\cdot\hat{L}\hat{R}+h.c.)
\nonumber\\
&&\hspace{2.5cm}+(a_{d}\hat{H}_{1}\cdot\hat{Q}\hat{D}_{R}+h.c.)
+(a_{u}\hat{H}_{2}\cdot\hat{Q}\hat{U}_{R}+h.c.)]
\\\nonumber\\
&&{\cal L}_{GMT}=\int d^{4} \theta \left[ \frac{1}{2}   
\left(M_{G}{\cal G}^{\alpha}{\cal G}_{\alpha}
+ M_{w}W^{\alpha }W_{\alpha} + M_YW^{Y\alpha} W^{Y}_{\alpha} 
+ M_B W^{B\alpha} W^{B}_{\alpha}   \right)
+h.c.\right] \delta^{4}(\theta,\bar{\theta})
\nonumber\\
\end{eqnarray}

\section*{Appendix C: The $O^{\chi}$ matrix}

\begin{eqnarray}
O^{\chi}_{11}&=&\frac{v_{2}v_{S}}{\sqrt{v_{1}^{2}v_{2}^{2}+v_{S}^{2}v^{2}}},
\nonumber\\
O^{\chi}_{12}&=&\frac{v_{1}v_{S}}{\sqrt{v_{1}^{2}v_{2}^{2}+v_{S}^{2}v^{2}}},
\nonumber\\
O^{\chi}_{13}&=&\frac{v_{1}v_{2}}{\sqrt{v_{1}^{2}v_{2}^{2}+v_{s}^{2}v^{2}}},
\nonumber\\
O^{\chi}_{14}&=&0,
\nonumber\\
O^{\chi}_{21}&=&-\frac{v_1 (f_1-2 B_{H_{1}} g_B x_B+\sqrt{f_1^2+4g^2x_B^2})}{2x_B}
\sqrt{\frac{1}{8}-\frac{f_1}{8\sqrt{f_1^2+4 g^2 x_B^2}} },
\nonumber\\
O^{\chi}_{22}&=&\frac{v_2 (f_1+2 B_{H_{2}} g_B x_B+\sqrt{f_1^2+4g^2x_B^2})}{2x_B}
\sqrt{\frac{1}{8}-\frac{f_1}{8\sqrt{f_1^2+4 g^2 x_B^2}} },
\nonumber\\
O^{\chi}_{23}&=&B_S g_B v_S
\sqrt{\frac{1}{8}-\frac{f_1}{8\sqrt{f_1^2+4 g^2 x_B^2}} },
\nonumber\\
O^{\chi}_{24}&=&2 M_{st}
\sqrt{\frac{1}{8}-\frac{f_1}{8\sqrt{f_1^2+4 g^2 x_B^2}} },
\nonumber\\
O^{\chi}_{31}&=&-\frac{v_1 (f_1-2 B_{H_{1}} g_B x_B-\sqrt{f_1^2+4g^2x_B^2})}{2x_B}
\sqrt{\frac{1}{8}+\frac{f_1}{8\sqrt{f_1^2+4 g^2 x_B^2}} },
\nonumber\\
O^{\chi}_{32}&=&\frac{v_2 (f_1+2 B_{H_{2}} g_B x_B-\sqrt{f_1^2+4g^2x_B^2})}{2x_B}
\sqrt{\frac{1}{8}+\frac{f_1}{8\sqrt{f_1^2+4 g^2 x_B^2}} },
\nonumber\\
O^{\chi}_{33}&=&B_S g_B v_S
\sqrt{\frac{1}{8}+\frac{f_1}{8\sqrt{f_1^2+4 g^2 x_B^2}} },
\nonumber\\
O^{\chi}_{34}&=&2 M_{st}
\sqrt{\frac{1}{8}+\frac{f_1}{8\sqrt{f_1^2+4 g^2 x_B^2}} },
\nonumber\\
O^{\chi}_{41}&=&\frac{2 M_{st} v_1 v_2^2}
{\sqrt{(v_1^2 v_2^2+ v^2 v_S^2)[B_S^2 g_B^2(v_1^2 v_2^2+ v^2 v_S^2)+4 M_{st}^2 v^2]}},
\nonumber\\
O^{\chi}_{42}&=&\frac{2 M_{st} v_2 v_1^2}
{\sqrt{(v_1^2 v_2^2+ v^2 v_S^2)[B_S^2 g_B^2(v_1^2 v_2^2+ v^2 v_S^2)+4 M_{st}^2 v^2]}},
\nonumber\\
O^{\chi}_{43}&=&-\frac{2 M_{st} v_S v^2}
{\sqrt{(v_1^2 v_2^2+ v^2 v_S^2)[B_S^2 g_B^2(v_1^2 v_2^2+ v^2 v_S^2)+4 M_{st}^2 v^2]}},
\nonumber\\
O^{\chi}_{44}&=&\frac{B_S g_B\sqrt{v_1^2 v_2^2+ v^2 v_S^2}}
{\sqrt{B_S^2 g_B^2(v_1^2 v_2^2+ v^2 v_S^2)+4 M_{st}^2 v^2}}.
\end{eqnarray}

\bibliographystyle{h-elsevier3}

%\bibliography{marconi_21_11}
\end{document}